\pdfoutput=1
\documentclass[aip,jcp,preprint,amsmath,amssymb,floatfix,nofootinbib]{revtex4-2}
\usepackage{bm}
\usepackage{booktabs}
\usepackage{array}
\usepackage{enumitem}
\usepackage{xcolor}
\usepackage{graphicx}
\usepackage[export]{adjustbox}   
\usepackage{tikz}   
\graphicspath{{figures/pdf/}{figures/png/}}
\usepackage{algorithm}
\usepackage{algpseudocode}
\usepackage{hyperref}
\usepackage[capitalise]{cleveref}   

\usepackage{placeins}
\makeatletter
\AtBeginDocument{%
  \let\LMRD@section\section
  \renewcommand{\section}{\FloatBarrier\LMRD@section}%
  \let\LMRD@subsection\subsection
  \renewcommand{\subsection}{\FloatBarrier\LMRD@subsection}%
}
\makeatother

\newcommand{\kT}{k_{\!B}T}

\newcommand{\Oh}{\mathcal{O}}

\newcommand{\Fex}{\mathcal{F}^{\mathrm{ex}}}

\newcommand{\dd}{\,\mathrm{d}}

\newcommand{\sT}{\sigma_{\mathrm{T}}}
\newcommand{\sB}{\sigma_{\mathrm{B}}}
\newcommand{\DT}{D_{\mathrm{T}}}
\newcommand{\DB}{D_{\mathrm{B}}}
\newcommand{\DL}{D_{\mathrm{L}}}
\newcommand{\phiB}{\phi_{\mathrm{B}}}
\newcommand{\rhoB}{\rho_{\mathrm{B}}}

\algrenewcommand\algorithmicrequire{\textbf{Init:}}
\algrenewcommand\algorithmicensure{\textbf{Loop:}}
\algnewcommand\algorithmicparfor{\textbf{parfor}}
\algnewcommand\algorithmicendparfor{\textbf{end}}
\algdef{SE}[PARFOR]{ParFor}{EndParFor}[1]{\algorithmicparfor\ #1\ \algorithmicdo}{\algorithmicendparfor}
\algrenewcommand\algorithmiccomment[1]{\hfill\textcolor{gray}{$\triangleright$\ \textit{#1}}}

\makeatletter
\def\SItocHide{\let\SI@lsec\l@section\let\SI@lsub\l@subsection\let\SI@lsubsub\l@subsubsection
  \let\l@section\@gobble@tw@\let\l@subsection\@gobble@tw@\let\l@subsubsection\@gobble@tw@}
\def\SItocShow{\let\l@section\SI@lsec\let\l@subsection\SI@lsub\let\l@subsubsection\SI@lsubsub}
\makeatother
\begin{document}
\makeatletter\immediate\write\@auxout{\string\@writefile{toc}{\string\setcounter{tocdepth}{-10}\string\SItocHide}}\makeatother

\title{A free-energy approach for stochastic reaction-drift-diffusion lattice simulations of crowded biological systems}

\author{Ron Michael Acda}
\affiliation{NSF Science and Technology Center for Quantitative Cell Biology, University of Illinois at Urbana-Champaign, Urbana, IL, USA}
\affiliation{Center for Biophysics and Quantitative Biology, University of Illinois at Urbana-Champaign, Urbana, IL, USA}

\author{Alfia Parvez}
\affiliation{NSF Science and Technology Center for Quantitative Cell Biology, University of Illinois at Urbana-Champaign, Urbana, IL, USA}

\author{Abner T. Apsley}
\affiliation{NSF Science and Technology Center for Quantitative Cell Biology, University of Illinois at Urbana-Champaign, Urbana, IL, USA}

\author{Hyun Youk}
\email{youk@illinois.edu}
\affiliation{NSF Science and Technology Center for Quantitative Cell Biology, University of Illinois at Urbana-Champaign, Urbana, IL, USA}
\affiliation{Center for Biophysics and Quantitative Biology, University of Illinois at Urbana-Champaign, Urbana, IL, USA}
\affiliation{Department of Physics, University of Illinois at Urbana-Champaign, Urbana, IL, USA}
\affiliation{The Grainger College of Engineering, University of Illinois at Urbana-Champaign, Urbana, IL, USA}

\author{Zaida Luthey-Schulten}
\email{zan@illinois.edu}
\affiliation{NSF Science and Technology Center for Quantitative Cell Biology, University of Illinois at Urbana-Champaign, Urbana, IL, USA}
\affiliation{Department of Physics, University of Illinois at Urbana-Champaign, Urbana, IL, USA}
\affiliation{The Grainger College of Engineering, University of Illinois at Urbana-Champaign, Urbana, IL, USA}
\affiliation{Beckman Institute for Advanced Science and Technology, University of Illinois at Urbana-Champaign, Urbana, IL, USA}
\affiliation{Department of Chemistry, University of Illinois at Urbana-Champaign, Urbana, IL, USA}

\date{\today}

\begin{abstract}
Molecular size alters spatial distributions, reaction equilibria, and transport in crowded mixtures,
but resolving individual particle encounters can make large stochastic simulations computationally
costly. We developed a lattice-based framework for simulating reaction--drift--diffusion systems with volume exclusion captured as a free energy functional in each voxel. Changes in this free
energy, together with external one-body potentials, modify diffusion and reaction propensities. Our framework accommodates mixtures of particle sizes without
resolving each particle's position within a voxel. We implemented our framework in an operator splitting scheme and compared our predictions with off-lattice
interacting-particle simulations. Inclusion of the local free energy functional improved spatial
occupancies and shifts in reaction equilibria relative to point-particle simulations. With separate approximations for larger spheres and polymers, we simulated crowding-induced slowing and spatial
segregation. We then modeled motor proteins' traffic jams on a structured track representing a microtubule. Our results demonstrate how local composition can inform stochastic reaction and transport rates without resolving individual encounters.
\end{abstract}

\maketitle

\section{Introduction}
Molecular sizes affect the behavior of crowded mixtures. Inserting a molecule or changing its size
through a reaction changes the space available to its neighbors. Even purely repulsive interactions
can alter chemical equilibria, spatial distributions, and
transport.\citep{minton1981excluded,minton2001influence,zhou2008macromolecular} Capturing these
effects becomes challenging when simulations must follow large molecular populations and many
reaction channels.\citep{roberts2013lattice,hallock2014simulation,thornburg2026bringing} How much
detail must a model retain to describe crowding without tracking every molecular encounter?

The reaction--diffusion master equation (RDME) offers a natural starting point. It divides space
into subvolumes (voxels), tracks molecular populations within each one, and represents diffusion
as hops between neighboring subvolumes.\citep{roberts2013lattice,isaacsonpeskin2006} However, it does not account for particle-size-based volume exclusion on its own. Earlier lattice models addressed this limitation through explicit exclusion or restrictions on site occupancy (e.g., upper
bound on occupancy numbers), which changed reaction equilibria and population
fluctuations.\citep{arjunan2010new,cianci2016molecular} One could also mimic volume exclusion in the RDME by augmenting it with 
hook-based routines in moving particles larger than a subvolume and imposing diffusion rules, as done for ribosomes in a whole-cell model of a minimal bacterial cell. \citep{thornburg2026bringing}
Physics-based approaches using scaled particle theory used molecular sizes and local composition to estimate insertion probabilities and describe diffusion in crowded,
including spatially heterogeneous, environments.\citep{cianci2017capturing,smithcianci2017crowding}
These motivate a practical question of how we can incorporate molecular size into both reaction and
transport of mixtures of differently-sized particles.

Reversible reactions add further challenges because a model must connect its reaction rules to the
equilibrium distribution it intends to sample. Previous free-energy-based descriptions related
reaction kinetics to non-ideal chemical potentials,\citep{lutsko2016nonideal,miangolarra2023nonideal}
while interacting-particle methods resolved pair interactions and used reaction proposals and
acceptance rules designed to preserve a specified equilibrium
distribution.\citep{schoneberg2013readdy,frhner2018reversible,hoffmann2019readdy2} Convergent
lattice reaction--drift--diffusion models incorporated finite-range reactions in external
potentials,\citep{isaacsonzhang2024crddme} and a recent extension included explicit pair
interactions.\citep{heldman2026crddme} These approaches highlight the need to specify the reaction
rates, proposal probabilities, and statistical weights of molecular count states alongside any
energetic acceptance factor.

Here we present our computationally efficient framework that links reaction and transport rates to
the excess free energy of a multicomponent hard-sphere fluid. We assigned each voxel the homogeneous
bulk limit of a free-energy functional from fundamental measure theory (FMT), evaluated at its current
composition.\citep{rosenfeld1989free,roth2002whitebear,mansoori1971equilibrium} Removing a particle
from one voxel and inserting it into another changes this free energy as does changing the species
populations through a reaction. We used these energy differences to modify 
diffusion and reaction propensities without calculating individual collisions. Our approximation retains the effects of particle size
and local composition at finite voxel resolution, while omitting the spatially nonlocal density
structure of full FMT.

\begin{figure}[htbp]
    \centering
    \includegraphics[max width=\linewidth,max height=0.80\textheight]{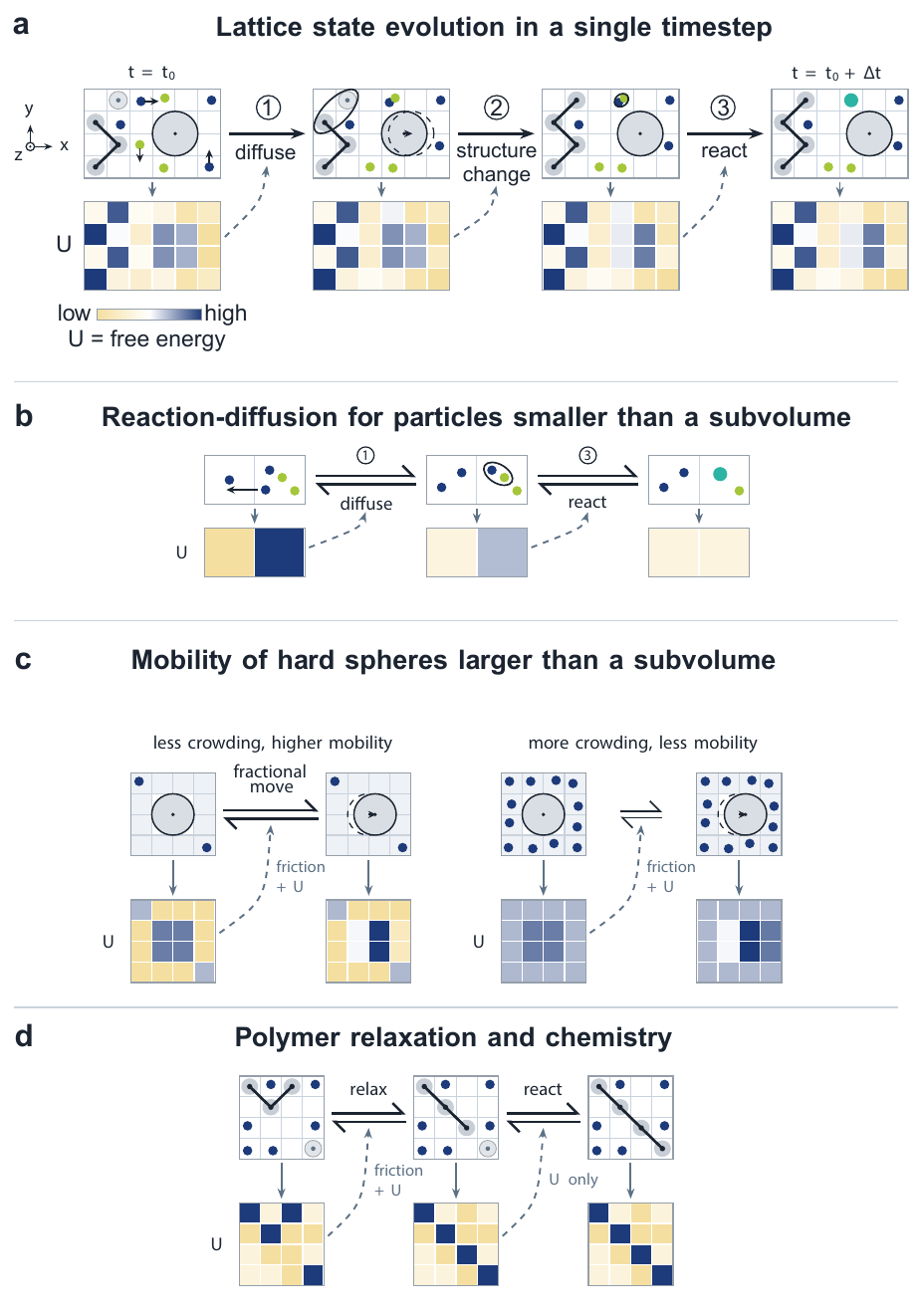}
    \caption{Every operator reads the free-energy landscape $U$ that the current lattice state
    generates (grids, light for low and dark for high $U$); dashed arrows point to the reading operator.
    (a) One timestep. (b) Particles smaller than a subvolume. (c) Spheres larger than a subvolume:
    crowding captured by a friction term. (d) Polymers: relaxation reads friction term and $U$, chemistry reads $U$ only.}
    \label{fig:schematic}
\end{figure}

We implemented our construction in an extension of Lattice Microbes,\citep{roberts2013lattice}
combining Wang--Peskin--Elston drift rates\citep{wang2003robust,mauro2014first} with operator-split
diffusion and reaction updates (see Fig.~\ref{fig:schematic}). We tested population distributions, shifts in reaction equilibria, and
selected dynamical observables against interacting-particle simulations. To represent lattice polymers and spheres larger
than a subvolume, we introduced separate mobility approximations captured by a crowding-induced friction term. We
then explored polymer segregation and biological motor traffic as applications. Together, our
comparisons and demonstrations identified what the framework captured, where it improved on
point-particle models, and which observables require further refinement.

\section{Methods}

\subsection{Reaction and diffusion propensities on a Cartesian lattice}\label{sec:rddme}

We divided the domain into cubic subvolumes of side length $h$ and represented the system state $\mathbf
X$ by the counts $x_v^s$ of every species $s$ in each subvolume $v$. We confined reactions to individual
subvolumes (i.e., voxels), with each reaction having baseline reaction propensities
$a_v^r(\mathbf X)$ dependent on the state $\mathbf{X}$. We assumed that each voxel was sufficiently well-mixed so that we could use these
local propensities uniformly within a voxel.\citep{gillespie1977exact,roberts2013lattice}

We denoted $\boldsymbol\nu_{vr}$ to denote reaction stoichiometry of reaction $r$ in subvolume $v$.
We used $\boldsymbol\nu^s_{vw}$ for the change that removes one particle of species $s$ from subvolume $v$
and adds it to a neighboring subvolume $w$. A continuous-time jump process with reaction propensity
$\widetilde a_v^r$ and per-particle diffusion propensity $d_{vw}^s$ obeys the master equation

\begin{equation}
 \partial_t P(\mathbf X,t)=(\mathcal R+\mathcal D)P(\mathbf X,t),
\end{equation}

where

\begin{align}
 \mathcal R P(\mathbf X,t)&=\sum_{v,r}\big[
 \widetilde a_v^r(\mathbf X-\boldsymbol\nu_{vr})P(\mathbf X-\boldsymbol\nu_{vr},t)
 -\widetilde a_v^r(\mathbf X)P(\mathbf X,t)\big],\label{eq:rxnOperator}\\
 \mathcal D P(\mathbf X,t)&=\sum_{v,w,s}\big[
 d_{vw}^s(\mathbf X-\boldsymbol\nu^{s}_{vw})(x_v^s+1)
 P(\mathbf X-\boldsymbol\nu^{s}_{vw},t)\nonumber\\
 &\hspace{42mm}-d_{vw}^s(\mathbf X)x_v^sP(\mathbf X,t)\big].
 \label{eq:driftDiffOperator}
\end{align}

We only considered diffusion hops for the six nearest-neighbor subvolumes  in Eq.~\eqref{eq:driftDiffOperator}, so the destination subvolume $w$ has a displacement of $\pm \hat{x}, \pm \hat{y}, \pm \hat{z}$ along the Cartesian directions. We retained the full state in the rate expressions because
a hop rate can depend on the populations at both endpoints.

For species represented by subvolume counts, we defined the dimensionless energetic correction

\begin{equation}
 U(\mathbf X,t)=\sum_{v,s}x_v^s\Phi_s^{\mathrm{ext}}(\mathbf r_v,t)
                  +\sum_v\beta F_v(\mathbf x_v),\qquad \beta=(k_BT)^{-1},
 \label{eq:totalU}
\end{equation}

where $\mathbf r_v$ is the voxel's centroid, $\Phi_s^{\mathrm{ext}}$ is a one-body potential energy in units of
$k_BT$, and $F_v$ is the excess free energy defined in the next section.

We used the Wang--Peskin--Elston rate\citep{wang2003robust,mauro2014first}

\begin{equation}
 d_{vw}^s(\mathbf X)=\frac{D_{vw}^s}{h^2}\mathcal B(\Delta U_{vw}^s),
 \qquad \Delta U_{vw}^s=U(\mathbf X+\boldsymbol\nu^{s}_{vw})-U(\mathbf X),
 \label{eq:diffusionPropensity}
\end{equation}

with

\begin{equation}
 \mathcal B(u)=\begin{cases}u/(e^u-1),&u\ne0,\\1,&u=0.\end{cases}
 \label{eq:wpe-factor}
\end{equation}

In the absence of excess interactions,

\begin{equation}
 \Delta U_{vw}^s=\Phi_s^{\mathrm{ext}}(\mathbf r_w)-\Phi_s^{\mathrm{ext}}(\mathbf r_v).
 \label{eq:hopEnergyOneBody}
\end{equation}

The identity $\mathcal B(u)/\mathcal B(-u)=e^{-u}$ gives the energetic factor in the ratio of forward and
reverse rates. To establish detailed balance, the prefactor must also remain unchanged when the
state transition reverses. The function $\mathcal B$ multiplies the rate and exceeds one for downhill
moves. It therefore does not serve as an acceptance probability.

For each reaction proposal, we defined the energy change and acceptance factor, respectively, as

\begin{equation}
 \Delta U_{vr}=U(\mathbf X+\boldsymbol\nu_{vr})-U(\mathbf X),\qquad
 A_{vr}(\mathbf X)=\min(1,e^{-\Delta U_{vr}}).
 \label{eq:reactionEnergyChange}
\end{equation}

Thinning baseline reaction proposals with this acceptance factor gives the rate relation
$\widetilde a_v^r=a_v^r A_{vr}$. Our finite-step sampler tested at most one reaction candidate per
subvolume (Algorithm~\ref{alg:rddme}) and assigned the probability

\begin{equation}
 q_{vr}(\mathbf X;\Delta t)=\frac{a_v^r(\mathbf X)}{a_{v,\mathrm{tot}}(\mathbf X)}
 [1-e^{-a_{v,\mathrm{tot}}(\mathbf X)\Delta t}]A_{vr}(\mathbf X),
 \label{eq:effectiveReactionPropensity}
\end{equation}

where $a_{v,\mathrm{tot}}=\sum_r a_v^r$.

For a reversible reaction pair, the acceptance factors at the two connected states (i.e., product and reactant state counts) obey

\begin{equation}
 \frac{A_{vr}(\mathbf X)}{A_{v\bar r}(\mathbf X+\boldsymbol\nu_{vr})}
 =e^{-\Delta U_{vr}}.\label{eq:acceptanceRatio}
\end{equation}

If the baseline forward and reverse rates satisfy detailed balance with respect to $\pi_0(\mathbf X)$,
the corrected continuous-time rates satisfy detailed balance with respect to $\pi(\mathbf
X)\propto\pi_0(\mathbf X)e^{-U(\mathbf X)}$ for a time-independent $U$. The baseline measure includes the
count-state combinatorial factors and chemical equilibrium constants. This result applies to the
continuous-time rates under the stated conditions. It does not by itself establish detailed balance for
the finite-step update with split operators or for the separate large-particle and polymer extensions.

\subsection{A local free energy defined via bulk hard-sphere approximation}\label{sec:fmt}

For a voxel with species counts $N_s$ and hard-sphere diameters $\sigma_s$, we defined the total particle
count as $N=\sum_s N_s$. We introduced the following quantities

\begin{equation}
 M_n=\frac{1}{N}\sum_s N_s\sigma_s^n,\qquad
 \eta=\frac{\pi}{6h^3}\sum_sN_s\sigma_s^3,\qquad
 \lambda=\frac{M_1M_2}{M_3},\quad \gamma=\frac{M_1^3}{M_3},
 \label{eq:fmtMoments},
\end{equation}
which correspond to the composition moment, crowder density, and two dimensionless measures of size polydispersity, respectively. 

We assigned the homogeneous bulk limit of the White Bear fundamental measure functional to each subvolume
separately.\citep{roth2002whitebear} Under this functional, the excess free energy is

\begin{align}
 \beta F_v(\{N_s\})=N\bigg\{&-\ln(1-\eta)+\frac{3\eta}{1-\eta}\lambda\nonumber\\
 &+\left[\frac{\eta}{(1-\eta)^2}+\ln(1-\eta)\right]\frac{\lambda^3}{\gamma}\bigg\}.
 \label{eq:betaFV}
\end{align}

For an empty subvolume, we set $F_v=0$. For a nonempty mixture with positive particle volumes, the
expression applies when $\eta<1$ and becomes singular as $\eta\to1$. For a single species, the excess
free energy per particle reduces to the Carnahan--Starling expression
$\eta(4-3\eta)/(1-\eta)^2$.\citep{carnahan1969equation} For a mixture, it gives the
Boublík--Mansoori--Carnahan--Starling--Leland bulk equation of state.\citep{mansoori1971equilibrium}
Section~\ref{sec:si-fmt} provides the derivation.

For each hop, we removed a particle from its source and inserted it at its destination. Denoting $\mathbf{N}_m$ as the occupancy vector at subvolume $m$, and $\mathbf{1}_s$ the unit vector indicating species $s$, it follows that the change in excess free energy of a particle $s$ moving from subvolume $v$ to $w$ is 

\begin{align}
 \Delta_{vw}\beta F_{\mathrm{ex}}={}&
 [\beta F_w(\mathbf N_w+\mathbf1_s)-\beta F_w(\mathbf N_w)]\nonumber\\
 &+[\beta F_v(\mathbf N_v-\mathbf1_s)-\beta F_v(\mathbf N_v)],
 \label{eq:dFexHop}
\end{align}

Combining this with the external one-body potential energies gives

\begin{equation}
 \Delta U^s_{vw}=\Phi_s^{\mathrm{ext}}(\mathbf r_w)-\Phi_s^{\mathrm{ext}}(\mathbf r_v)
                         +\Delta_{vw}\beta F_{\mathrm{ex}}.
 \label{eq:hopDeltaU}
\end{equation}

For reaction $r$ in subvolume $v$, we instead used the excess term $\beta F_v(\mathbf N_v+\mathbf
\nu_{vr})-\beta F_v(\mathbf N_v)$. These finite differences retain the dependence on particle sizes and
composition. A decrease in particle number alone does not determine the sign of the excess free-energy
change.

Our local approximation neglected spatial correlations across voxel boundaries and density structure
within each voxel. A bulk interpretation is most applicable when a voxel contains many particles, their
diameters are small compared with $h$, and composition varies slowly over that distance. These physical
requirements supplement the numerical timestep constraints below. Our approximation does not guarantee convergence at the $h\to0$ limit at fixed molecular sizes and density. We therefore used tests with particles
whose dimensions were comparable to side length $h$ to assess how far we could extend this local description.

To account for large bodies reducing the space available to smaller particles, we evaluated the bulk
density in the unoccupied fraction of each subvolume, using the free-volume approximation in
Sec.~\ref{sec:si-freevol}. For an excluded fraction $w_v$, we divided the bath (i.e., occupied subvolumes surrounding a large body) densities by $1-w_v$ and
multiplied the resulting bulk excess free-energy density by the available volume $h^3(1-w_v)$. This
expression specifies the excess term and does not include the ideal accessible-volume contribution. We
updated the body's own motion separately, without using the change in bath free energy
(Sec.~\ref{sec:large-bodies}).

\subsection{The finite-timestep update}\label{sec:algorithm}

In simulating the master equation, we used a Lie-Trotter operator splitting scheme. In each timestep, we first applied the $x$,
$y$, and $z$ diffusion passes, then updated large bodies and polymers, and finally sampled reactions.
Each operator used the state produced by its predecessor. All decisions within an operator used its
starting state (Sec.~\ref{sec:si-splitting}).\citep{roberts2013lattice,trotter1959product}. We called our simulation scheme multiparticle-diffusion reaction-drift-diffusion master equation (MPD-RDDME), based on the MPD-RDME framework used in previous work Lattice Microbes~\citep{roberts2013lattice}. Further details of the GPU implementation are provided in Sec~\ref{sec:si-gpu}.

In addition to the Lie--Trotter splitting, our sampler involved additional approximations. Specifically, during each operator
update, we held the field and occupancies used to compute rates fixed, allowed each particle at most one
hop along the current axis, and tested at most one reaction candidate per subvolume. These approximations
require timestep-convergence tests of the complete sampler. The splitting order alone does not determine its accuracy.

We tested fixed and randomized operator orders for reversible binding and ternary-mixture relaxation
(Sec.~\ref{sec:si-splitting}). For binding, reducing $\Delta t$ from $0.08$ to $0.01\,\mu\mathrm{s}$
reduced the reported early-time product-fraction error from $2.3\%$ to $0.4\%$, relative to a linear
extrapolation to zero timestep. For the ternary mixture, the large-body mean-squared displacement at
$\tau=4$ varied within its $3\%$ statistical error over timesteps from $50$ to $6.25\,\mu\mathrm{s}$.
Fixed and randomized orders agreed within approximately one standard error over 16 replicates. These
tests quantified errors in the sampled observables. They did not establish exact sampling at finite
timestep.

While the volume exclusion free energies diverge at the limit $\eta \to 1$, penalizing diffusion into highly-crowded subvolumes, our source code implementation allocated storage for at most $64$ particles in each subvolume.
If an update exceeded this capacity, our
implementation tracked the excess particles and relocated each to the nearest available subvolume of
the same site type within a Euclidean distance of $4h$. A run terminated if relocation failed or if more
than 2048 particles overflowed in one step. This procedure allowed displacements beyond the
nearest-neighbor hops in Eq.~\eqref{eq:diffusionPropensity}. 

\begin{figure}[htbp]
\par\smallskip\noindent\hrulefill
\refstepcounter{algorithm}\label{alg:rddme}
\par\noindent\textbf{Algorithm \thealgorithm:} Schematic MPD-RDDME update
\par\noindent\hrulefill
\begingroup\small
\begin{algorithmic}[1]
\While{$t<t_{\mathrm{end}}$}
 \For{Cartesian direction $d\in\{x,y,z\}$}
  \State Freeze the current state and field for this directional update
  \ParFor{particle of species $s$ initially in subvolume $v$}
   \State Compute $\Delta U_\pm$ for $v\to v\pm\widehat d$ using Eq.~\eqref{eq:hopDeltaU}
   \State $P_\pm\gets(D_s\Delta t/h^2)\mathcal B(\Delta U_\pm)$
   \State Require $P_++P_-\leq1$
   \State Draw one outcome: hop forward, hop backward, or remain
  \EndParFor
  \State Commit the directional update
 \EndFor
 \State Update large bodies and lattice polymers 
 \ParFor{subvolume $v$}
  \State Compute baseline propensities and $a_{v,\mathrm{tot}}=\sum_r a_v^r$
  \If{$a_{v,\mathrm{tot}}>0$ and $u_1<1-e^{-a_{v,\mathrm{tot}}\Delta t}$}
   \State Select $r$ with probability $a_v^r/a_{v,\mathrm{tot}}$
   \State Compute $\Delta U_{vr}$ from the proposed composition change
   \If{$u_2<\min(1,e^{-\Delta U_{vr}})$}
    \State Apply the reaction stoichiometry
   \EndIf
  \EndIf
 \EndParFor
 \State Update the time-dependent external field as required
 \State $t\gets t+\Delta t$
\EndWhile
\end{algorithmic}
\endgroup
\noindent\hrulefill
\end{figure}

We drew $u_1$ and $u_2$ as independent uniform random numbers. If we rejected a reaction candidate, we
left the subvolume unchanged for that reaction update and did not attempt a second candidate.

\subsection{Large spheres and lattice polymers}\label{sec:large-bodies}

A count in one subvolume cannot adequately represent a particle larger than $h$. We represented these
particles by geometric footprints spanning several subvolumes and placed each sphere's center on a finer
grid of spacing $\delta=h/Q$, with $Q$ powers of two. We used the rate (see Sec. ~\ref{sec:si-bodymotion} for further details)\,

\begin{equation}
 p(\mathbf v\to\mathbf v')=\frac{D_T}{\delta^2}
 m(\mathbf v,\mathbf v')\mathcal B(\Delta U_{\mathrm{own}}),
 \label{eq:structure-hop-propensity}
\end{equation}

where $\Delta U_{\mathrm{own}}$ includes the stated external and structure interaction energies. Here, $\mathbf{v}'$ is the destination subvolume obtained by displacing $\mathbf{v}$ by $\delta \hat{e}$. 
We used the bath-dependent mobility factor as derived in Sec. ~\ref{sec:si-bodymotion}.

\begin{align}
 m(\mathbf v,\mathbf v')&=\left[1+\sum_s c_s\bar\phi_s(\mathbf v,\mathbf v')\right]^{-1},
 \nonumber\\
 c_s&=\frac12\frac{D_T}{D_T+D_s}
       \frac{(1+\sigma_s/\sigma_T)^3}{(\sigma_s/\sigma_T)^3}.
 \label{eq:frictionFactor}
\end{align}

We averaged the local bath volume fractions over a shell and symmetrized them between the two endpoints.
Sections~\ref{sec:si-diffusion-multi-site} and \ref{sec:si-bodymotion} describe the dilute-bath
derivation and its extension to finite concentrations. Our approximation accounts for mobility without
resolving bath collisions or hydrodynamic interactions.

We modeled polymers with discrete moves on a face-centered cubic (fcc) lattice, using excluded sites,
fixed bond lengths, and bending energies. We allowed chains to grow and shrink through reactions at both tips. Our
move set included full-kink moves, end-bond rotations, segment translations, and whole-chain translations
(Secs.~\ref{sec:structures} and \ref{sec:fcc-moveset}). We assigned whole-chain mobility using a Rouse approximation $D_T(N)=D_1/N$, where $N$ is the number of subunits~\citep{rouse1953theory}. Internal moves also change bead positions
and can contribute to center-of-mass motion. Their attempt frequencies, together with the whole-chain
proposals, set the kinetics of our polymer model. We therefore assessed static chain statistics and dynamical observables separately.

\subsection{Spatial resolution and timestep requirements}\label{sec:constraints}

Choosing an appropriate subvolume size requires a balance between spatial resolution, local mixing on the reaction
timescale, and the validity of the bulk fluid FMT approximation. 

For a smooth, dimensionless one-body potential, we expressed the spatial-resolution criteria as

\begin{equation}
 \varepsilon_{\mathrm{WM}}^{(n)}=h^n\|D^n\Phi\|\ll1,\qquad n=1,2,\ldots,
 \label{eq:spatialFlatnessCondition}
\end{equation}

where $\|D^n \Phi\|$ indicates the magnitude of the $n$th tensor derivative of the potential $\Phi$. For a smooth local density and a symmetric cell, a centroid
quadrature approximation, which we used in our algorithm, has a leading error of order $h^2$. This quadrature result does not imply that the entire
evolving density is uniform within a subvolume. Further details are shown in Sec.~\ref{sec:si-constraints}.

We expressed temporal-resolution criteria in terms of the energy differences that enter the rates.
Sufficient conditions for these differences to vary little during a timestep are

\begin{equation}
 (\Delta t)^m\|\partial_t^m\Delta U\|\ll1,\qquad m=1,2,\ldots,
 \label{eq:temporalFlatnessCondition}
\end{equation}

where we held occupancy fixed when differentiating the imposed field. For a hop in a smooth field, the
leading spatial contribution gives

\begin{equation}
 h\Delta t\|\partial_t\nabla\Phi\|\ll1.
 \label{eq:temporalFlatnessHop}
\end{equation}

Our directional hop sampler required $P_++P_-\leq1$. A sufficient bound for each species is

\begin{equation}
 \frac{D_s\Delta t}{h^2}\max_{v,w}\mathcal B(\Delta U^s_{vw})
 \leq\frac12.\label{eq:samplingBound}
\end{equation}

Exceeding this sufficient bound does not necessarily violate the sum constraint, but a sum above one
makes the directional draw invalid. Because $\mathcal B(u)\sim|u|$ for large negative $u$, the zero-field
diffusion bound alone is insufficient. Valid probabilities also do not guarantee accurate sampling.

Our same-subvolume bimolecular reactions do not include the finite-range, inter-subvolume reaction
construction of the convergent RDME and RDDME\citep{isaacson2013convergent,isaacsonzhang2024crddme}. Together with the bulk fluid subvolume approximation of the FMT, our method is not expected to converge with any spatially-continuous model at the limit $h\to 0$. 

\section{Results}

\subsection{Relaxation of a mixture of finite-sized particles}\label{sec:finitesize}

\begin{figure}[htbp]
    \centering
    \includegraphics[max width=\linewidth,max height=0.63\textheight,trim=0.4in 0.4in 0.4in 0.4in,clip]{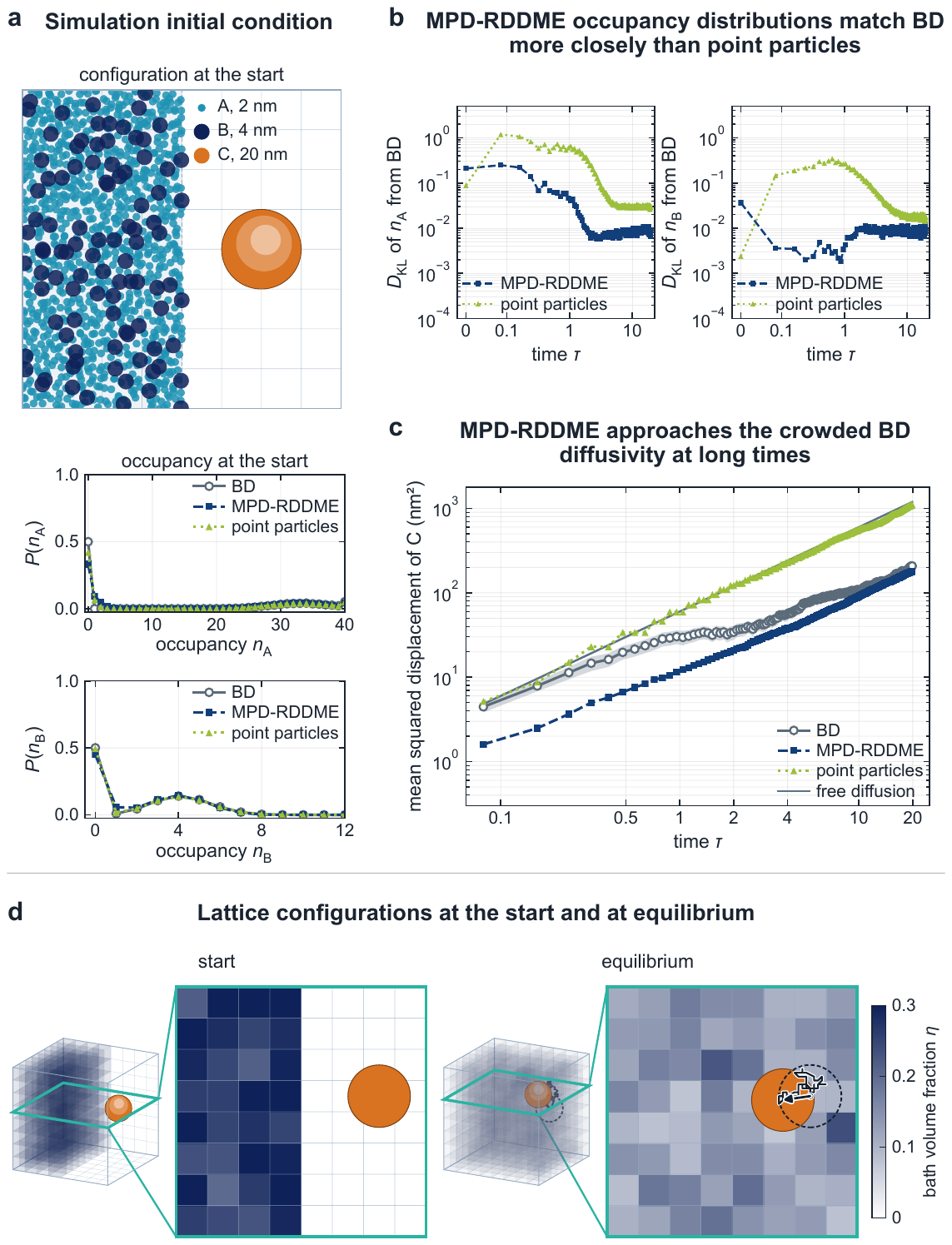}
    \caption{Relaxation dynamics of a purely diffusing ternary mixture. Particle diameters are $\sigma_A = 2$, $\sigma_B = 4$, and $\sigma_C = 20\ \text{nm}$. Box-average bath volume fractions were $0.07$ per species, giving $0.14$ per species in the initially occupied half. Bare diffusion coefficients were $D_\mathsf{A}=0.1$, $D_\mathsf{B}=0.05$, and $D_\mathsf{C}=0.01~\mu\mathrm{m}^2/\mathrm{s}$. (a) Initial configuration in an $8$-nm slab through $\mathsf{C}$ and initial bath-occupancy distributions. (b) Kullback--Leibler divergence $D_\mathrm{KL}$ of the occupancy distributions of $\mathsf{A}$ and $\mathsf{B}$ from the BD reference over time. (c) Mean-squared displacement of $\mathsf{C}$, with $\tau=D_\mathsf{A}t/h^2$. The solid reference line is $6D_\mathsf{C}t$. We pooled 40 ReaDDy runs and 10 lattice runs per arm. The ReaDDy band shows one standard error across pooled $\mathsf{C}$-particle squared displacements. BD denotes the off-lattice Brownian-dynamics reference. (d) Lattice snapshots at $\tau=0$ and $20$, with an enlarged slice through one $80$-nm cell. Color shows bath volume fraction. The dashed outline and path mark the large sphere's starting position and projected trajectory.}
    \label{fig:three-relax-readdy}
\end{figure}

We first tested how our framework captured the relaxation of a crowded mixture containing a particle
larger than a voxel (Fig.~\ref{fig:three-relax-readdy}). We compared our lattice simulations with and
without volume-exclusion free energies against off-lattice particle simulations in
ReaDDy.\citep{schoneberg2013readdy,hoffmann2019readdy2} ReaDDy is an open-source simulator that tracks
individual particles as they diffuse, interact, and react off-lattice. It combines Brownian
dynamics (BD) with interaction potentials and stochastic reaction rules. 
Since ReaDDy does not natively support hard-sphere interactions, we used harmonic repulsion between two contacting spheres separated by distance $d_{ij}$, $U(r) = (k/2)(r-d_{ij})^2$ for $r<d_{ij}$, with $k= 50 k_B T/\mathrm{nm^2}$. Then, we chose $d_{ij}$ such that the Barker--Henderson effective diameter \citep{barker1967square,barker1967successful}, $\int_0^{d_{ij}} \dd r [1-e^{-\beta U(r)}]$, equals the hard-sphere contact 
distance $\sigma_{ij} = (\sigma_i + \sigma_j)/2$, where $\sigma_i$ and $\sigma_j$ are diameters of particles $i$ and $j$.

We placed bath species $\mathsf{A}$ and $\mathsf{B}$, with diameters $\sigma_\mathsf{A}=2$ nm and
$\sigma_\mathsf{B}=4$ nm, randomly without overlap in one half of the box. Each species had a box-average
volume fraction of $\eta=0.07$, corresponding to 0.14 in the initially occupied half. We placed a freely
moving particle of species $\mathsf{C}$, with diameter $20$ nm (twice the lattice spacing), at the center
of the empty half, one per $(80\ \mathrm{nm})^{3}$ (Fig.~\ref{fig:three-relax-readdy}a). We assigned bare
diffusion coefficients $D_\mathsf{A}=0.1$, $D_\mathsf{B}=0.05$, and
$D_\mathsf{C}=0.01~\mu\mathrm{m}^2/\mathrm{s}$, inversely proportional to particle diameter as in the
Stokes--Einstein relation. The maximum imposed timestep for the lattice simulations is $\Delta t = 50\, \mu\text{s}$. However, to ensure that we do not violate Eq.~\ref{eq:samplingBound}, our framework allowed for adaptive time stepping. The resulting time steps taken range from $12$ to $42\,\mu{\text{s}}$. Particle $\mathsf{C}$ moves at a sublattice grid with $Q=8$.

We ran the simulations to $\tau = D_\mathsf{A}t/h^2 = 20$. At long times ($\tau\gtrsim10$), the
mean-squared displacement of $\mathsf{C}$ approached the off-lattice reference
(Fig.~\ref{fig:three-relax-readdy}c). A fit over $4\leq\tau\leq20$ gave a diffusion coefficient of $0.0015\
\mathrm{\mu m^2/s}$ for $\mathsf{C}$ on the lattice, compared with $0.0013 \ \mathrm{\mu m^2/s}$ in the
reference, a difference of approximately $14\%$. Our approximation thus captured the scale of the
long-time slowing in this mixture.

The MPD-RDDME simulations with volume exclusion also have voxel occupancy distributions closer to the off-lattice reference
(Fig.~\ref{fig:three-relax-readdy}b). We measured the Kullback--Leibler (KL) divergence of the occupancy
distribution of $\mathsf{A}$ and of $\mathsf{B}$ from the reference throughout the entire trajectory. After the
start, the volume-excluded lattice stayed below the point-particle control for both species. During the
transient ($0<\tau<1$), its median KL divergence was $9$ times lower for $\mathsf{A}$ and $91$ times lower for
$\mathsf{B}$. For $\tau\geq1$, its median divergence was $8.3\times10^{-3}$ for $\mathsf{A}$ and
$8.6\times10^{-3}$ for $\mathsf{B}$, $3.7$ and $2.3$ times lower than the point-particle control. Exclusion
also produced anti-correlated species occupancies (Sec.~\ref{sec:si-rho}), and the occupancy distributions
departed from the near-Poisson statistics of point particles (Sec.~\ref{sec:si-three-eq}).
Figure~\ref{fig:three-relax-readdy}d shows snapshots at the two endpoints.

The comparisons also distinguished which observables our approximation captured most closely. The
occupancy correlation was $-0.51$ on the lattice and $-0.37$ in the reference (Sec.~\ref{sec:si-rho}),
and separate tests across particle-size ratios revealed larger differences in population fluctuations
(Sec.~\ref{sec:si-twospecies}). Thus, agreement in long-time dynamics did not imply the same accuracy for
local fluctuations or throughout the transient. Table~\ref{tab:cost} reports the computational costs.

\begin{table}[htbp]
  \centering
  \caption{Wall-clock time per numerical step, in milliseconds, for MPD-RDDME and ReaDDy. We timed MPD-RDDME on one
NVIDIA B200 GPU and on one CPU core using the implementations described below. We timed ReaDDy on one
core and on the 56 cores of one Xeon Platinum 8570 socket. For the ternary mixture ($38\,503$ particles)
and reversible reaction tests ($13\,038$ particles), we used a $32\times8\times8$-voxel box. For polymer
segregation ($999\,661$ particles), we used a $64^{3}$-voxel box in the state reached at $10$ s, with
approximately $2\,650$ chains containing $45\,500$ of the $60\,000$ subunits. These timings compare the
cost of numerical steps. Assessing speedup per unit of simulated time at matched accuracy also requires
the timestep and error of each method. Sections~\ref{sec:si-performance} and \ref{sec:si-advantage}
provide further details.}
  \label{tab:cost}
  \vspace{6pt}
  \renewcommand{\arraystretch}{1.25}%
  \linespread{1}\selectfont
  \setlength{\tabcolsep}{9pt}%
  \begin{tabular}{@{}lcccc@{}}
  \toprule
   & \multicolumn{2}{c}{MPD-RDDME} & \multicolumn{2}{c}{ReaDDy} \\
  \cmidrule(lr){2-3}\cmidrule(l){4-5}
  system & GPU & 1 core & 1 core & 56 cores \\
  \midrule
  ternary mixture (Sec.~\ref{sec:finitesize})    & $0.355$ & $6.37$ & $319$      & $45.1$ \\
  reversible binding (Sec.~\ref{sec:reversible}) & $0.133$ & $3.25$ & $23.9$     & $19.3$\textsuperscript{a} \\
  polymer demixing (Sec.~\ref{sec:phase-seg})   & $3.98$  & $268$ & $10\,730$  & $599$ \\
  \bottomrule
  \end{tabular}

  \vspace{4pt}
  \parbox{0.9\linewidth}{\footnotesize\raggedright \textsuperscript{a}ReaDDy's multicore kernel lacked the detailed-balance reaction handler used for the single-core measurement. We therefore used its standard handler for this entry, which represents a different reaction model.}
\end{table}

For the ternary-mixture and polymer timings, we used a dedicated single-core C++ port of our GPU engine
that included large-body and chain updates. The median costs over three runs were $6.37$ and $308$ ms per
step, respectively. For reversible binding, we used a separate single-file CPU reference implementation.
The repository describes the CPU port's adaptations to sequential execution and its differences from our
GPU implementation.

\subsection{Reversible bimolecular reactions in an external field}\label{sec:reversible}

\begin{figure}[htbp]
    \centering
    \includegraphics[max width=\linewidth,max height=0.61\textheight,trim=0.4in 0.4in 0.4in 0.4in,clip]{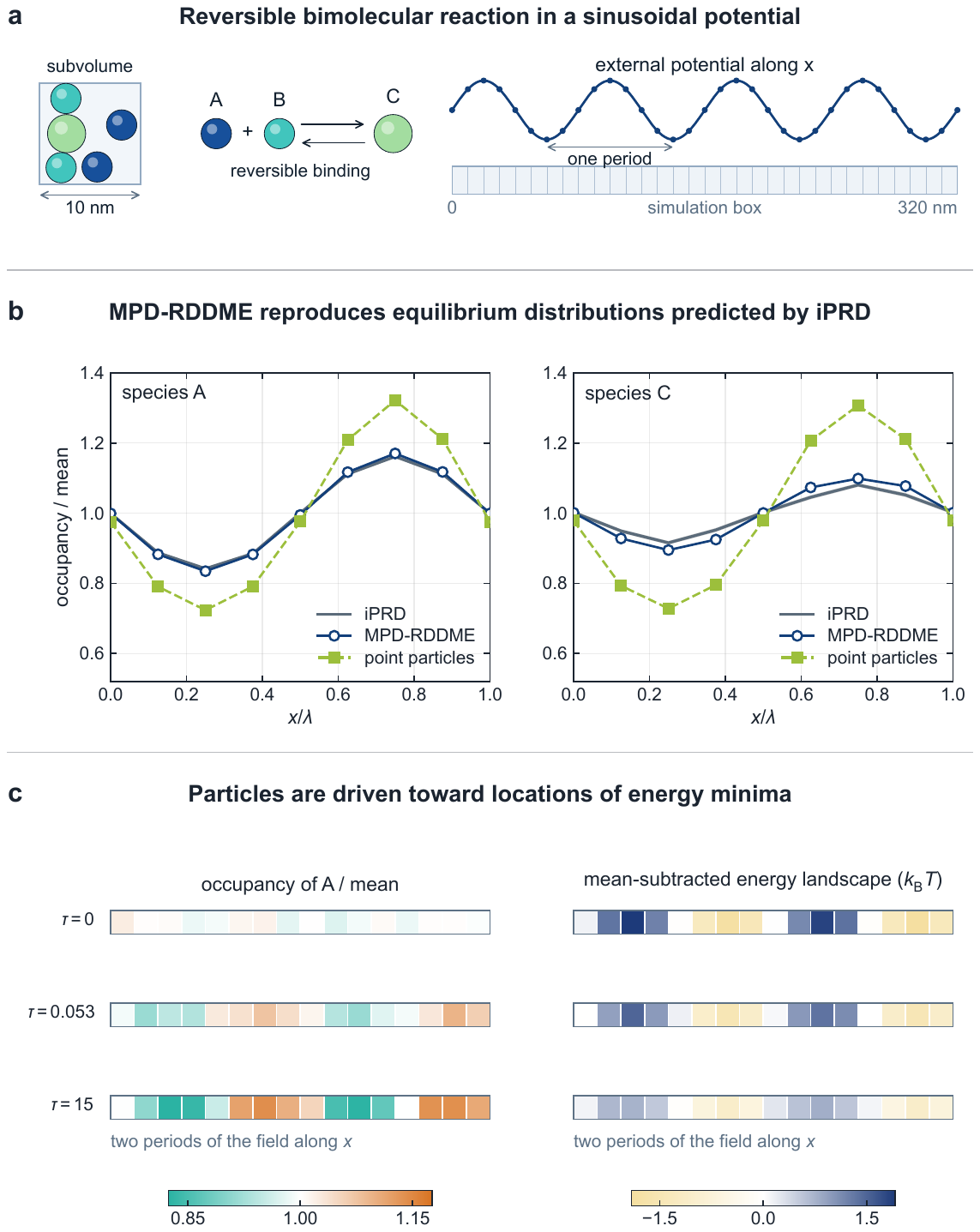}
    \caption{Reversible binding $\mathsf{A}+\mathsf{B}\rightleftharpoons\mathsf{C}$ in the static field $\Phi_s=\Phi_0\sin(2\pi x/\Lambda)$. (a) Particle sizes relative to a $10$-nm voxel, the reaction, and the field along the $320$-nm box. We chose $\sigma_\mathsf{A}=\sigma_\mathsf{B}=3$ nm and $\sigma_\mathsf{C}=3.78$ nm to approximately conserve occupied volume on binding. (b) Steady-state occupancy profiles, folded onto one field period and divided by each species' spatial mean. (c) Two field periods at $\tau=0$, $0.053$, and $15$. The left maps show reactant occupancy relative to its spatial mean. The right maps show the dimensionless voxel energy $\sum_s\bar{x}_v^s\Phi_s+\beta F_v(\bar{\mathbf x}_v)$ with its spatial mean subtracted. We evaluated this energy at occupancies averaged over planes perpendicular to the field, rather than averaging instantaneous energies. Parameters were $\eta=0.088$, $\Phi_0=0.3$, $\Lambda=80$ nm, $K=200$ nm$^3$, $k_{\mathrm{off}}=5\times10^5$ s$^{-1}$, and $D=0.1~\mu\mathrm{m}^2/\mathrm{s}$ for every species, with $\tau=Dt/h^2$.}
    \label{fig:rx-field-kl}
\end{figure}

Next, we simulated reversible binding, $\mathsf{A}+\mathsf{B}\rightleftharpoons\mathsf{C}$, to test how
crowding affected both composition and spatial organization (Fig.~\ref{fig:rx-field-kl}). We assigned
species $\mathsf{A}, \mathsf{B}, \mathsf{C}$ diameters
$\sigma_\mathsf{A},\sigma_\mathsf{B},\sigma_\mathsf{C}$ of 3, 3, and 3.78 nm, respectively. We compared
our lattice simulations with ReaDDy's Doi volume-reactivity model and detailed-balance reaction
handler,\citep{schoneberg2013readdy,hoffmann2019readdy2} using the same baseline equilibrium constant $K$
and dissociation rate $k_{\mathrm{off}}$. 

With volume-exclusion free energies, we found that the apparent equilibrium constant $K_{\mathrm{eff}} =
\langle N_\mathsf{C}\rangle V/(\langle N_\mathsf{A}\rangle\langle N_\mathsf{B}\rangle)$ increased, and closely followed the shift in the off-lattice interacting-particle reference
(Sec.~\ref{sec:si-rxK}).\citep{hoffmann2019readdy2} For these particle sizes and this composition, the
excess free-energy correction favored association.  

To test spatially biased reaction--diffusion dynamics, we imposed the static sinusoidal potential $\Phi_s
= \Phi_0\sin(2\pi x/\Lambda)$, with dimensionless amplitude $\Phi_0=0.3$ (an energy amplitude of
$0.3k_BT$) and wavelength $\Lambda=80$ nm, or $8h$. This choice gave
$\varepsilon^{(1)}_{\mathrm{WM}}=0.24$ and $\varepsilon^{(2)}_{\mathrm{WM}}=0.19$. Because these values
were below one but not asymptotically small, we used the off-lattice reference simulations to assess accuracy at this
spatial resolution.

We used separate lattice runs from the same initial condition to resolve three observation windows. We
used $\Delta t=1$ ns for runs to $30~\mu\mathrm{s}$, $10$ ns for runs to $600~\mu\mathrm{s}$, and $100$
ns for runs to $15$ ms. We used one seed per model for each of the first two windows and four seeds for
the longest window. For ReaDDy, we used $\Delta t=0.16~\mu\mathrm{s}$ and 48 seeds per window.

With $k_{\mathrm{off}}=5\times10^5$ $\mathrm{s}^{-1}$, our lattice timesteps gave $k_{\mathrm{off}}\Delta
t=5\times10^{-4}$, $5\times10^{-3}$, and $0.05$. These values resolved the single-molecule dissociation
timescale. As discussed in Sec.~\ref{sec:constraints}, the single-candidate reaction update also requires
checks of the total reaction activity and convergence of the measured observables.

Our free-energy correction reduced particle accumulation near the potential minima and brought the
steady-state occupancy profiles close to the reference (Fig.~\ref{fig:rx-field-kl}b). At $x/\Lambda =
3/4$, the point-particle occupancies of $\mathsf{A}$ and $\mathsf{C}$ reached $1.32$ and $1.31$ times
their respective spatial means. With volume exclusion, these values fell to $1.17$ and $1.10$, compared with
$1.16$ and $1.08$ in the reference.

Exclusion also shifted the composition of the whole box toward the product, consistent with the increase
in $K_{\mathrm{eff}}$ (Fig.~\ref{fig:rx-field-counts}). At steady state, the point-particle lattice gave
$N_\mathsf{A}/N_\mathsf{A}(0) = 0.700$ and $N_\mathsf{C}/N_\mathsf{A}(0) = 0.300$. Our volume-excluded
lattice gave $0.630$ and $0.370$, compared with $0.636$ and $0.364$ in the reference. The time-dependent
marginal distributions likewise approached the reference more closely with volume exclusion
(Fig.~\ref{fig:rx-field-kl-time}), as quantified by their reduced KL divergence. This
comparison assessed the measured marginals and did not test the full path distribution or every
relaxation timescale.

\subsection{Polymer statistics dynamics and spatial segregation}\label{sec:phase-seg}

We then tested polymer conformations, dynamics, and segregation using chains with $10$-nm subunits and
$h=10$ nm. Our fcc representation allowed more bond orientations than a simple cubic lattice while
retaining fixed bond lengths of $\sqrt{2}h$. We assigned the bending energy $\kappa(1-\cos\theta)$ and
used the moves described in Sec.~\ref{sec:si-fcc}. Because the allowed angles remain discrete, matching
the bending constant alone does not make an fcc chain equivalent to a off-lattice chain.

We first compared the mean-squared end-to-end distance with exact phantom-chain results and self-avoiding
reference chains (Fig.~\ref{fig:polymer}a). We tested $M=7$, $15$, $31$, and $63$ bonds at $\kappa=1$ and
$4k_BT$, and $M=10$ bonds over $\kappa=0$ to $8k_BT$. For each lattice point, we averaged $32$ chains
sampled for $200$ timesteps with $200$ interior relaxation attempts per timestep. We generated the
self-avoiding reference by rejection sampling of $4\times10^5$ phantom chains.

These comparisons assessed the sampled conformations separately from the mapping of attempted moves to
physical time. The exact phantom-chain result provided an analytical baseline against which we could
examine the effect of self-avoidance on chain statistics.

We next simulated polymer dynamics in crowded environments (Fig.~\ref{fig:polymer}b). We used an
$11$-subunit chain with bead diffusion coefficient $D=0.3~\mu\mathrm{m}^2/\mathrm{s}$, surrounded by
$3$-nm bath particles with $D=0.3~\mu\mathrm{m}^2/\mathrm{s}$ or $20$-nm crowders with
$D=0.04~\mu\mathrm{m}^2/\mathrm{s}$ at volume fractions $\eta=0.05$, $0.10$, and $0.15$. For the
bath-particle runs, we used a lattice timestep of $\Delta t=20~\mu\mathrm{s}$, placed $8$ chains in a
$32^3$-voxel box, and averaged $8$ runs of $5$ s. For the $20$-nm crowders, we used $8$ chains in a
$64^3$-voxel box and averaged $3$ runs of $10$ s.

In these dynamics runs, we used four internal move attempts per timestep and a nominal whole-chain
translation probability of approximately $0.033$ per step for an $11$-subunit chain. This schedule
supplied approximately 120 internal attempts per whole-chain translation proposal before rejection of
moves.

For the off-lattice reference, we used ReaDDy with $\Delta t=50$ ns and one chain in a periodic
$140$-nm box. We matched the bond and angle potentials as described in Sec.~\ref{sec:si-chain-match} and
averaged $4$ runs of $0.3$ s for the bath and dilute-chain ($\eta = 0$) conditions, and $3$ runs of $2$ s
for the $20$-nm crowders.

Both descriptions showed that crowding slowed center-of-mass diffusion $D_{\mathrm{cm}}$ and the first
chain (Rouse) relaxation mode. The comparison also identified center-of-mass mobility as a more demanding test of
our polymer approximation. For $20$-nm crowders at $\eta=0.15$, the normalized diffusion coefficient was
approximately $0.15$ on the lattice and $0.62$ in the original reference. The original reference used
bending stiffness $\kappa=2.31$. Matching the equilibrated persistence length instead gave
$\kappa=2.11$. With this correction, repeat simulations of the dilute chain and densest-crowder
condition changed the reference diffusion ratio from $0.62\pm0.05$ to $0.64\pm0.01$ (standard errors),
leaving the difference largely unchanged (Sec.~\ref{sec:si-chain-match}).

These tests identified translational mobility as a target for refining our polymer update, while leaving
the specific cause of the difference unresolved. We therefore assessed the normalized relaxation rate
$\tau_1(0)/\tau_1$ and translational diffusion separately. Sections~\ref{sec:si-fcc-exact} and
\ref{sec:si-chain-match} give the fcc-chain statistics, Rouse relaxation time, and parameter matching to
the off-lattice model.

\begin{figure}[htbp]
    \centering
    \includegraphics[max width=\linewidth,max height=0.60\textheight,trim=0.4in 0.4in 0.4in 0.4in,clip]{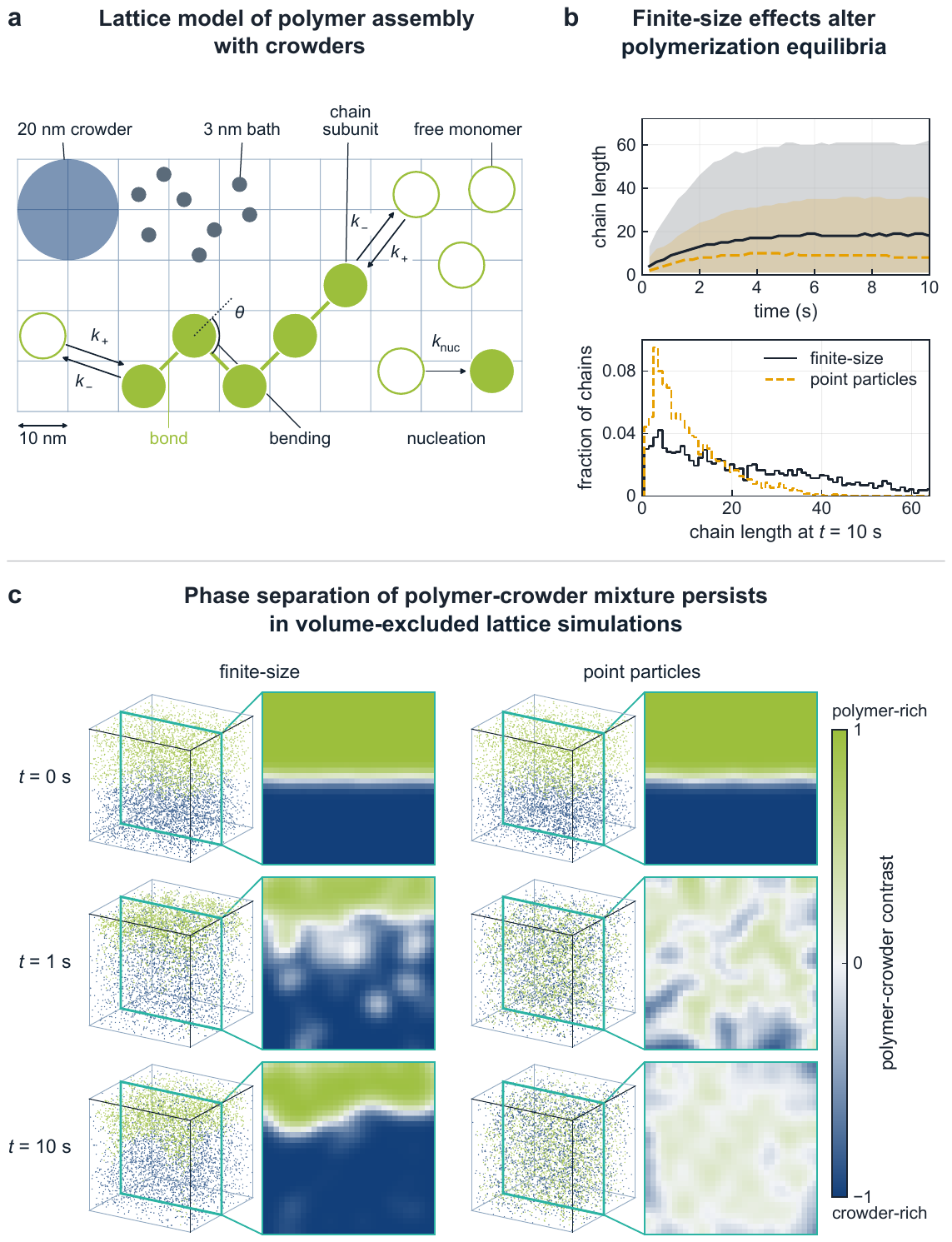}
    \caption{Spatial segregation of growing polymers and crowders. (a) Model geometry at $h=10$ nm, with $20$-nm crowders, $3$-nm bath particles, free monomers, and an fcc chain. Each chain subunit occupies one voxel, bonds have length $\sqrt{2}h$, and bending costs $\kappa(1-\cos\theta)$. Chains nucleated at rate $k_{\mathrm{nuc}}$ and gained or lost tip subunits at rates $k_+$ and $k_-$. Bare diffusion coefficients were $0.04$, $0.15$, and $0.3~\mu\mathrm{m}^2/\mathrm{s}$ for crowders, bath particles, and free monomers, respectively. (b) Median chain length over time, with bands spanning the 1st--99th percentiles, and chain-length distributions at $10$ s. (c) Snapshots from an initially separated state, with monomers above and crowders below. Green and blue points show subsampled polymer subunits and crowders. Enlarged slices show $(p-c)/(p+c)$, where $p$ and $c$ are local amounts normalized by their respective slice means [Eq.~(\ref*{eq:localAmount})]. Polymer amounts count free monomers at $t=0$ and chain subunits thereafter. Chains could not cross the reflecting box faces. The snapshots demonstrate segregation over the simulated interval.}
    \label{fig:phase}
\end{figure}

As a larger-scale demonstration, we simulated the spatial segregation of growing polymers and crowders
(Fig.~\ref{fig:phase}). We used a $64^3$-voxel lattice ($640$ nm per side), initially containing
$60\,000$ free monomers at volume fraction $0.12$, $12\,516$ crowders of diameter $20$ nm at $\phi=0.20$,
and $927\,145$ bath particles of diameter $3$ nm at $\eta=0.05$. We assigned bare diffusion coefficients
of $D=0.3~\mu\mathrm{m}^2/\mathrm{s}$ to monomers, $0.04~\mu\mathrm{m}^2/\mathrm{s}$ to crowders, and
$0.15~\mu\mathrm{m}^2/\mathrm{s}$ to bath particles.

We allowed chains to nucleate from single monomers at $k_{\mathrm{nuc}}=0.02~\mathrm{s}^{-1}$ per
monomer, grow at either tip at $k_{+}=20~\mathrm{s}^{-1}$ per accessible monomer, and lose tip subunits
at $k_{-}=3~\mathrm{s}^{-1}$ per tip. We did not allow bonds to form or break within chain interiors. We
displaced crowders on the finer grid $\delta=h/2$ and used timesteps of $\Delta t=20~\mu\mathrm{s}$. Assigned bending stiffness is $\kappa = 4$. 

We initially placed monomers randomly in one half of the box, crowders in the other half, and bath
particles uniformly throughout. We imposed no external bias. Without volume-exclusion free energies, the
two halves mixed within $1$ s and remained mixed (Fig.~\ref{fig:phase}c). With volume exclusion, the mixture
remained separated, with $95\%$ of the chain subunits in the initial monomer half at $1$ s and $98\%$ at
$10$ s, compared with $53\%$ and $50\%$ in the point-particle control.

This segregation was qualitatively consistent with the behavior of polymer--colloid mixtures with
excluded-volume interactions.\citep{asakura1954interaction,asakura1958interaction,lekkerkerker1992phase}
To test whether it depended on the initial separation, we also started from a uniform mixture and from a
polymer droplet (Sec.~\ref{sec:si-demix-starts}). With volume exclusion, the initially uniform mixture demixed
and the droplet held together. The point-particle controls mixed or remained mixed.

We also found fewer and longer chains in the simulations with volume exclusion, with $2161$ chains of median length $18$ subunits
at $10$ s, compared with $3204$ chains of median length $8$ in the point-particle control
(Fig.~\ref{fig:phase}b). Together, the different initial conditions supported segregation within our
simulated geometry. We used reflecting boundaries for chains, which could not cross the box faces and
accumulated near them in the slab-start runs. Establishing bulk phase coexistence would require tests
with chains crossing periodic boundaries, which we did not perform.

\subsection{Motor traffic on a structured track}\label{sec:motors}

Kinesin-8 (Kip3) motors can form traffic jams on microtubules, with a dense, slow-moving region growing
backward from the plus end. A totally asymmetric exclusion process with attachment and detachment
reproduces this behavior.\citep{leduc2012jams,parmeggiani2003} We used this system to test whether our
free-energy approach could reproduce collective directed transport on a structured track.

\begin{figure}[htbp]
    \centering
    \includegraphics[max width=\linewidth,max height=0.62\textheight,trim=0.4in 0.4in 0.4in 0.4in,clip]{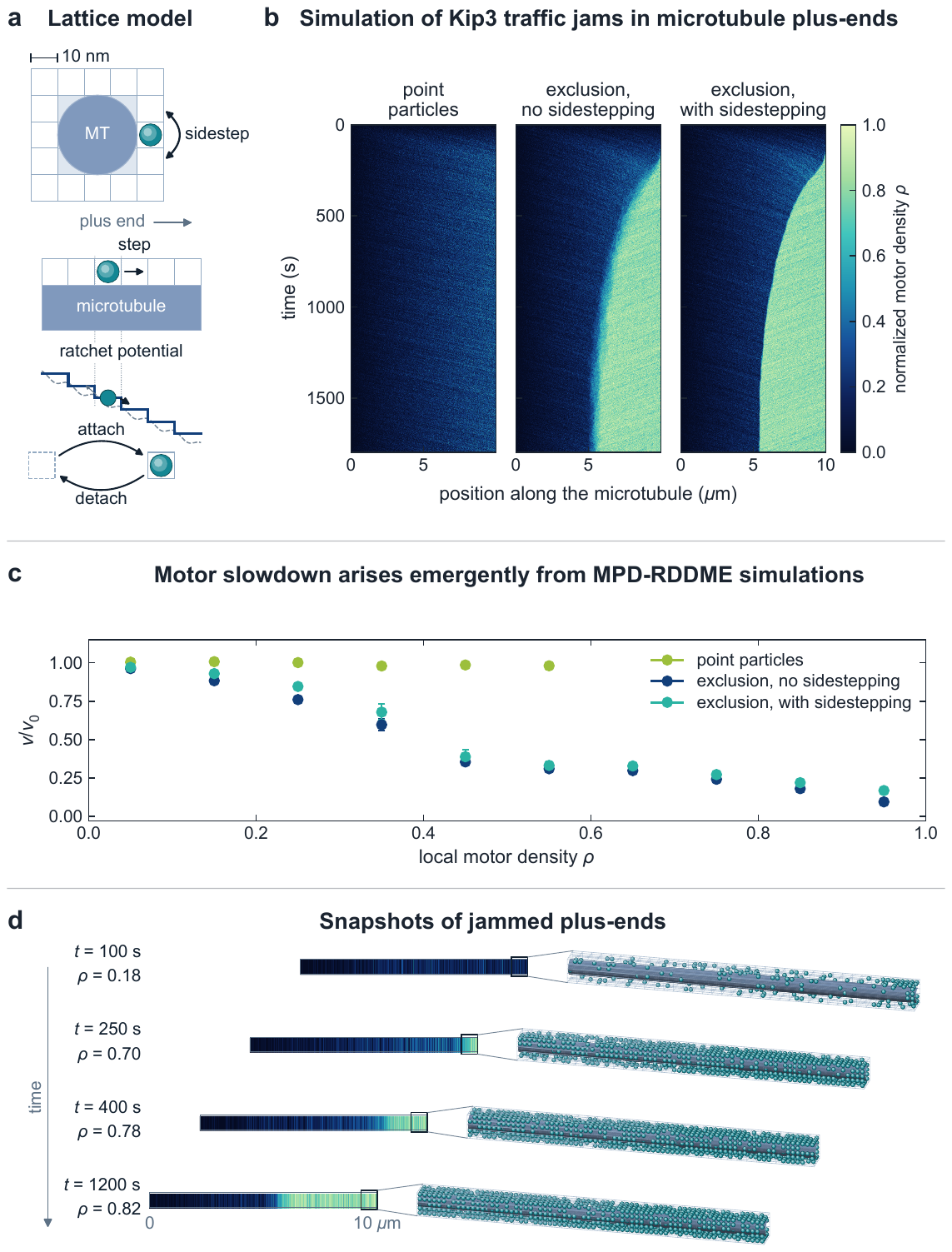}
    \caption{Motor traffic on a lattice microtubule. (a) Track geometry at $h=10$ nm, with a $30$-nm-wide microtubule, forward steps at $k_+=v_0/h$, sidesteps at $k_s=0.3$ s$^{-1}$, and a bias $\Phi(s)=-\alpha s$ giving backward rate $k_+e^{-\alpha}$. We used $v_0=56$ nm\,s$^{-1}$ and $\alpha=10$. Motors attached at $k_{\mathrm{on}}$ per empty site and detached at $k_{\mathrm{off}}$ along the track or $k_{\mathrm{end}}$ at its plus end. (b) Lane-averaged occupancy for point particles and for free-energy-based exclusion without or with sidesteps. Parameters were $\bar{k}_{\mathrm{on}}=3.2\times10^{-4}$, $\bar{k}_{\mathrm{end}}=0.2$, and $\rho_s=k_{\mathrm{on}}/(k_{\mathrm{on}}+k_{\mathrm{off}})=0.91$, with $\bar{k}_r=k_r/k_+$. The $N=1000$ sites span $10~\mu$m. These dimensionless parameters correspond to the $3.2$-nM condition in Ref.~\citep{leduc2012jams}. (c) Mean tracer speed divided by $v_0$, binned by local density for $t\geq1200$ s. Error bars show standard deviations divided by the square root of the number of tracer-window measurements in each bin. (d) Enlarged snapshots near the jam front, alongside the full track with the occupancy color scale of (b).}
    \label{fig:kip3-traffic}
\end{figure}

We represented the track as a $3\times3$ block of occupied voxels, surrounded by sixteen lanes for
spherical motors of diameter $h=10$ nm (Fig.~\ref{fig:kip3-traffic}a). Our free-energy correction
suppressed multiple motor occupancy without an explicit one-motor-per-site limit. The implementation
retained its 64-particle storage cap, but the volume exclusion free energies forbid occupancies where $\eta \geq 1$. Nevertheless, transient multiple occupancies can occur because of the operator splitting scheme of the simulation framework.

We imposed the dimensionless bias $\Phi(s)=-\alpha s$, with $\alpha=10$ and $s$ measured in lattice
steps. This represents a ratchet potential directing the motion of the motor proteins along the microtubule. This gave a backward-to-forward rate ratio of $e^{-\alpha}=4.5\times10^{-5}$. We set the forward
rate to $k_+=v_0/h$, with $v_0=56$ nm $\mathrm{s}^{-1}$. Motors attached at $k_{\mathrm{on}}$ per empty
site and detached at $k_{\mathrm{off}}$ along the track or $k_{\mathrm{end}}$ at the plus end. We used
the dimensionless rate combinations for $3.2$ nM Kip3 in Ref.~\citenum{leduc2012jams} and allowed motors
to sidestep between neighboring lanes.\citep{bugiel2015} Our bias represented directed motion
phenomenologically and did not resolve an ATP-dependent mechanochemical cycle.

With volume exclusion, the jam reached $4.56\pm0.03~\mu\mathrm{m}$ at $1800$ s over three runs
(Fig.~\ref{fig:kip3-traffic}b). Tracer speeds decreased from $0.90v_0$ at upstream density $\rho=0.09$ to
$0.19v_0$ at jam density $\rho=0.81$, corresponding to approximately $50.4$ and $10.6$ nm
$\mathrm{s}^{-1}$, respectively. The experimental speeds were $35$--$38$ and $6.5$--$10$ nm
$\mathrm{s}^{-1}$.\citep{leduc2012jams} Our model captured a similar relative slowdown, with a
jam-to-upstream speed ratio of $0.21$, compared with approximately $0.22$ experimentally, although our
upstream absolute speed was higher.

Motor speed decreased with density over the measured range
(Fig.~\ref{fig:kip3-traffic}c), and the jam front advanced along the track
(Fig.~\ref{fig:kip3-traffic}d). The point-particle control did not form a jam. Thus, our free-energy
correction reproduced the collective buildup and relative slowing of motor traffic without imposing a
strict single-occupancy rule.

\section{Discussion}
We introduced crowding corrections to reaction and diffusion propensities in lattice simulations of the reaction-drift-diffusion master equation that uses molecular
sizes and voxel populations without resolving individual encounters. Our
correction improved spatial occupancies (Sec.~\ref{sec:finitesize}) and shifts in reaction equilibria (Sec.~\ref{sec:reversible}) relative to point-particle lattice simulations. Local friction corrections captured
substantial crowding-induced dynamics of lattice polymers and spheres larger than a subvolume (Sec.~\ref{sec:phase-seg}), absent in a point-particle description. We also demonstrated a biological application of our simulation framework by implementing a microtubule motor traffic model (Sec.~\ref{sec:motors}). With this model, exclusion and
directed transport generated motor protein traffic jams, while also allowing for motor sidestepping.

Together, our comparisons helped delineate the range of validity of our approximation. Local
composition supported improved occupancy and reaction equilibria predictions, while discrepancies
in fluctuation statistics, position correlations, and crowded-polymer diffusion identified targets for
further model refinement. Our approximation thus provides a useful test of how much molecular detail
each observable requires. Agreement indicates where local populations suffice for the tested
conditions. Disagreement motivates targeted tests of spatial correlations, mobility, and polymer
dynamics. Although our comparisons do not isolate every cause, they turn the limits of a coarse
description into specific questions for improving it.

Our results also clarified two distinct choices in using our framework. Smaller voxels resolve
spatial variation better but compromise the bulk-fluid approximation of FMT, as well as convergence to a spatially-continuous limit in the CRDME and CRDDME formulations.
Shorter timesteps improve numerical resolution within that physical approximation. Our convergence
tests showed why these choices require checks of the observable being predicted. 

The lower measured costs per timestep make our approach promising for simulations of crowded
mixtures with many reacting species. Comparing costs at matched accuracy remains necessary to
quantify that advantage. More broadly, our framework provides both a practical model of selected
crowding effects and a way to examine which details a coarse description must retain.

\section*{Supplementary Material}
The supplementary material contains derivations of the spatial and temporal
resolution criteria, the bulk hard-sphere free energy, the large-body mobility
approximation, and the lattice-polymer statistics. It also contains the specifics of the GPU implementation of the framework. Reference-model
parameter matching, occupancy distributions and fluctuation statistics for the presented test systems, convergence, timing tests,
alternative initial conditions for segregation, and an additional biological example describing immunological synapse dynamics, are also provided.

\begin{acknowledgments}
The National Institute of General Medical Sciences of the National Institutes of Health
supported H.Y. through Award No.~R35GM147508. The National Science Foundation supported
A.T.A., H.Y., and Z.L.-S. in part through the Science and Technology Center for Quantitative
Cell Biology (Grant No.~DBI-2243257) and supported Z.L.-S. in part through Grant
No.~MCB-2221237.
\end{acknowledgments}

\section*{Author Declarations}
\subsection*{Conflict of Interest}
The authors have no conflicts to disclose.

\subsection*{Author Contributions}
\textbf{Ron Michael Acda}: Conceptualization (lead), Methodology (lead), Software (lead),
Investigation (lead), Formal analysis (lead), Visualization (lead), Writing -- original draft
(lead), Writing -- review \& editing (equal).

\textbf{Alfia Parvez}: Software (supporting), Investigation (supporting).

\textbf{Abner T. Apsley}: Investigation (supporting), Writing -- review \& editing (supporting).

\textbf{Hyun Youk}: Conceptualization (supporting), Supervision, Funding acquisition, Resources, Writing -- review \& editing (equal).

\textbf{Zaida Luthey-Schulten}: Conceptualization (supporting), Supervision, Funding acquisition, Resources,
Writing -- review \& editing.

\section*{Data Availability Statement}

The simulation outputs used to construct the figures, analysis and plotting scripts, and archived simulation drivers are available at \url{https://github.com/quarkron/mpd-rddme-data}. The repository documents the software dependencies and the available run provenance.

\bibliographystyle{aipnum4-2}
\bibliography{refs}

\clearpage
\begin{center}
\textbf{\large Supplemental Material}
\end{center}
\renewcommand{\thesection}{S\arabic{section}}
\renewcommand{\thefigure}{S\arabic{figure}}
\numberwithin{equation}{section}
\renewcommand{\theequation}{\thesection.\arabic{equation}}
\renewcommand{\theHsection}{S\arabic{section}}
\renewcommand{\theHfigure}{S\arabic{figure}}
\renewcommand{\theHequation}{\thesection.\arabic{equation}}
\setcounter{section}{0}\setcounter{figure}{0}\setcounter{equation}{0}
%
\setcounter{page}{1}
\renewcommand{\thepage}{S\arabic{page}}
\makeatletter
\newwrite\SI@auxfile
\immediate\openout\SI@auxfile=SIrefs.aux
\immediate\write\SI@auxfile{\relax}
\immediate\write\SI@auxfile{\string\bibstyle{aipnum4-2}}
\immediate\write\SI@auxfile{\string\bibdata{refs}}
\gdef\@extra@b@citeb{@SI}
\immediate\write\@auxout{\gdef\string\@extra@binfo{@SI}}
\let\SI@origcitep\citep
\let\SI@origcite\cite
\DeclareRobustCommand\SIcitep[1]{\begingroup\let\@auxout\SI@auxfile\SI@origcitep{#1}\endgroup}
\DeclareRobustCommand\SIcite[1]{\begingroup\let\@auxout\SI@auxfile\SI@origcite{#1}\endgroup}
\let\citep\SIcitep
\let\cite\SIcite
\makeatother
\addtocontents{toc}{\protect\setcounter{tocdepth}{2}\protect\SItocShow}
\tableofcontents
\clearpage


\section[Constraints for Lattice Sizes and Timesteps]{Constraints for Lattice Sizes and Timesteps}\label{sec:si-constraints}
This section derives the well-mixing and quadrature results quoted in \Cref{sec:constraints}.
In the absence of a potential field i.e., in a pure reaction-diffusion scheme, the equilibrium distribution of the particles in a subvolume is uniform, and so subvolume-centroid quadratures to represent subvolume-averaged values introduce no errors. However, when a potential is present, the equilibrium distribution is no longer necessarily uniform.

Let \(\Phi\) be a smooth, non-dimensionalized potential energy defined in a domain \(\Omega\subset\mathbb{R}^3\) arising from an interaction of a specie with a time-independent potential. Under the RDDME framework, we discretize \(\Omega\) into Cartesian subvolumes with lattice spacing \(h\). Each lattice site \(\mathbf{x}_v\in h\mathbb{Z}^3\) is the centroid of a cubic subvolume \(C_v^3\) of side length \(h\).
We quantify the deviation from local well-mixing by estimating how much the equilibrium distribution as measured by the particle density $\rho (\Phi(\mathbf{x}))$, of a specie interacting with the field, deviates from the uniform distribution within \(C_v^3\).
Let $\mathbf{x}=\mathbf{x}_v+\boldsymbol{\delta}$, with $\boldsymbol{\delta}\in[-h/2,h/2]^3$ the displacement from the subvolume centroid. For \(m\geq 2\), Taylor expansion of the potential about the subvolume centroid gives
\begin{align}
    \Phi(\mathbf{x}_v+\boldsymbol{\delta})
    &=
    \Phi(\mathbf{x}_v)
    +
    \nabla\Phi(\mathbf{x}_v)\cdot\boldsymbol{\delta}
    +
    \frac{1}{2}\boldsymbol{\delta}^{T}D^2\Phi(\mathbf{x}_v)\boldsymbol{\delta}
    + \cdots
    +
    \frac{1}{m!}D^m\Phi(\mathbf{x}_v)
    \!\left[\boldsymbol{\delta}^{\otimes m}\right]
    +
    \mathcal{O}\!\left(
        h^{m+1}\|D^{m+1}\Phi\|
    \right),
    \label{eq:potentialTaylorGeneral}
\end{align}
where \(D^n\Phi\) denotes the \(n\)-th spatial derivative tensor of \(\Phi\), and
\(D^n\Phi(\mathbf{x}_v)[\boldsymbol{\delta}^{\otimes n}]\) denotes its tensor contraction with \(n\) copies of \(\boldsymbol{\delta}\).
Thus, when the flatness condition of Eq.~\ref{eq:spatialFlatnessCondition} holds, that is, when the potential doesn't vary appreciably inside the subvolume i.e., sufficiently ``spatially'' flat, the subvolume-centroid quadrature approximation of the potential is a good coarse-grained approximation to the potential of the subvolume. Indeed,
\begin{align}
    \frac{\rho(\mathbf{x}_v + \boldsymbol{\delta})}{\rho(\mathbf{x}_v)} &= \exp[\Phi(\mathbf{x}_v) -\Phi(\mathbf{x}_v + \boldsymbol{\delta})] \approx 1 - \nabla \Phi(\mathbf{x}_v)\cdot \boldsymbol{\delta} + \frac{1}{2}( \nabla \Phi(\mathbf{x}_v)\cdot \boldsymbol{\delta})^2 - \frac{1}{2}\boldsymbol{\delta}^{\top}D^2\Phi(\mathbf{x}_v)\boldsymbol{\delta} + \dots
    \label{eq:well-mixingApprox}
\end{align}
approach unity i.e., well-mixing conditions when Eq. \ref{eq:spatialFlatnessCondition} is satisfied. From this, it can be read that within a subvolume, the shape of the probability density function of a particle's position changes with leading error $\mathcal{O}(h)$. Additionally, MPD-RDDME implicitly replaces the average probability density of finding a particle in a subvolume of volume $h^3$
\begin{align}
    \langle \rho \rangle_V = \frac{1}{h^3}\int_{C_v^3} \rho(\mathbf{x}_v + \boldsymbol{\delta})\, \dd^3 \boldsymbol{\delta}
    \label{eq:voxelAverage}
\end{align}
by the subvolume-centroid quadrature value $\rho(\mathbf{x}_v)$. As derived in \Cref{sec:si-quadrature},
\begin{align}
    \frac{\rho(\mathbf{x}_v)}{\langle \rho \rangle_V} &\approx 1 + \frac{h^2}{24}\bigg[\mathrm{tr}\, D^2\Phi(\mathbf{x}_v) - \|\nabla \Phi(\mathbf{x}_v) \|^2\bigg] \label{eq:quadratureErrorPDF}
\end{align}
for a cubic subvolume, so centroid quadratures in the MPD-RDDME algorithm introduce a $\mathcal{O}(h^2)$ leading error in probability densities of particle locations.

\subsection[The Dragged Harmonic Potential as a Test Problem]{The Dragged Harmonic Potential as a Test Problem}\label{sec:si-testproblem}
To further examine well-mixing deviations of point-particle species due to the presence of an external potential, we examine the response 
of point particles subject to a time-dependent harmonic potential. We solve the Fokker-Planck equation
\begin{align}
    \frac{\partial \rho(\mathbf{x},t)}{\partial t} = \nabla \cdot D[\nabla \rho(\mathbf{x},t) + \rho(\mathbf{x}) \nabla \Phi(\mathbf{x},t)],
    \label{eq:fokkerPlanck}
\end{align}
for a time-dependent harmonic potential
\begin{align}
    \Phi(\mathbf{x},t) = \frac{k}{2}|\mathbf{x} - \mathbf{c}(t)|^2, \label{eq:draggedHarmonicPotential}
\end{align}
where we set $\mathbf{c(t)} = c\sin (\omega t)\cdot(1,\ 1,\ 1)$.
The drift in Eq.~\ref{eq:fokkerPlanck} is linear in $\mathbf{x}$ for this potential. We try a Gaussian ansatz, inspired 
by the Ornstein-Uhlenbeck process.
Writing $\rho$ as a Gaussian of mean $\mathbf{u}(t)$ and isotropic
variance $\sigma^2(t)$ and matching the first two moments gives
\begin{align}
    \frac{\dd \mathbf{u}}{\dd t} = -Dk\left[\mathbf{u} - \mathbf{c}(t)\right],
    \qquad
    \frac{\dd \sigma^2}{\dd t} = 2D\left[1 - k\sigma^2\right].
    \label{eq:gaussianMoments}
\end{align}
So that the full nonequilibrium steady state is
\begin{align}
    \rho(\mathbf{x},t) = \left(\frac{1}{\sigma\sqrt{2\pi}}\right)^{3}
    \exp\left[-\frac{1}{2\sigma^2}\left|\mathbf{x}-\mathbf{u}(t)\right|^2\right],
    \qquad \mathbf{u}(t) = u(t)\cdot(1,\ 1,\ 1).
    \label{eq:gaussianNESS}
\end{align}
With
$\mathbf{c}(t) = c\sin(\omega t)\cdot(1,\ 1,\ 1)$ the mean satisfies a linear equation
under sinusoidal forcing. Setting $\mathbf{u}(t) = u(t)\cdot(1,\ 1,\ 1)$ and matching
amplitude and phase yields
\begin{align}
    u(t) = \frac{cDk}{\sqrt{(Dk)^2+\omega^2}}\,\sin\left(\omega t - \delta\right),
    \qquad \tan \delta = \frac{\omega}{Dk}.
    \label{eq:gaussianMean}
\end{align}
Both the amplitude
reduction and the phase lag can be tuned by the dimensionless parameter
$\omega/Dk$. 

For this potential the well-mixing error parameters of \Cref{sec:si-constraints} take the closed
forms
\begin{align}
    \varepsilon^{(1)}_{\mathrm{WM}}(\mathbf{x}_v) = hk\|\mathbf{x}_v - \mathbf{c}\|, \qquad \varepsilon^{(2)}_{\mathrm{WM}} = h^2 k,
\end{align}
with higher-order terms vanishing. The intra-subvolume density given by Eq. \ref{eq:well-mixingApprox} is then exact at second-order:
\begin{align}
   \rho(\mathbf{x}_v+\boldsymbol\delta)
&=\frac{1}{h^3}\Big[\,
1
- k\,(\mathbf{x}_v-\mathbf{c})\cdot\boldsymbol\delta
- \frac{k}{2}\lvert\boldsymbol\delta\rvert^2
+ \frac{k^2}{2}\big((\mathbf{x}_v-\mathbf{c})\cdot\boldsymbol\delta\big)^2
+ \frac{h^2 k}{8}
- \frac{h^2 k^2}{24}\lvert\mathbf{x}_v-\mathbf{c}\rvert^2 + \mathcal{O}(h^3)\Big].
\end{align}

\subsection{Temporal Flatness for a Time-Dependent Potential}\label{sec:si-temporal}
When $\Phi$ depends on $t$, we note that a second approximation enters. We note that the algorithm evaluates every 
configurational energy at the start of that step, 
so it evaluates $\Phi(\mathbf{x},t)$ where the exact process (continuous implementation) would evaluate the
average of $\Phi(\mathbf{x},t+s)$ over $s\in[0,\tau)$. Expanding about the start of the step,
\begin{align}
    \Phi(\mathbf{x},t+s) = \Phi(\mathbf{x},t)
    + s\,\partial_t\Phi(\mathbf{x},t)
    + \frac{s^2}{2}\,\partial_t^2\Phi(\mathbf{x},t)
    + \mathcal{O}\!\left(\tau^3\|\partial_t^3\Phi\|\right),
    \qquad 0\le s<\tau,
    \label{eq:potentialTaylorTime}
\end{align}
so the per-step discrepancy in the evaluated energy is $\tfrac{1}{2}\tau\,\partial_t\Phi + \mathcal{O}(\tau^2)$.
The field is a good quasi-static representation of the potential over the step exactly when the
parameters of Eq.~\ref{eq:temporalFlatnessCondition} are small. Eq.~\ref{eq:potentialTaylorTime} is
the time-direction counterpart of Eq.~\ref{eq:potentialTaylorGeneral}.

For the dragged harmonic potential of Eq.~\ref{eq:draggedHarmonicPotential} with
$\mathbf{c}(t)=c\sin(\omega t)(1,1,1)$ the two derivatives are
\begin{align}
    \partial_t\Phi = -k\,(\mathbf{x}-\mathbf{c})\cdot\dot{\mathbf{c}},
    \qquad
    \partial_t^2\Phi = k\,\|\dot{\mathbf{c}}\|^2 - k\,(\mathbf{x}-\mathbf{c})\cdot\ddot{\mathbf{c}},
    \label{eq:draggedHarmonicTimeDerivs}
\end{align}
with $\dot{\mathbf{c}}=c\omega\cos(\omega t)(1,1,1)$ and $\ddot{\mathbf{c}}=-c\omega^2\sin(\omega t)(1,1,1)$.
At the phase $\omega t=0$, where $\partial_t\Phi$ attains its largest magnitude, the $\ddot{\mathbf{c}}$
term vanishes and the second derivative collapses to $\partial_t^2\Phi = 3kc^2\omega^2$, independent
of position, in the same way that $D^2\Phi=k\mathbb{I}$ is independent of position in the spatial
expansion. Writing $r=\|\mathbf{x}_v-\mathbf{c}\|$ and taking the subvolume to lie along the drag
diagonal, so that $|\sum_i (x_i-c_i)|=\sqrt{3}\,r$,
\begin{align}
    \varepsilon^{(1)}_{\tau}(r) = \sqrt{3}\,\tau k c\, \omega\, r,
    \qquad
    \varepsilon^{(2)}_{\tau} = 3\,\tau^2 k c^2 \omega^2 .
    \label{eq:draggedHarmonicTemporalEps}
\end{align}
Note that $\varepsilon^{(2)}_{\tau} = \Bigl(\varepsilon^{(1)}_{\tau}\big|_{r=\sigma}\Bigr)^{2}$ since $r=\sigma = k^{-1/2}$.
The second-order parameter is therefore the square of the first, so it is automatically small
whenever the first-order condition holds.

\subsection{Validation of the Well-Mixing Conditions under a One-Body Potential}\label{sec:results-wm}
We now measure the previously derived conditions against the analytic solution of the dragged harmonic potential in
\Cref{sec:si-testproblem}.

\begin{figure}[htbp]
    \centering
    \includegraphics[max width=\linewidth,max height=0.80\textheight]{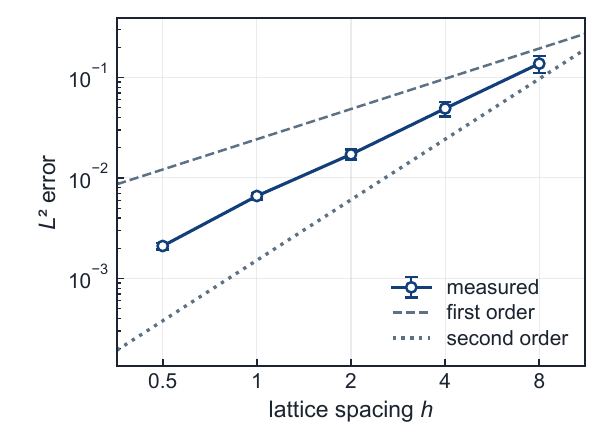}
    \caption{Convergence of the simulated one-dimensional marginal density to the analytic 
    dragged-harmonic solution as a function of lattice spacing $h$, with $h=1$ representing $10$ nm. Dashed and dotted lines are first- and second-order guides, $\propto h$ and $\propto h^2$.}
    \label{fig:wm-convergence}
\end{figure}

We simulated with periodic boundary conditions on a cube $[-L, L]^3$, with lattice spacings $h$, 
resulting to a system size of $N= (2L/h)^3$ subvolumes. 
Here, we took $L=256$ and $h \in \{0.5,1,2,4,8\}$. The resulting $L^2$ error in the predicted marginalized distribution of the MPD-RDDME algorithm with the 
analytic solutions for the time-independent case is shown in Figure \ref{fig:wm-convergence}.

\begin{figure}[htbp]
    \centering
    \includegraphics[max width=\linewidth,max height=0.80\textheight]{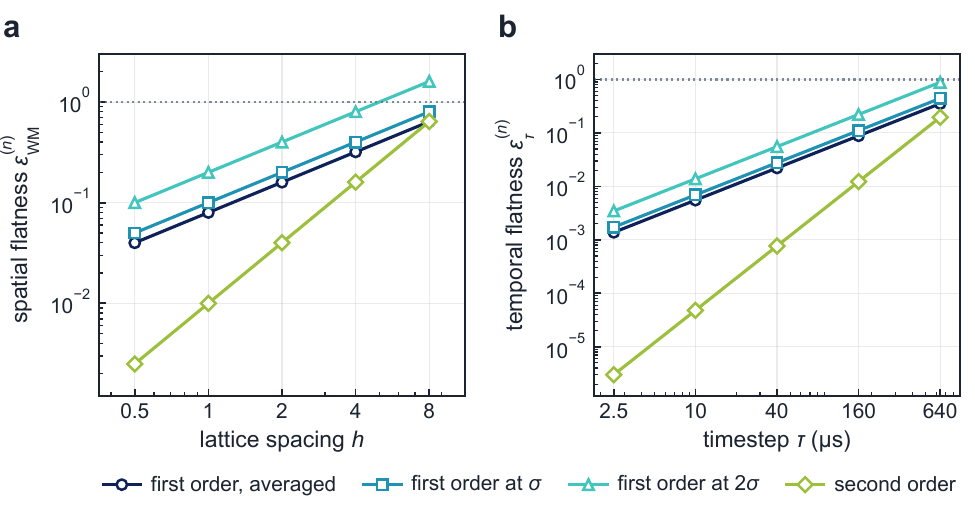}
    \caption{\textit{A priori} flatness parameters for the dragged harmonic potential
    Eq.~\ref{eq:draggedHarmonicPotential}. (a) Spatial,
    $\varepsilon^{(n)}_{\mathrm{WM}}=h^n\|D^n\Phi\|$, against lattice spacing, evaluated at three
    distances from the well center and averaged over the equilibrium Gaussian. (b) Temporal, $\varepsilon^{(n)}_{\tau}=\tau^n\|\partial_t^n\Phi\|$, against timestep. The second-order parameters are $\varepsilon^{(2)}_{\mathrm{WM}}=h^2k$ and $\varepsilon^{(2)}_{\tau}=3\tau^2kc^2\omega^2$. The dotted line marks
    unity. The five markers are the same five runs in both panels, since
    Eq.~\ref{eq:samplingBound} ties the timestep to the lattice spacing through
    $\tau=\varepsilon^{\mathrm{samp}}h^2/D$.
    Parameters: $D=10^{-12}\ \mathrm{m^2\,s^{-1}}$, $k=0.01$ in inverse squared box units so
    $\sigma=10$ box units, drive amplitude $c=40$ box units, drive frequency $\omega=Dk=100\
    \mathrm{s^{-1}}$, periodic cube $[-L,L]^3$ with $L=256$ box units, one box unit being $10$ nm.}
    \label{fig:wm-flatness-parameters}
\end{figure}

Next we assessed the flatness parameters of the same potential against lattice spacing
and timestep in \Cref{fig:wm-flatness-parameters}. Panel (a) shows that only the coarsest
lattice, $h=8$, violated spatial flatness. Panel (b) shows that the timestep the sampling bound assigns to each of those
lattice spacings kept the potential quasi-static at $\omega=Dk$. 

Finally we measured the two timestep conditions directly. We tested different driving
frequencies $w$ at a fixed lattice spacing, and tested the agreement with the predicted amplitude and phase lag. Then we tested
the timestep at two lattice spacings, which separated the two conditions because
$\varepsilon^{(n)}_{\tau}$ is not a function of $h$ while $\varepsilon^{\mathrm{samp}}$ falls as
$h^{-2}$.

\begin{figure}[htbp]
    \centering
    \includegraphics[max width=\linewidth,max height=0.80\textheight]{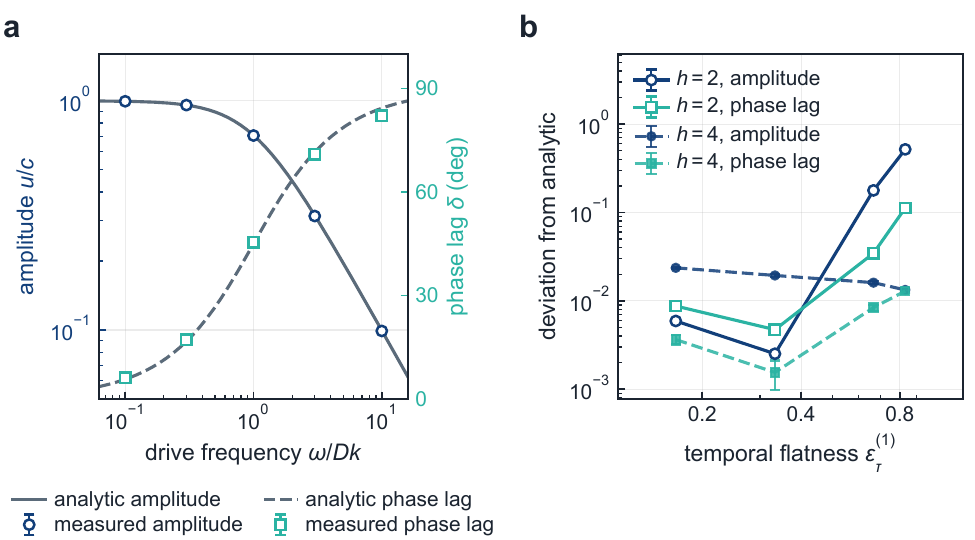}
    \caption{Measured response to the dragged harmonic potential against the analytic
    nonequilibrium steady state, at the parameters of \Cref{fig:wm-flatness-parameters}.
    (a) Frequency response at $h=2$ and $\tau=40\ \mu$s. Lines are the analytic amplitude $Dk/\sqrt{(Dk)^2+\omega^2}$ and phase lag $\arctan(\omega/Dk)$, markers the measured ones. Amplitude and phase lag come from a least-squares fit of $\mathrm{COM}(t)=a\sin\omega t+b\cos\omega t$ over the last two of three
    drive periods. (b) Timestep response at fixed $\omega=3Dk$, swept over
    $\tau\in\{40,80,160,200\}\ \mu$s at two lattice spacings. The abscissa is shared by the two
    series by construction; the sampling parameter is not: $\varepsilon^{\mathrm{samp}}=0.10$, $0.20$, $0.40$ and $0.50$ at $h=2$ against $0.025$, $0.05$, $0.100$ and $0.125$ at $h=4$. The temporal flatness is evaluated at $|x_v-c|=2\sigma$. Every point is the mean of ten independent replicates and the error
    bars are the standard error of that mean.}
    \label{fig:wm-omega}
\end{figure}

Panel (a) of \Cref{fig:wm-omega} reproduces the steady state over two decades in $\omega/Dk$,
the amplitude to better than $0.7\%$ and the lag to about $1\%$ near $\omega\sim Dk$. Panel (b)
separates the two timestep conditions. At matched $\varepsilon^{(1)}_{\tau}$, the $h=2$ data deviated by $18\%$ and $52\%$ in amplitude while the $h=4$ series stays flat. Hence, the binding constraint on the
timestep is Eq.~\ref{eq:samplingBound}.

\subsection[Subvolume-Average Quadrature Error]{Subvolume-Average Quadrature Error}\label{sec:si-quadrature}
The algorithm evaluates the centroid value $\rho(\mathbf{x}_v)$ in place of the subvolume average
$\langle\rho\rangle_V$ of Eq.~\ref{eq:voxelAverage}. We derive here the size of the resulting error,
which is Eq.~\ref{eq:quadratureErrorPDF}. Suppose $\mathbf{g}=\nabla\Phi(\mathbf{x}_v)$ and
$\mathbf{H}=D^2\Phi(\mathbf{x}_v)$ for the gradient and the Hessian of the potential at the
centroid, both evaluated once per subvolume.

The equilibrium density at a displacement $\boldsymbol{\delta}$ from the centroid is
\begin{align}
\rho(\mathbf{x}_v+\boldsymbol{\delta})
= \rho(\mathbf{x}_v)\exp\left[\Phi(\mathbf{x}_v)-\Phi(\mathbf{x}_v+\boldsymbol{\delta})\right].
\label{eq:quadBoltzmann}
\end{align}
Truncating Eq.~\ref{eq:potentialTaylorGeneral} at second order, we have
\begin{align}
\Phi(\mathbf{x}_v)-\Phi(\mathbf{x}_v+\boldsymbol{\delta})
= -\,\mathbf{g}\cdot\boldsymbol{\delta}
-\tfrac{1}{2}\boldsymbol{\delta}^{T}\mathbf{H}\boldsymbol{\delta}
+\Oh(h^3).
\label{eq:quadExponent}
\end{align}
We can then expand the exponential in this quantity, which is itself of order $h$, yielding
\begin{align}
\frac{\rho(\mathbf{x}_v+\boldsymbol{\delta})}{\rho(\mathbf{x}_v)}
= 1 - \mathbf{g}\cdot\boldsymbol{\delta}
- \tfrac{1}{2}\boldsymbol{\delta}^{T}\mathbf{H}\boldsymbol{\delta}
+ \tfrac{1}{2}\left(\mathbf{g}\cdot\boldsymbol{\delta}\right)^{2}
+ \Oh(h^3),
\label{eq:quadDensityExpansion}
\end{align}
which is Eq.~\ref{eq:well-mixingApprox}. The quadratic term of the exponential contributes only
$(\mathbf{g}\cdot\boldsymbol{\delta})^2/2$ at this order, since every other product is already of
order $h^3$.

Next, we calculate the average over the subvolume. We take advantage of the symmetries of the Cartesian lattice. 
Averaging Eq.~\ref{eq:quadDensityExpansion} over the subvolume requires only the first two moments
of $\boldsymbol{\delta}$ over the cube $[-h/2,h/2]^3$, and because the integrand factorizes over the
three orthogonal Cartesian directions. The first moment vanishes, 
\begin{align}
\left\langle \delta_i \right\rangle
= \frac{1}{h^3}\int_{C_v^3} \delta_i \dd^3\boldsymbol{\delta}
= \frac{1}{h}\int_{-h/2}^{h/2} \delta_i \dd\delta_i = 0 .
\label{eq:quadFirstMoment}
\end{align}
For the second moment the off-diagonal terms
factorize into a product of two such integrals and vanish, while the diagonal terms give
\begin{align}
\left\langle \delta_i^2 \right\rangle
= \frac{1}{h}\int_{-h/2}^{h/2} \delta_i^{2} \dd\delta_i
= \frac{1}{h}\cdot\frac{2}{3}\left(\frac{h}{2}\right)^{3}
= \frac{h^2}{12},
\qquad\text{so}\qquad
\left\langle \delta_i \delta_j \right\rangle = \frac{h^2}{12}\,\delta_{ij}.
\label{eq:voxelMoments}
\end{align}
The two quadratic forms of Eq.~\ref{eq:quadDensityExpansion} follow by contracting
Eq.~\ref{eq:voxelMoments} with the constant tensors they carry,
\begin{align}
\left\langle \boldsymbol{\delta}^{T}\mathbf{H}\boldsymbol{\delta}\right\rangle
&= \sum_{i,j} H_{ij}\left\langle \delta_i\delta_j\right\rangle
= \frac{h^2}{12}\sum_i H_{ii}
= \frac{h^2}{12}\,\mathrm{tr}\,\mathbf{H},
\nonumber \\
\left\langle \left(\mathbf{g}\cdot\boldsymbol{\delta}\right)^{2}\right\rangle
&= \sum_{i,j} g_i g_j \left\langle \delta_i\delta_j\right\rangle
= \frac{h^2}{12}\,\|\mathbf{g}\|^{2}.
\label{eq:quadContractions}
\end{align}
Collecting these and dropping the term linear in $\boldsymbol{\delta}$, which averages to zero by
Eq.~\ref{eq:quadFirstMoment},
\begin{align}
\frac{\langle \rho \rangle_V}{\rho(\mathbf{x}_v)}
= 1 - \frac{h^2}{24}\,\mathrm{tr}\,\mathbf{H} + \frac{h^2}{24}\|\mathbf{g}\|^{2} + \Oh(h^4)
= 1 + \frac{h^2}{24}\left[\|\mathbf{g}\|^2 - \mathrm{tr}\,\mathbf{H}\right] + \Oh(h^4).
\label{eq:voxelAverageExpansion}
\end{align}
The remainder is $\Oh(h^4)$ and not $\Oh(h^3)$ because the cube is symmetric under
$\boldsymbol{\delta}\to-\boldsymbol{\delta}$, so every odd moment vanishes and the third-order terms
of Eq.~\ref{eq:quadDensityExpansion}, all of which are odd in $\boldsymbol{\delta}$, average to zero
along with the linear one. Inverting Eq.~\ref{eq:voxelAverageExpansion} to the same order, which
amounts to reversing the sign of the bracket since it is already $\Oh(h^2)$, gives
Eq.~\ref{eq:quadratureErrorPDF}.

\begin{figure}[htbp]
    \centering
    \includegraphics[max width=\linewidth,max height=0.80\textheight]{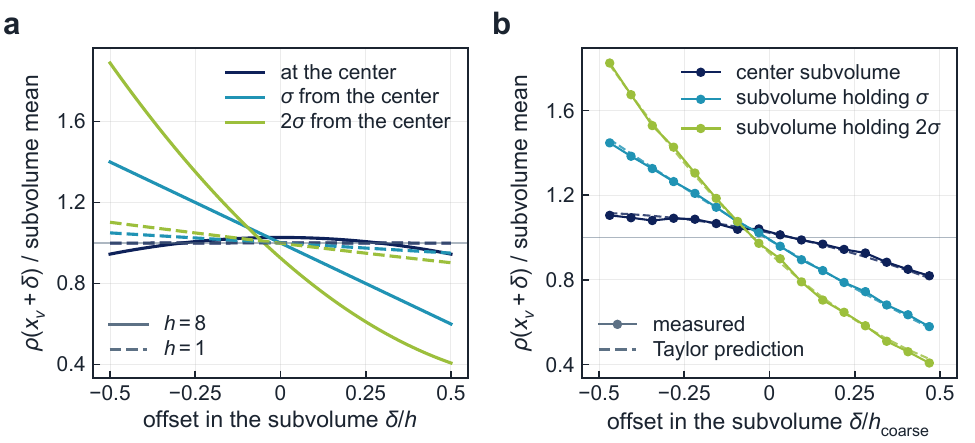}
    \caption{ 
    Deviations of the true subvolume-averaged distribution $\langle \rho \rangle_V$ from the measured distribution at a displacement $\delta$ from the subvolume centroid $\mathbf{x}_v$, 
    $\rho(x_v + \delta)$ as a function of $\delta/h_{\text{coarse}}$. 
    This measures intra-subvolume well-mixing deviations introduced by the MPD-RDDME quadrature approximation of the potential Eq.~\ref{eq:draggedHarmonicPotential}. 
    Curves closer to the unity $\rho(\mathbf{x}) / \langle \rho \rangle_V = 1$ represent conditions where well-mixing in the presence of the potential holds well. 
    Panel (a) shows the analytical predictions using Eq. ~\ref{eq:quadratureErrorPDF} at lattice spacings $h=8$ and $h=1$ for subvolumes at the well center and at $\sigma$ and $2\sigma$ from it ($|x_v - c|$), where $x_v$ is the subvolume centroid and $c$ is the well center.
    Panel (b) shows the resulting deviations as obtained by MPD-RDDME, obtained by comparing the coarse-grained results for $h=8$ with the fine-grained ones for $h=0.5$, for the subvolumes whose centroids lie at $|x_v-c|=4$, $12$ and $20$ ($0.4\sigma$, $1.2\sigma$ and $2\sigma$, with $\sigma=10$). 
    The dashed line in (b) is the analytical prediction from Eq.~\ref{eq:quadratureErrorPDF} at lattice spacing $h=8$.}
    \label{fig:wm-intravoxel}
\end{figure}

To illustrate the effects of well-mixing deviation of the subvolume introduced by the centroid quadrature approximation and the potential, 
we took the deviation of the particle density obtained using a coarse-grained subvolume with spacing $h_{\text{coarse}}=8$ 
with fine-grained subvolumes with spacings $h=0.5$ and $h=1$, as done in Fig.~\ref{fig:wm-intravoxel}. 
More precisely, if $\rho(\mathbf{x}_v+\boldsymbol{\delta})$ is the particle density 
measured in the fine-grained subvolume at a displacement of $\boldsymbol{\delta}$ from the centroid $\mathbf{x}_v$ of the coarse subvolume, 
we then plotted simulated values of $\rho(\mathbf{x}_v+\boldsymbol{\delta})/\rho(\mathbf{x}_v)$ as a function of $\delta/h_{\text{coarse}}$ 
which lies in the interval $[-0.5, 0.5]$. We then compared to the analytical prediction of Eq. \ref{eq:quadratureErrorPDF}.

\subsection{Errors in Reaction and Diffusion Propensities due to Quadrature Approximation}\label{sec:propensity}
Errors in the reaction and diffusion propensities would also be introduced by a subvolume-centroid quadrature approximation. 
Here, we derive corrections to these propensities under the small Doi reaction radius limit $r_b \ll h$. Note that 
this assumes that reactants are independently distributed, so no finite-size particle effects are considered. 

Consider a general chemical reaction $r$ involving $s$ reactant species, each with 
stoichiometric coefficient $\nu_{r,s}$ and reaction order $R = \sum_{s} \nu_{r,s}$. MPD-RDDME limits reaction order up to two. 
Assume that the reactant species are independently distributed within the subvolume with no two-body interactions between them, 
and that the Doi reaction radius satisfies $r_b \ll h$. Then in the quasi-static regime prescribed by Eq.~\ref{eq:temporalFlatnessCondition} 
the reaction propensity of $r$ in subvolume $v$, up to second order, is
\begin{align}
    \tilde{a}_r(v,t) &= a_r(v)\left[1+ \frac{h^2}{12} \sum_{i<j} \nabla \Phi_{s_i}(\mathbf{x}_v,t) \cdot \nabla \Phi_{s_j}(\mathbf{x}_v,t) + \mathcal{O}(h^4)\right],
    \label{eq:correctedReactionPropensities}
\end{align}
where
\begin{align}
    a_r(v) \approx \frac{k_r}{h^{3(R-1)}} \prod_{s} \frac{1}{\nu_{r,s}!}\frac{n_s(v)!}{(n_s(v) - \nu_{r,s})!} \notag
\end{align}
is the reaction propensity in the purely diffusive (zero potential) case,
$k_r$ is the time-independent macroscopic rate constant, and $\Phi_{s_i}(\mathbf{x}_v, t)$ is the potential energy arising from the interaction of specie $s_i$ with the time-dependent field, so that $\nabla\Phi_{s_i}$ is its gradient. For unimolecular reactions, the second-order correction vanishes. For bimolecular reactions $A + B \to C$, the second-order correction is
$h^2 \nabla \Phi_{A}(\mathbf{x}_v,t) \cdot \nabla \Phi_{B}(\mathbf{x}_v,t)/12$. The probability that $r$ fires in the interval 
$[t, t+\tau)$ follows from Eq.~\ref{eq:effectiveReactionPropensity}.

Subvolume-centroid quadrature would also introduce errors in the WPE diffusion propensities. This is because, from Eq. \ref{eq:potentialTaylorGeneral}, expanding the Boltzmann factor about the centroid gives
\begin{align}
    e^{-\Phi(\mathbf{x})} = e^{-\Phi(\mathbf{x}_v)} \left[1 - \nabla \Phi(\mathbf{x}_v) \cdot \boldsymbol{\delta} - \frac{1}{2} \boldsymbol{\delta}^T D^2 \Phi(\mathbf{x}_v) \boldsymbol{\delta} + \frac{1}{2}(\nabla \Phi(\mathbf{x}_v) \cdot \boldsymbol{\delta})^2 + \mathcal{O}(h^3)\right],
\end{align}
and the diffusion propensities are a function of the potential difference $\Phi_j(\mathbf{x}_v) - \Phi_i(\mathbf{x}_v)$ between the source and destination subvolumes $i$ and $j$, respectively.

\section[Subvolume Bulk-Fluid Approximation of the Hard-Sphere Functional]
{Subvolume Bulk-Fluid Approximation of the Volume Exclusion Functional}\label{sec:si-fmt}
This section derives Eq.~\ref{eq:betaFV} from fundamental measure theory~\citep{rosenfeld1989free} based on the 
White-Bear formulation~\citep{roth2002whitebear} and 
records the single-component and continuum limits quoted in \Cref{sec:fmt}.

By fundamental measure theory~\citep{rosenfeld1989free, roth2002whitebear}, we can write the excess free energy of a
hard-sphere mixture as a local functional of a set of weighted densities,
\begin{align}
    \beta\Fex[\{\rho_s\}] = \int_\Omega f_{\mathrm{HS}}\!\left(\{n_\alpha(\mathbf{x})\}\right)\, \dd^3\mathbf{x},
    \qquad
    n_\alpha(\mathbf{x}) = \sum_s \int \rho_s(\mathbf{x}')\, w^{(s)}_\alpha(\mathbf{x}-\mathbf{x}')\,\dd^3\mathbf{x}',
    \label{eq:fmtFunctional}
\end{align}
where the weight functions $w^{(s)}_\alpha$ encode the so-called fundamental geometric measures of a sphere
of radius $R_s = \sigma_s/2$. We have four scalars
\begin{align}
    w^{(s)}_3(\mathbf{r}) = \Theta(R_s - r), \quad
    w^{(s)}_2(\mathbf{r}) = \delta(R_s - r), \quad
    w^{(s)}_1 = \frac{w^{(s)}_2}{4\pi R_s}, \quad
    w^{(s)}_0 = \frac{w^{(s)}_2}{4\pi R_s^2},
    \label{eq:fmtWeights}
\end{align}
together with two vector weights that vanish in a homogeneous fluid. Here, $\Theta$ and $\delta$ are the Heaviside step function and 
Dirac-delta function, respectively. We use the form of $f_{\mathrm{HS}}$
given by Roth et al.~\citep{roth2002whitebear},
\begin{align}
    f_{\mathrm{HS}} = -n_0\ln(1-n_3) + \frac{n_1 n_2}{1-n_3}
    + \frac{n_2^3\left[\,n_3 + (1-n_3)^2\ln(1-n_3)\,\right]}{36\pi\, n_3^2 (1-n_3)^2},
    \label{eq:fHSbulk}
\end{align}

The subvolume bulk-fluid approximation treats each subvolume as a homogeneous fluid independent of the composition of neighboring 
subvolunes.
For a subvolume of volume $h^3$ holding $N_s$ particles each of diameter $\sigma_s$, the number densities are
$\rho_s = N_s/h^3$. With the weights of Eq.~\ref{eq:fmtWeights}, the integrals in Eq. \ref{eq:fmtFunctional} simplify to
\begin{align}
    n_0 = \rho, \qquad
    n_1 = \tfrac{1}{2}\rho M_1, \qquad
    n_2 = \pi \rho M_2, \qquad
    n_3 = \tfrac{\pi}{6}\rho M_3 = \eta,
    \label{eq:weightedBulk}
\end{align}
with $\rho = N/h^3$ and $M_n$, $\eta$ as defined in Eq.~\ref{eq:fmtMoments}. Substituting
Eq.~\ref{eq:weightedBulk} into Eq.~\ref{eq:fHSbulk} and multiplying by the subvolume volume,
$\beta F_V = h^3 f_{\mathrm{HS}}$, gives Eq.~\ref{eq:betaFV} term by term. The first is
$-h^3 n_0 \ln(1-\eta) = -N\ln(1-\eta)$. For the remaining two it is convenient to eliminate the
density through $\pi\rho = 6\eta/M_3$, which follows from the last equality of
Eq.~\ref{eq:weightedBulk}. The second term is then
\begin{align}
    h^3\,\frac{n_1n_2}{1-\eta} = \frac{N}{1-\eta}\cdot\frac{\pi\rho\, M_1 M_2}{2}
    = N\,\frac{3\eta}{1-\eta}\cdot\frac{M_1M_2}{M_3} = N\,\frac{3\eta\lambda}{1-\eta},
\end{align}
and the third, using $(\pi\rho)^2 = 36\eta^2/M_3^2$,
\begin{align}
    h^3\,\frac{n_2^3\left[\,\eta + (1-\eta)^2\ln(1-\eta)\,\right]}{36\pi\, \eta^2(1-\eta)^2}
    = N\,\frac{M_2^3}{M_3^2}\left[\frac{\eta}{(1-\eta)^2} + \ln(1-\eta)\right].
\end{align}
The polydispersity ratios of Eq.~\ref{eq:fmtMoments} satisfy
$\lambda^3/\gamma = (M_1M_2/M_3)^3 (M_3/M_1^3) = M_2^3/M_3^2$, so the last prefactor is
$\lambda^3/\gamma$ and the three terms together are Eq.~\ref{eq:betaFV}. Grouping the two logarithms
recovers the Boubl\'ik-Mansoori-Carnahan-Starling-Leland (BMCSL) free energy~\citep{mansoori1971equilibrium}, which is the bulk equation of state that Eq.~\ref{eq:fHSbulk}
was constructed to reproduce.

For a single species $M_n = \sigma^n$, so $\lambda = \gamma = 1$, the two logarithms cancel, and
\begin{align}
    \frac{\beta F_V}{N} = \frac{3\eta}{1-\eta} + \frac{\eta}{(1-\eta)^2}
    = \frac{\eta\,(4-3\eta)}{(1-\eta)^2},
    \label{eq:csLimit}
\end{align}
recovering the Carnahan-Starling excess free energy per particle~\citep{carnahan1969equation}. 

The subvolume bulk-fluid approximation is not valid at every lattice spacing. A subvolume of volume $h^3$ in
a fluid of number density $\rho$ holds
\begin{align}
    \langle N \rangle = \rho\, h^3 = \left(\frac{h}{\rho^{-1/3}}\right)^{3}
    \label{eq:voxelOccupancy}
\end{align}
particles on average, so the requirement $h \gg \rho^{-1/3}$ of \Cref{sec:fmt} is
$\langle N \rangle \gg 1$ written as a length.

In this approximation, the convolution in Eq.~\ref{eq:fmtFunctional} is approximated by replacing the weighted densities
\begin{align}
    n_\alpha(\mathbf{x}) = \sum_s \int \rho_s(\mathbf{x}')\, w^{(s)}_\alpha(\mathbf{x}-\mathbf{x}')\, \dd^3\mathbf{x}',
    \label{eq:weightedConvolution}
\end{align}
where the weight functions $w^{(s)}_\alpha$ of Eq.~\ref{eq:fmtWeights} have range $\sigma_s$, so a weighted density $n_\alpha(\mathbf{x})$ samples the
fluid over a particle diameter about $\mathbf{x}$, whereas Eq.~\ref{eq:weightedConvolution} replaces it by a
value from the occupancy of the single subvolume containing $\mathbf{x}$. That replacement is
valid only when the density varies slowly over a diameter, which for a density held
piecewise-constant on subvolumes means $h \gg \sigma$. 

Hecne, this approximation precludes the limit $h \to 0$. At fixed physical density the occupancy of
Eq.~\ref{eq:voxelOccupancy} vanishes almost everywhere as the lattice is refined, and the local free
energy does not converge to the functional of Eq.~\ref{eq:fmtFunctional}. The volume-exclusion model
therefore has no $h \to 0$ limit. In such cases, we resort to an RDDME with 
two-body volume-exclusion interactions, which has just been recently developed~\citep{heldman2026crddme}.

\section{Multi-Lattice Bonded Structures}\label{sec:structures}
In this section,
we elaborate on the structure moves, where large particles and polymers are represented 
with a multi-lattice representation.

Let a polymer $\mathcal{P}$ be an ordered set of occupied subvolumes
$\{v_1,\dots,v_N\}$, with a per-unit internal state $q_i$, and a set of bonds between
consecutive units. We can then assign any arbitrary Hamiltonian that describes the
interactions between the units of the polymer and the surrounding environment. For polymers, 
a natural choice would be the sum of the one-body potential energies of its units and bond 
energies,
\begin{align}
    U_{\mathcal{P}} = \sum_{i=1}^{N} \Phi_{s_i}(\mathbf{x}_{v_i}) + \sum_{\langle i,j\rangle} u_{ij}(q_i,q_j)
    {} + \kappa \sum_{\langle i,j,k\rangle} \left(1 - \cos\theta_{ijk}\right),
    \label{eq:structPotential}
\end{align}
where $u_{ij}$ is the bond energy between bonded units $i$ and $j$. Volume
exclusion free energies are accounted for by the first term $\Phi_{s_i}$.
The third sum runs over consecutive triples of bonded units, with $\theta_{ijk}$ the turn between
the two bond vectors meeting at unit $j$ and $\kappa$ the bending stiffness. This term is what gives
the polymer a persistence length as we 
will derive later for a certain class of lattice polymers. 

To capture a large set of possible bond angles in a discrete lattice, 
one might naively consider connecting subunits to the nearest twenty six neighbors in a cubic lattice
In this case, there are three possible bond lengths ($h, h\sqrt{2}, h\sqrt{3}$), and on the full set a bond's direction and its
length are one degree of freedom. This means that the bond and bend terms of Eq.~\ref{eq:structPotential} do not
control independent properties. Restricting the bond set to a single neighbor orbit (edge, face, body orbits) 
removes the coupling. Inspired by group theory, we use the term orbit to refer to a set of equivalent
bond vectors under the action of the cubic point group $O_h$. The three orbits in a cubic lattice are
the three cubic Bravais sublattices of the grid, the six face directions giving simple cubic (face orbit), the
eight body diagonals body-centered cubic (body orbit), and the twelve edge diagonals face-centered cubic (edge orbit). In each orbit 
every bond has the same length, so the bond energy is an additive constant that cancels
from every conformational energy difference and $\kappa$ is the only parameter left. Face-centered
cubic is the useful choice, as twelve is the largest coordination number available in three dimensions
at a single bond length~\citep{schutte1953problem,conway1999sphere}. This allows us to
keep a wider spectrum of semiflexible chains not available to face and body orbits, while preserving the independence of the bond 
and stretch terms.

Polymer moves are proposed by a kinetic Monte Carlo scheme at a size-dependent rate and accepted with the
WPE factor of the total potential energy change. For simplicity, we use the Rouse model~\citep{rouse1953theory}
\begin{align}
    D_T(N) \approx \frac{D_1}{N},
    \label{eq:rouseMobility}
\end{align}
with $D_1$ the translational diffusion coefficient of a single unit.

Algorithm~\ref{alg:structmove} describes one transport step for a polymer.

\begin{algorithm}[H]
\caption{Structure drift-diffusion hop}\label{alg:structmove}
\begin{algorithmic}[1]
\ParFor{polymer $\mathcal{P}$}
  \State $N \gets$ number of units in $\mathcal{P}$
  \State $D_T \gets D_1/N$ \Comment{Rouse model}
  \State draw a candidate move: a unit translation along a random axis.
  \State $\Delta U_{\mathcal{P}} \gets U_{\mathcal{P}}^{\mathrm{new}} - U_{\mathcal{P}}^{\mathrm{old}}$ \Comment{summed over occupied subvolumes and bonds}
  \State $M \gets D_T/h^2$
  \State $P \gets M\,\tau\,\mathcal{B}(\Delta U_{\mathcal{P}})$ 
  \If{the destination lies in the confining region and $\text{uniformRand}() \le P$}
    \State apply the move to all units of $\mathcal{P}$
  \EndIf
\EndParFor
\end{algorithmic}
\end{algorithm}

Polymers grow, shorten, sever, and anneal through reactions that add or remove a
bond. Each reaction is proposed at its base rate and accepted with the
Metropolis-Hastings factor of the bond-energy change, in the same manner as the
reaction operator of \Cref{sec:algorithm}. Hence, the bonded and unbonded states
satisfy detailed balance with respect to the bond energies of
Eq.~\ref{eq:structPotential}. Algorithm~\ref{alg:structbond} details one such 
reaction step for polymers.

\begin{algorithm}[H]
\caption{Structure reactions}\label{alg:structbond}
\begin{algorithmic}[1]
\ParFor{polymer $\mathcal{P}$}
  \For{each allowable reaction $c$ at an end bond or interior bond}
    \State $a_c \gets$ base attempt rate (propensity) of reaction $c$
    \If{$\text{uniformRand}() \le 1 - e^{-a_c \tau}$}
      \State $\Delta u \gets$ bond-energy change of firing $c$
      \If{$\text{uniformRand}() \le \min(1, e^{-\Delta u})$} \Comment{Metropolis-Hastings acceptance}
        \State apply $c$: add or remove the bond and update the occupied subvolume set
      \EndIf
    \EndIf
  \EndFor
\EndParFor
\end{algorithmic}
\end{algorithm}

Internal rearrangement of a linear polymer at fixed bond topology can then be sampled to replicate continuum Rouse dynamics, 
by using lattice movesets such as the Verdier-Stockmayer moveset~\citep{verdier1962monte,binder1997monte,shaffer1994effects},
with each move accepted by the same factor in the total energy change of
Eq.~\ref{eq:structPotential}. We adopt in this work the kink and end-bond flip moves from the Verdier-Stockmayer moveset, 
adapted for lattice polymers with edge-orbit bonds i.e., face-centered cubic lattice chains. We propose an alternative to 
the crankshaft move that forbids intra-chain strand crossings, exploiting the symmetries of the FCC lattice chain and show 
ergodicity by exhaustive AI-agent assisted computer enumeration for $N \leq 8$ subunits (see \Cref{sec:si-fcc}).

\section{Free-Volume Functional for a Body Larger than the Subvolume}\label{sec:si-freevol}
A body larger than the subvolume has an excluded volume spanning multiple lattice sites through a 
stencil of occupied sites. To the bath (particles smaller than the subvolumes), a body larger than the subvolume is a hard wall: it removes volume from the reach of
their centers and does not raise their packing (fraction of occupied volume)~\citep{reiss1959statistical,rosenfeld1989free}. Letting $w_v$ be the
fraction of subvolume $v$ excluded to the bath's centers, the bath occupies the free volume $V(1-w_v)$, and its excess free energy is
\begin{align}
    \beta F_{\mathrm{fv}}(v) = V(1-w_v)\,
    f_{\mathrm{HS}}\!\left(\left\{\frac{n^{\mathrm{bath}}_\alpha(v)}{1-w_v}\right\}\right).
    \label{eq:freevol}
\end{align}
Here $f_{\mathrm{HS}}$ is the volume-exclusion free-energy density of Eq.~\ref{eq:fHSbulk}, and the
$n^{\mathrm{bath}}_\alpha(v)$ for $\alpha=0,1,2,3$ are the four scalar weighted densities of
Eq.~\ref{eq:weightedBulk}, evaluated on the bath species of subvolume $v$ alone with the body's own
occupancy excluded from the sums. Dividing each by $1-w_v$ rescales the bath to the volume
available to it. We emphasize that the body's own motion is not coupled to the
bath through an energy; its friction is the collisional factor of \Cref{sec:si-diffusion-multi-site}. 
We elaborate in the next section.

\section{Friction factor for Particles Larger than the Subvolume}\label{sec:si-diffusion-multi-site}
A body larger than the subvolume is slowed by the surrounding particles it moves through.
Instead of using an energy functional, we derive the additional friction a hard-sphere bath exerts. 

Let a particle of diameter $\sT > h$ (which we call the tracer) and bare diffusion coefficient $\DT$ move among bath spheres of
diameter $\sB < h$, bare diffusion coefficient $\DB$, number density $\rhoB$ and volume fraction
$\phiB=\pi\rhoB\sB^3/6$, with hard-core interactions only and no hydrodynamic interactions. Its
long-time self-diffusion coefficient to first order in $\rhoB$ is
\begin{align}
\frac{\DL}{\DT} = 1 - \frac{1}{2}\,\frac{\DT}{\DT+\DB}\;\rhoB\,\frac{4\pi}{3}a^3 + \Oh(\rhoB^2),
\qquad a=\frac{\sT+\sB}{2},
\label{eq:frictionFirstOrder}
\end{align}
with $a$ the contact distance. We derive this using a derivation in the spirit
of Smoluchowski's treatment of a diffusing sphere among
others~\citep{smoluchowski1917koagulation}. In this, the tracer perturbs the bath around it and the perturbed
bath pushes back with an equivalent force. Force balance equations would then give the mobility. Throughout, long-time means after the tracer has moved far enough to have rearranged
its neighbors, which is the regime in which the friction is a
constant~\citep{cichocki1988longtime,medinanoyola1988longtime}.

Apply a weak constant force $\bm F$ to the tracer and compute its steady drift velocity $\bm v$.
The long-time mobility is $\mu_{\mathrm L}=v/F$ and the long-time self-diffusion coefficient follows
from the Einstein relation $\DL=\kT\mu_{\mathrm L}$, which holds for the linear response of a
Brownian tracer in a bath at equilibrium~\citep{dhont1996introduction,nagele1996dynamics}. To first
order in $\rhoB$ it is enough to treat one bath particle at a time, so consider the tracer at
$\bm r_{\mathrm T}$ and one bath sphere at $\bm r_{\mathrm B}$. Both are overdamped, so their
Langevin equations of motion are
\begin{align}
\frac{\dd \bm r_{\mathrm T}}{\dd t} = \beta \DT\left(\bm F + \bm F_{\mathrm{int}}\right) + \bm\xi_{\mathrm T}(t),
\qquad
\frac{\dd \bm r_{\mathrm B}}{\dd t} = -\beta \DB\,\bm F_{\mathrm{int}} + \bm\xi_{\mathrm B}(t),
\label{eq:frictionLangevin}
\end{align}
with $\bm F_{\mathrm{int}}$ the hard-core repulsive force the bath sphere exerts on the tracer. Note
independent white noises of strengths $\langle \xi_{{\mathrm T},i}(t)\xi_{{\mathrm T},j}(t')\rangle
= 2\DT\delta_{ij}\delta(t-t')$ and likewise for the bath. In the absence of hydrodynamic interactions
, the white noise terms are independent. Subtracting the two lines of Eq.~\ref{eq:frictionLangevin}, the relative coordinate
$\bm r=\bm r_{\mathrm B}-\bm r_{\mathrm T}$ therefore diffuses with
\begin{align}
D_{\mathrm r} = \DT + \DB
\label{eq:frictionDrel}
\end{align}
because the variances of independent noises add, and it carries a deterministic drift $-\bm u$ with
\begin{align}
\bm u = \beta \DT \bm F
\label{eq:frictionFreeVelocity}
\end{align}
the velocity the tracer would have in an empty bath. The hard core condition enters as a reflecting boundary
at $|\bm r| = a$.

Let $\rhoB\,g(\bm r)$ be the stationary bath density at separation $\bm r$ from the tracer. Its flux
in the tracer's frame has an advective part from the drift and a diffusive part,
$\bm j=-\bm u\,g-D_{\mathrm r}\nabla g$, and stationarity, impenetrability at contact and the
unperturbed far field give
\begin{align}
D_{\mathrm r}\nabla^2 g+\bm u\cdot\nabla g=0\quad(r>a),\qquad
\hat{\bm r}\cdot\bm j\big|_{r=a}=0,\qquad g\to 1\quad(r\to\infty),
\label{eq:pairStationary}
\end{align}
respectively.
Under a weak-force (weak tracer perturbation limit), we can linearize in $u$. Writing $g=1+g_1$ with $g_1=\Oh(u)$, the advective
term becomes $\bm u\cdot\nabla g = \bm u\cdot\nabla g_1 = \Oh(u^2)$ and drops out, leaving Laplace's
equation,
\begin{align}
\nabla^2 g_1 = 0 \quad (r>a), \qquad g_1 \to 0 \quad (r\to\infty).
\label{eq:frictionLaplace}
\end{align}
We use the drift in the boundary condition expression. Writing it out at $r=a$ and truncating tofirst order,
\begin{align}
\hat{\bm r}\cdot\bm j\big|_{r=a}
= -\left(\bm u\cdot\hat{\bm r}\right)\left(1+g_1\right) - D_{\mathrm r}\,\partial_r g_1\big|_{a}
= 0
\qquad\Longrightarrow\qquad
D_{\mathrm r}\,\partial_r g_1\big|_{a} = -\,u\cos\theta ,
\label{eq:frictionBC}
\end{align}
with $\theta$ the angle between $\bm r$ and $\bm u$, since
$\bm u\cdot\hat{\bm r}=u\cos\theta$ and the product $(\bm u\cdot\hat{\bm r})\,g_1$ is second order.

The solutions of Eq.~\ref{eq:frictionLaplace} that decay at infinity are the exterior solid
harmonics~\citep{jackson1999classical}
\begin{align}
g_1(r,\theta) = \sum_{\ell\ge 0} A_\ell\,\frac{P_\ell(\cos\theta)}{r^{\,\ell+1}} ,
\label{eq:frictionHarmonics}
\end{align}
with $P_\ell$ the Legendre polynomials. The boundary data in
Eq.~\ref{eq:frictionBC} is proportional to $\cos\theta = P_1(\cos\theta)$, and the Legendre
polynomials are orthogonal on $[-1,1]$, so every coefficient except $A_1$ vanishes and the response
is a pure dipole,
\begin{align}
g_1(r,\theta) = A\,\frac{\cos\theta}{r^{2}} .
\label{eq:frictionDipoleForm}
\end{align}
Differentiating, $\partial_r g_1 = -2A\cos\theta/r^{3}$, and substituting into
Eq.~\ref{eq:frictionBC} at $r=a$,
\begin{align}
D_{\mathrm r}\left(-\frac{2A\cos\theta}{a^{3}}\right) = -\,u\cos\theta
\qquad\Longrightarrow\qquad
A = \frac{u\,a^{3}}{2D_{\mathrm r}},
\label{eq:frictionAmplitude}
\end{align}
the $\cos\theta$ canceling on both sides, which is the consistency check that the dipole alone
satisfies the condition. The contact value of the pair distribution is therefore
\begin{align}
g(a,\theta)=1+\frac{u\,a}{2D_{\mathrm r}}\cos\theta .
\label{eq:frictionDipole}
\end{align}
The bath piles up ahead of the moving tracer and thins behind it, by a fraction set by how fast the
tracer moves compared with how fast the pair separation diffuses over the contact distance.

For hard spheres the only force between tracer and bath is the contact force, and its mean is the
contact pressure $\kT$ times the contact density, integrated over the sphere $r=a$ with the inward
normal~\citep{hansen2013theory},
\begin{align}
\langle\bm F_{\mathrm B}\rangle
= -\,\kT\,\rhoB\,a^{2}\oint g(a,\theta)\,\hat{\bm r}\,\dd\Omega .
\label{eq:frictionContactIntegral}
\end{align}
The isotropic part of $g$ contributes nothing, since $\oint \hat{\bm r}\,\dd\Omega = 0$. For the
dipole part, take $\hat{\bm u}$ along $z$ and write
$\hat{\bm r}=(\sin\theta\cos\varphi,\ \sin\theta\sin\varphi,\ \cos\theta)$. The $x$ and $y$
components vanish on the $\varphi$ integral, and with $\mu=\cos\theta$ the surviving component is
\begin{align}
\oint \cos\theta\,\hat{\bm r}\,\dd\Omega
= \hat{\bm u}\,2\pi\!\int_{-1}^{1}\!\mu^{2}\,\dd\mu
= \frac{4\pi}{3}\,\hat{\bm u}.
\label{eq:frictionAngular}
\end{align}
Substituting Eq.~\ref{eq:frictionDipole} and Eq.~\ref{eq:frictionAngular} into
Eq.~\ref{eq:frictionContactIntegral},
\begin{align}
\langle\bm F_{\mathrm B}\rangle
= -\,\kT\,\rhoB\,a^{2}\cdot\frac{u\,a}{2D_{\mathrm r}}\cdot\frac{4\pi}{3}\,\hat{\bm u}
= -\frac{2\pi}{3}\,\kT\,\rhoB\,a^{3}\,\frac{u}{D_{\mathrm r}}\,\hat{\bm u}.
\label{eq:frictionMeanForce}
\end{align}
The force opposes the motion, as a friction must.

The tracer's steady velocity is its mobility times the total force it feels,
\begin{align}
\bm v = \beta\DT\left(\bm F+\langle\bm F_{\mathrm B}\rangle\right).
\label{eq:frictionVelocity}
\end{align}
Substituting Eq.~\ref{eq:frictionMeanForce}, using $\beta\kT=1$ and dividing by the free velocity
$u=\beta\DT F$ of Eq.~\ref{eq:frictionFreeVelocity},
\begin{align}
\frac{v}{u}
= 1 - \beta\DT\,\frac{1}{u}\cdot\frac{2\pi}{3}\,\kT\,\rhoB\,a^{3}\,\frac{u}{D_{\mathrm r}}
= 1-\frac{2\pi}{3}\,\rhoB\,a^{3}\,\frac{\DT}{\DT+\DB},
\label{eq:frictionBalance}
\end{align}
and since the mobility is linear in the force at this order, $\DL/\DT=\mu_{\mathrm L}/\mu_0=v/u$.
Writing $(2\pi/3)a^3 = \tfrac12(4\pi/3)a^3$ recovers Eq.~\ref{eq:frictionFirstOrder}. 
The coefficient follows by writing both factors in terms of the size ratio $\lambda=\sB/\sT$. From
$\sT=\sB/\lambda$ the contact distance is $a=(\sT+\sB)/2=(\sB/2)(1+\lambda)/\lambda$, so
\begin{align}
\rhoB\,\frac{4\pi}{3}a^{3}
= \rhoB\,\frac{4\pi}{3}\cdot\frac{\sB^{3}}{8}\cdot\frac{(1+\lambda)^{3}}{\lambda^{3}}
= \underbrace{\frac{\pi\rhoB\sB^{3}}{6}}_{=\,\phiB}\cdot\frac{(1+\lambda)^{3}}{\lambda^{3}} ,
\label{eq:frictionCountFactor}
\end{align}
the bracket being the bath volume fraction by definition. With Stokes-Einstein scaling of the bare
coefficients, $D\propto1/\sigma$, we have $\DT/\DB=\sB/\sT=\lambda$ and therefore
\begin{align}
\frac{\DT}{\DT+\DB} = \frac{\lambda}{1+\lambda} .
\label{eq:frictionShareFactor}
\end{align}
Multiplying Eq.~\ref{eq:frictionCountFactor} by Eq.~\ref{eq:frictionShareFactor} and by the one
half, the powers of $(1+\lambda)$ and $\lambda$ combine as
$(1+\lambda)^{3}\lambda/\lambda^{3}(1+\lambda) = (1+\lambda)^{2}/\lambda^{2}$, so
\begin{align}
\frac{\DL}{\DT} = 1 - c\,\phiB,\qquad c = \frac{(1+\lambda)^2}{2\lambda^2}.
\label{eq:frictionC}
\end{align}
For equal spheres $\lambda=1$ and $c=2$, the classical result $\DL=\DT(1-2\phi)$ of Smoluchowski
theory without hydrodynamic
interactions is obtained~\citep{hanna1982self,ackerson1982correlations,lekkerkerker1984calculation,cichocki1988longtime}.
For a large tracer $c\to(\sT/\sB)^2/2$, so the friction from a free-draining bath grows as the square
of the size ratio. The same first-order structure for unequal spheres appears in Batchelor's
polydisperse theory in the limit of no hydrodynamic interaction~\citep{batchelor1983diffusion} and in
the matched-asymptotic derivation of the cross-diffusion equations for hard spheres of two sizes by
Bruna and Chapman~\citep{bruna2012excluded,bruna2012multiple}. 

Eq.~\ref{eq:frictionFirstOrder} is a statement about the friction to first
order, $\zeta_{\mathrm L}=\zeta_{\mathrm T}(1+c\phiB)$. Writing the diffusion coefficient as the
inverse of an additive friction,
\begin{align}
\frac{\DL}{\DT}=\frac{1}{1+c\,\phiB},
\label{eq:frictionResummed}
\end{align}
agrees with Eq.~\ref{eq:frictionC} to first order and stays positive at every $\phiB$.

We can extend this to several bath species. At first order in the densities the mean force
of Eq.~\ref{eq:frictionContactIntegral} is a sum of independent one-particle contributions, one per
species, each with its own contact distance and its own relative diffusion coefficient, so the
frictions add and
\begin{align}
\frac{\DL}{\DT}=\Bigl[1+\sum_s c_s\,\phi_s\Bigr]^{-1},
\qquad
c_s=\frac{1}{2}\,\frac{\DT}{\DT+D_s}\,\frac{(1+\lambda_s)^3}{\lambda_s^3},
\qquad \lambda_s=\sigma_s/\sT,
\label{eq:frictionPoly}
\end{align}
with the bare coefficients left general, since a lattice model need not tie $D$ to $\sigma$.

Eq.~\ref{eq:frictionResummed} is the free-draining limit, in which every
bath particle in the body's way must diffuse out of it. 

In our lattice-based framework, we opt for the free-draining limit. Since the
comparisons of \Cref{sec:finitesize} are against a Brownian reference without hydrodynamics, of the
kind whose self-diffusion has been checked against Monte Carlo dynamics for hard
bodies~\citep{sanz2010dynamic,behringer2013brownian}, the
free-draining coefficients of Eq.~\ref{eq:frictionPoly} are the ones used throughout. 

\section{Motion of a Body Larger than the Subvolume}\label{sec:si-bodymotion}

We found that in a previous version of the code (simulation results not shown), 
a body whose diameter exceeds the lattice spacing cannot be confined to lattice sites without quantizing its displacement at its own radius. Its center
therefore lives on a sub-lattice of spacing $\delta=h/Q$ with $Q$ chosen to be powers of two, and its
hard-sphere ``footprint'' is re-drawn at the new center after every accepted step. The bath sees
that footprint through the free-volume functional of \Cref{sec:si-freevol}, so the coupling from body
to bath is an exclusion, exactly as for any other occupant of a subvolume.

For a step from $x$ to $x'=x+\delta\hat e$ the probability per unit time (propensity) is
\begin{align}
p(x\to x') = \frac{\DT}{\delta^{2}}\; m(x,x')\;
\mathcal{B}\bigl(\Delta U_{\mathrm{own}}(x\to x')\bigr),
\qquad
m(x,x') = \Bigl[1+\sum_s c_s\,\overline{\phi}_s(x,x')\Bigr]^{-1},
\label{eq:bodyRate}
\end{align}
with $\mathcal{B}$ the Wang-Peskin-Elston factor of \Cref{sec:rddme} and $c_s$ the coefficients of
Eq.~\ref{eq:frictionPoly}. The energy
$\Delta U_{\mathrm{own}}$ contains the body's own terms only, which are its hard core against other
bodies, against walls and against bonded polymers, together with any external one-body potential,
and it contains nothing from the bath. The bath enters solely through $m$, a mobility factor on the
rate, so the body is slowed by the bath but not pushed by it.

The average $\overline{\phi}_s(x,x')$ is the packing
fraction of species $s$ over the layer of subvolumes immediately outside the body's footprint, taken
symmetrically for the step. In our implementation, the mean of the two endpoint
values is taken. The subvolumes the body itself covers is not included in the packing fraction calculation. They are depleted by the body's own
excluded volume, to a small fraction of the bulk value, whereas the closure of
\Cref{sec:si-diffusion-multi-site} was derived for the bulk density at contact, which on this lattice is first
found one layer further out. For a $20$ nm sphere on $10$ nm subvolumes that layer is the
$5^3-3^3=98$ subvolumes surrounding the body's own $3^3$ block. 

One can then ask if this preserves detailed balacne, and here we show that it does. 
The factor $m$ is symmetric on each step, $m(x,x')=m(x',x)$, and the
Wang-Peskin-Elston factor satisfies $\mathcal{B}(u)/\mathcal{B}(-u)=e^{-u}$. Hence
\begin{align}
\frac{p(x\to x')}{p(x'\to x)}
=\frac{\mathcal{B}(\Delta U_{\mathrm{own}})}{\mathcal{B}(-\Delta U_{\mathrm{own}})}
=e^{-\Delta U_{\mathrm{own}}},
\label{eq:bodyDetailedBalance}
\end{align}
and the stationary distribution of the body is proportional to $e^{-U_{\mathrm{own}}}$ whatever the
field $m(x)$. In a homogeneous bath with no external potential the body is uniformly distributed, as
it must be, since a friction that varies in space does not by itself produce a density gradient.

The continuum limit of a walk with
symmetric step factors and a position-dependent rate is
\begin{align}
\partial_t\rho=\nabla\cdot\bigl[D(x)\,\bigl(\nabla\rho+\rho\nabla U_{\mathrm{own}}\bigr)\bigr],
\qquad D(x)=\DT\,m(x),
\label{eq:bodyFokkerPlanck}
\end{align}
which is the Fokker-Planck form whose equilibrium is the Boltzmann distribution independent of
$D(x)$. 

\section{Face-centered Cubic Lattice Chains}\label{sec:si-fcc}
\Cref{sec:structures} gives the reason for restricting the twenty six neighbors of the cubic grid to
a single orbit, and for preferring the twelve edge diagonals. In this section, we provide further 
justifiactions on using fcc lattice chains to model polymer dynamics. We also describe the 
ergodicity-preserving Rouse dynamics moves 
for the fcc chain. One convenient property of FCC chains is that, sans the dynamics, they cannot pass through one another
(intra-chain strand crossings) unless they lie on different halves of the lattice.

We preface with some definitions. A (FCC) bond is one of the twelve permutations of $(\pm1,\pm1,0)$, each of
length $\sqrt{2}\,h$, so direction is the only freedom a bond has and the bond energy cancels from
every conformational energy difference. Each such step changes $x+y+z$ by $0$ or $\pm2$, so the
parity of $x+y+z$ is conserved along a chain, and every unit of one chain therefore sits on the same
half of the grid. We call that half the chain's parity class. Each class, with these twelve vectors
as its nearest neighbors, is a face-centered cubic lattice.

We now show that two FCC chains can cross only if they occupy opposite parity classes. 
Consider a unit square with corners $\mathbf{x}$, $\mathbf{x}+\mathbf{e}_i$,
$\mathbf{x}+\mathbf{e}_i+\mathbf{e}_j$ and $\mathbf{x}+\mathbf{e}_j$ for some $i \neq j$, with
$\mathbf{e}_i$ the standard basis vectors. Its two diagonals are
\begin{align}
D_1=\left\{\mathbf{x},\; \mathbf{x}+\mathbf{e}_i+\mathbf{e}_j\right\},
\qquad
D_2=\left\{\mathbf{x}+\mathbf{e}_i,\; \mathbf{x}+\mathbf{e}_j\right\},
\label{eq:fccDiagonals}
\end{align}
and writing $\pi(\mathbf{x})=(x+y+z)\bmod 2$ for the parity, both endpoints of $D_1$ have parity
$\pi(\mathbf{x})$, since their coordinate sums differ by two, while both endpoints of $D_2$ have
parity $\pi(\mathbf{x})+1$. The two diagonals of any unit square therefore lie in opposite parity
classes, so a crossing pair of edge bonds always has its two bonds in opposite classes. Since a
chain occupies a single class, no chain on this orbit can cross itself, and two chains can cross
only when they occupy different classes.

\subsection{FCC Lattice Chain Moveset}\label{sec:fcc-moveset}
In this section, we propose a moveset for FCC lattice chains that preserves the topology of each chain, 
eliminating the need for topology checks in the code for a chain on its own. 
The moveset borrows from the Verdier-Stockmayer algorithm~\citep{verdier1962monte,verdier1966montecarlo} but adapts it for FCC lattices 
without the use of the crankshaft. These moves capture the Rouse relaxation dynamics of polymers~\citep{ubertini2021computer}

\begin{figure}[htbp]
\centering
\includegraphics[max width=\linewidth,max height=0.80\textheight]{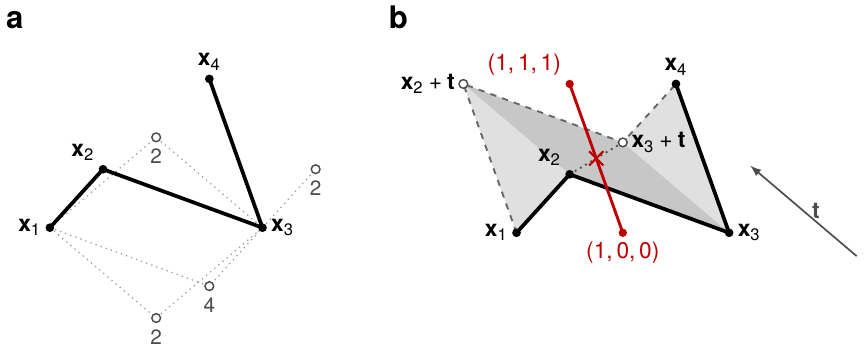}
\caption{Our moveset of \Cref{sec:fcc-moveset} acting on a four-unit chain (polymer). 
(a) Units $\mathbf x_1\ldots\mathbf x_4$; the three kink destinations of
$\mathbf x_2$ and one end-bond rotation of $\mathbf x_4$ are open circles labeled by their squared
chord length (see Fig.\ref{fig:si-apex}) in units of $h^2$, and the dotted lines are the new bonds. (b) The segment
$\{\mathbf x_2,\mathbf x_3\}$ translated by the edge vector $\mathbf t=(-1,0,1)$ to the open circles.
The anchor bonds sweep the two lighter equilateral triangles and the internal bond the darker
rhombus, whose short diagonal is the dotted segment from $\mathbf x_2$ to $(1,0,1)$. The red bond
is an example of a bond that is of an opposite parity class and crosses that diagonal at its midpoint.}
\label{fig:si-fccmoves}
\end{figure}

The moveset consists of four types of moves and is illustrated in \Cref{fig:si-fccmoves}.
\begin{enumerate}[leftmargin=2.2em,itemsep=2pt]
  \item \textbf{Full kink.} A unit with bonded neighbors $\mathbf{a}$ and $\mathbf{b}$ moves to any
        other site that is a common neighbor of both. The proposal is symmetric and detailed-balance preserving, 
        since the set of common
        neighbors depends only on the anchors, which the move leaves fixed.
  \item \textbf{End-bond rotation.} A terminal unit is given a new bond direction drawn uniformly
        from the $12$ bond directions.
  \item \textbf{Edge translation of a segment.} For two units of the same chain and an edge vector
        $\mathbf{t}$, every unit strictly between them is displaced by $\mathbf{t}$, admissible when
        the two bonds at the ends of the segment remain edge bonds.
  \item \textbf{Collective translation} of a whole chain by a face vector, a unit step along one
        coordinate axis. The three moves above displace units only by edge vectors, which preserve
        the parity of the coordinate sum, so on their own they confine a chain to the parity class
        in which it was placed. The face step is the one move that carries a chain between the two
        classes, and it is what makes the sampler irreducible over every position of the chain in
        the box. Because it changes the parity relation between two chains, the move set alone does
        not conserve the mutual topology of chains that share a box; where that topology must be
        kept, a crossing rule is required.
\end{enumerate}

It is quite easy to see that 
a full kink move alone is not ergodicity-preserving. Consider 
the chain connecting $(0,0,0)$, $(1,1,0)$, $(0,2,0)$ and $(-1,1,0)$, an example of a ``four-cycle''. This is a closed FCC lattice chain 
where all four bonds are edge vectors and all four sites are of the same parity.  At each unit, the flip move targets
the opposite corner of the square, which is occupied. Hence, every flip is blocked by self-avoidance. However,
other four-cycles exist, so a sampler using the flip alone is not ergodic.

We tested various combination of movesets that are ergodic by exhaustive computer enumeration. Self-avoiding cycles were counted up
to lattice translation, cyclic relabeling and reversal, giving $970$ distinct cycles at $N=6$ and
$42\,069$ at $N=8$, from $11\,640$ and $673\,104$ rooted ones. Taking these as the vertices of a
graph whose edges are the moves of a given set that respect self-avoidance,
\Cref{tab:si-moveclasses} counts its connected components.

\begin{table}[htb]
\caption{Connected components of the move graph on all self-avoiding cycles of $N$ units on the edge
orbit. Cycles identified up to translation, relabeling and orientation. The second column of each
pair is the size of the largest component, out of $970$ at $N=6$ and $42\,069$ at $N=8$. A move set
reaches every conformation only when the count is one.}
\label{tab:si-moveclasses}
\begin{ruledtabular}
\begin{tabular}{lcccc}
 & \multicolumn{2}{c}{$N=6$} & \multicolumn{2}{c}{$N=8$}\\
move set & components & largest & components & largest\\
\colrule
flip alone & 82 & 44 & 378 & 1\,226\\
crankshaft, spans two to six & 45 & 910 & 177 & 41\,577\\
flip and crankshaft & 21 & 910 & 25 & 41\,577\\
sideways kinks alone & 13 & 958 & 235 & 41\,571\\
full kink & 1 & 970 & 1 & 42\,069\\
full kink and crankshaft & 1 & 970 & 1 & 42\,069\\
full kink and edge translation & 1 & 970 & 1 & 42\,069\\
flip and edge translation & 1 & 970 & 1 & 42\,069\\
\end{tabular}
\end{ruledtabular}
\end{table}

\subsection{Construction of a Self-Avoiding Closed FCC Lattice Chain}\label{sec:construction-closed-fcc}
\begin{figure}[htbp]
\centering
\includegraphics[max width=\linewidth,max height=0.80\textheight]{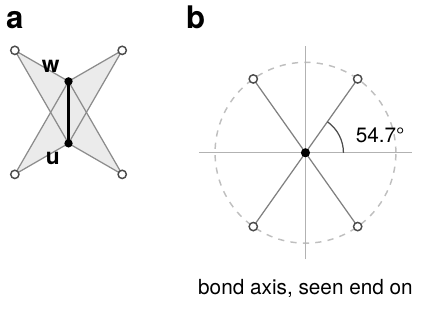}
\caption{Illustration of an apex. (a) Every edge bond belongs to exactly four triangles, and they are
equilateral: each apex is $\sqrt2\,h$ from both ends, the same distance the two ends are from each
other.
(b) Looking along the bond. All four apexes lie in its perpendicular bisector plane at the same
distance from the axis, so they sit on a circle.}
\label{fig:si-apex}
\end{figure}

For reference, we present an algorithm for construction of self-avoiding closed FCC lattice chains of any length $N \geq 3$. 
We preface with the definition of a bond ``apex''. Each edge bond has exactly four common
neighbors, which we call the four apexes i.e., the vertices of the four equilateral triangles shared by a bond. For example, 
for the bond $(1,1,0)$, the apexes are $(1,0,1)$, $(1,0,-1)$, $(0,1,1)$ and $(0,1,-1)$. \Cref{fig:si-apex} shows the four
triangles and the circle their apexes lie on.

Starting from an FCC bond, we can grow the chain by a move that cannot change its topology, adapting 
the midpoint displacement algorithmic construction of self-avoiding polymers (SAP) used to build lattice chromosome
models~\citep{gilbert2021chromosome}.
An insertion replaces the bond between two units by two bonds through one of its four
apexes, raising the unit count by one. Provided the apex is empty and inside the confining region,
the insertion preserves closure. It also preserves
self-avoidance, since no other unit moves. It cannot change the knot type either: the only surface the
chain sweeps is the triangle spanned by the old bond and the two new ones, and since those are edge
bonds every lattice site strictly inside that triangle belongs to the opposite parity class, so no
strand of the chain, which occupies one class only, can lie inside it.

The closed FCC chain growth procedure is then short. For an $N$-subunit chain,

\begin{enumerate}
    \item Place a triangle inside the region on the chosen parity class, which is planar. 
    \item Select a random bond on the closed chain, and identify the four apexes.
    \item Connect the endpoints of the bond to a selected apex that is not occupied or outside the region.
    \item Repeat until the desired chain length is reached.
\end{enumerate}
\begin{figure}[htbp]
\centering
\includegraphics[max width=\linewidth,max height=0.80\textheight]{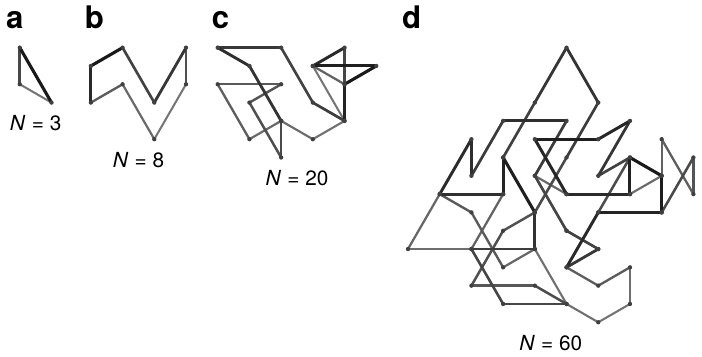}
\caption{Sample closed-chain generator output at four points of a single run, in
isometric projection, drawn back to front so that nearer strands overlay farther ones.}
\label{fig:si-growth}
\end{figure}

The chain is then equilibrated at the target stiffness. To assist in equilibriation, intermediate relaxation steps can be introduced 
during insertions. \Cref{fig:si-growth} follows one chain
through the procedure at four points of a single run.

\subsection{Exact Statistics of the Phantom Chain on the Edge Orbit}\label{sec:si-fcc-exact}
The statistics of the phantom FCC chains without excluded volume can be solved in closed form, from the bond
set to the mean squared end-to-end distance. In this section, we derive the exact formulas for the
mean cosine of the turn angle, the mean squared end-to-end distance, and the persistence length
of the phantom FCC chain. A calculation of the mean cosine of the turn angle for lattice chains was previously done
by Zhang et al.~\citep{zhang2019persistence}.

A configuration of a chain of $M$ subunits (an $M$-mer) is a sequence $(a_1,\dots,a_M)$ with each $a_j$ one of the twelve
edge directions. Its energy is the bending sum of Eq.~\ref{eq:structPotential} over the $M-1$
interior joints, and since the orbit is monodisperse (same length) the bond energy has already canceled, so with
energies in units of $\kT$ the Boltzmann weight of the sequence factors into one weight per joint,
\begin{align}
e^{-E(a_1,\dots,a_M)} = \prod_{j=1}^{M-1} w_{a_j a_{j+1}},
\qquad
w_{ab} \equiv e^{-\kappa\left(1-\cos\theta_{ab}\right)},
\label{eq:fccWeight}
\end{align}
where $\theta_{ab}$ is the turn between consecutive bond directions $a$ and $b$. Everything below
follows from one property of $w_{ab}$: the sum of the weights available to a joint does not depend
on the direction the chain arrived along i.e., Markovian. Writing that sum as
\begin{align}
z(a) \equiv \sum_{b} w_{ab} = \sum_{\theta} f_\theta\, e^{-\kappa(1-\cos\theta)},
\label{eq:fccOneJoint}
\end{align}
the multiplicities $f_\theta$ counting how many of the twelve directions make angle $\theta$ with
$a$. With the multiplicities $1,4,2,4,1$ at
$\theta = 0^\circ, 60^\circ, 90^\circ, 120^\circ, 180^\circ$ from \Cref{sec:si-fcc},
\begin{align}
z = 1 + 4e^{-\kappa/2} + 2e^{-\kappa} + 4e^{-3\kappa/2} + e^{-2\kappa},
\label{eq:fccRowsum}
\end{align}
which is the one-joint partition function.

The partition function of the whole chain follows by summing one bond at a time from the far end
inward. The last direction $a_M$ appears in a single factor, so
\begin{align}
Z_M = \sum_{a_1}\cdots\sum_{a_M} \prod_{j=1}^{M-1} w_{a_j a_{j+1}}
= \sum_{a_1}\cdots\sum_{a_{M-1}} \left(\prod_{j=1}^{M-2} w_{a_j a_{j+1}}\right)
\underbrace{\sum_{a_M} w_{a_{M-1}a_M}}_{=\,z}
= z\,Z_{M-1},
\label{eq:fccRecursion}
\end{align}
the inner sum telescoping to $z$ whatever $a_{M-1}$ is, which is exactly the property just
established. A chain of one bond has $Z_1=12$, so iterating Eq.~\ref{eq:fccRecursion} gives
\begin{align}
Z_M = 12\, z^{\,M-1} .
\label{eq:fccZclosed}
\end{align}
There are twelve ways to lay down the first bond, and every joint after it contributes an
independent factor $z$. The free energy $-\kT\ln Z_M$ is therefore extensive with $\ln z$ per joint
and no interaction between joints, exactly as for a chain of independent units.

Two consequences follow. First, the marginal distribution of an interior bond
is uniform. Fixing $a_j=a$ and summing over everything on either side of it, each side collapses one
bond at a time as in Eq.~\ref{eq:fccRecursion}, giving $z^{\,j-1}$ from the bonds before it and
$z^{\,M-j}$ from those after, neither depending on $a$, so
\begin{align}
P(a_j = a) = \frac{z^{\,j-1}\,z^{\,M-j}}{Z_M} = \frac{z^{\,M-1}}{12\,z^{\,M-1}} = \frac{1}{12}
\label{eq:fccMarginal}
\end{align}
at every stiffness and every position along the chain. The bending energy favors no lattice
direction, which is a symmetry statement and not an approximation. Second, the conditional law of
the next bond is the one-joint weight normalized, since the factors downstream of the joint cancel
between numerator and denominator,
\begin{align}
P(a_{j+1}=b \mid a_j = a)
= \frac{w_{ab}\; z^{\,M-j-1}}{\sum_{b'} w_{ab'}\; z^{\,M-j-1}}
= \frac{w_{ab}}{z} .
\label{eq:fccMarkov}
\end{align}
The direction sequence is therefore a Markov chain whose transition rule is the same at every joint.

The bond length average is trivial due to the monodispersity of the FCC chain.
\begin{align}
\langle b\rangle = \sum_a p_a\,\sqrt{2}\,h = \sqrt{2}\,h,
\qquad
\langle b^2\rangle = 2h^2 ,
\label{eq:fccMoments}
\end{align}
at every $\kappa$ and every chain length. 

The turns themselves are not merely Markov but independent. Given $a_j$, the probability that the
turn at joint $j$ equals $\theta$ is obtained by summing Eq.~\ref{eq:fccMarkov} over the $f_\theta$
directions making that angle with $a_j$, and by transitivity the result does not depend on $a_j$;
since the turn at the next joint is drawn the same way from $a_{j+1}$, it is independent of how the
chain reached there. The turns are therefore i.i.d., with
\begin{align}
P(\theta) = \frac{f_\theta\, e^{-\kappa(1-\cos\theta)}}{z} ,
\qquad
\langle\cos\theta\rangle
= \frac{\sum_\theta f_\theta \cos\theta\; e^{-\kappa(1-\cos\theta)}}{\sum_\theta f_\theta\, e^{-\kappa(1-\cos\theta)}} .
\label{eq:fccMeancos}
\end{align}
Substituting the angle table term by term, the numerator is
\begin{align}
\sum_\theta f_\theta \cos\theta\, e^{-\kappa(1-\cos\theta)}
&= (1)(1)(1) + (4)\left(\tfrac{1}{2}\right)e^{-\kappa/2} + (2)(0)e^{-\kappa}
\nonumber \\
&\qquad {}+ (4)\left(-\tfrac{1}{2}\right)e^{-3\kappa/2} + (1)(-1)e^{-2\kappa}
\nonumber \\
&= 1 + 2e^{-\kappa/2} - 2e^{-3\kappa/2} - e^{-2\kappa} \;\equiv\; z_1 ,
\label{eq:fccZ1}
\end{align}
so that, with $z$ from Eq.~\ref{eq:fccRowsum},
\begin{align}
r \;\equiv\; \langle\cos\theta\rangle \;=\; \frac{z_1}{z}
\;=\; \frac{1 + 2e^{-\kappa/2} - 2e^{-3\kappa/2} - e^{-2\kappa}}
{1 + 4e^{-\kappa/2} + 2e^{-\kappa} + 4e^{-3\kappa/2} + e^{-2\kappa}} .
\label{eq:fccRatio}
\end{align}
The continuum worm-like chain~\citep{kratky1949rontgenuntersuchung} is the same formula with the
twelve directions replaced by the sphere: sending $\sum_\theta f_\theta$ to
$\int\sin\theta\,\dd\theta$ turns Eq.~\ref{eq:fccMeancos} into $\coth\kappa-1/\kappa$, the Langevin
function. 

The tangent correlation is a single exponential, and we derive it here. The classical
freely-rotating-chain argument~\citep{flory1969statistical} needs one lattice-specific observation. 
That is, for any weight $u(\theta)$ depending only on the angle,
\begin{align}
\sum_b u(\theta_{ab})\,\mathbf{b}_b = c\,\mathbf{b}_a,
\qquad
c = \sum_b u(\theta_{ab})\cos\theta_{ab}.
\label{eq:fccLemma}
\end{align}
To see this, let $G_a$ be the subgroup of the octahedral group $O_h$ fixing $\mathbf{b}_a$. It permutes the
twelve bonds among themselves and preserves every angle to $\mathbf{b}_a$, so it leaves the vector
$\mathbf{v}=\sum_b u(\theta_{ab})\mathbf{b}_b$ invariant. For $\mathbf{b}_a=(1,1,0)$, $G_a$ contains
the mirror exchanging $x$ with $y$ and the mirror sending $z$ to $-z$; a vector fixed by both has
$v_x=v_y$ and $v_z=0$, and so lies along $(1,1,0)=\mathbf{b}_a$. Hence $\mathbf{v}=c\,\mathbf{b}_a$
for some scalar $c$, and taking the dot product of both sides with $\mathbf{b}_a$ and dividing by
$|\mathbf{b}_a|^2=2$ identifies $c$ as stated. Transitivity carries the result to every $a$.
Applying Eq.~\ref{eq:fccLemma} with $u(\theta_{ab})=w_{ab}/z$, which is the conditional law of
Eq.~\ref{eq:fccMarkov}, gives the conditional mean of the next bond,
\begin{align}
\mathbb{E}\!\left[\mathbf{b}_{j+1} \,\middle|\, \mathbf{b}_j = \mathbf{b}_a\right]
= \sum_b \frac{w_{ab}}{z}\,\mathbf{b}_b
= \left(\sum_b \frac{w_{ab}}{z}\cos\theta_{ab}\right)\mathbf{b}_a
= \frac{z_1}{z}\,\mathbf{b}_a
= r\,\mathbf{b}_a ,
\label{eq:fccCondmean}
\end{align}
the middle equality being Eq.~\ref{eq:fccLemma} and the next one the definition of $z_1$ in
Eq.~\ref{eq:fccZ1}. The perpendicular part of the next bond averages to zero and the parallel part
is reduced by exactly $r$. Iterating by the tower property of conditional expectation,
$\mathbb{E}[\mathbf{b}_n \mid \mathbf{b}_0] = r^{\,n}\,\mathbf{b}_0$, and therefore
\begin{align}
C(n) \equiv \left\langle \mathbf{b}_0\cdot\mathbf{b}_n\right\rangle
= \left\langle \mathbf{b}_0\cdot\mathbb{E}\!\left[\mathbf{b}_n \mid \mathbf{b}_0\right]\right\rangle
= r^{\,n}\left\langle b^2\right\rangle = 2h^2\,r^{\,n} .
\label{eq:fccCorr}
\end{align}

Writing
$C(n)=\langle b^2\rangle e^{-n\ln(1/r)}$ and measuring contour distance as $n\langle b\rangle$, the
decay length of the tangent correlation is
\begin{align}
L_p = -\frac{\langle b\rangle}{\ln r} = -\frac{\sqrt{2}\,h}{\ln r},
\qquad
\frac{L_p}{\langle b\rangle} = -\frac{1}{\ln r} ,
\label{eq:fccLp}
\end{align}
which is the freely-rotating-chain relation with the discrete mean cosine of Eq.~\ref{eq:fccRatio} in
place of the fixed cosine of the classical model. For the size, expand
$\mathbf{R}=\sum_{i=1}^{M}\mathbf{b}_i$ and count pairs. There are $M$ diagonal terms, and for each
separation $n$ there are exactly $M-n$ ordered pairs $(i,j)$ with $j-i=n$, so
\begin{align}
\left\langle R^2\right\rangle_M
&= \sum_{i}\left\langle \mathbf{b}_i^2\right\rangle
+ 2\sum_{i<j}\left\langle \mathbf{b}_i\cdot\mathbf{b}_j\right\rangle
= 2h^2 M + 2\sum_{n=1}^{M-1}(M-n)\,C(n)
\nonumber \\
&= 2h^2\left[M + 2\sum_{n=1}^{M-1}(M-n)\,r^{\,n}\right].
\label{eq:fccR2sum}
\end{align}
Using the finite geometric series and its derivative,
\begin{align}
\sum_{n=1}^{M-1}r^{\,n}=\frac{r-r^{M}}{1-r},
\qquad
\sum_{n=1}^{M-1}n\,r^{\,n}
= \frac{r\left[1-Mr^{M-1}+(M-1)r^{M}\right]}{(1-r)^2},
\label{eq:fccGeometric}
\end{align}
we get
\begin{align}
\sum_{n=1}^{M-1}(M-n)\,r^{\,n}
= M\,\frac{r-r^{M}}{1-r} - \frac{r\left[1-Mr^{M-1}+(M-1)r^{M}\right]}{(1-r)^2}
= \frac{r\left[M(1-r)-\left(1-r^{M}\right)\right]}{(1-r)^{2}} .
\label{eq:fccGeosum}
\end{align}
Substituting Eq.~\ref{eq:fccGeosum} into Eq.~\ref{eq:fccR2sum},
\begin{align}
\left\langle R^2\right\rangle_M
= 2h^2\left[ M + \frac{2r\left(M(1-r)-\left(1-r^{M}\right)\right)}{(1-r)^{2}}\right]
\;\xrightarrow[M\to\infty]{}\;
2h^2 M\,\frac{1+r}{1-r}.
\label{eq:fccR2}
\end{align}
This limit is the Kuhn result, an ideal chain of $M$ steps with an
effective segment inflated by $(1+r)/(1-r)$. Eq.~\ref{eq:fccR2} is exact at every $M$. 
At large $\kappa$ every term of Eq.~\ref{eq:fccRatio} beyond the
$60^\circ$ one is negligible against $e^{-\kappa/2}$, and keeping the first two terms of numerator
and denominator gives $r \simeq (1+2e^{-\kappa/2})/(1+4e^{-\kappa/2}) \simeq 1-2e^{-\kappa/2}$,
whence $L_p/\langle b\rangle \to \tfrac{1}{2}e^{\kappa/2}$. The persistence length is exponential
rather than linear in the stiffness, and the continuum relation $L_p=\kappa\langle b\rangle$ does not
hold at any stiffness.

\section{Comparison of FCC Chain Dynamics with an Off-Lattice Reference}\label{sec:si-chain-match}

\begin{figure}[htbp]
    \centering
    \includegraphics[max width=\linewidth,max height=0.72\textheight]{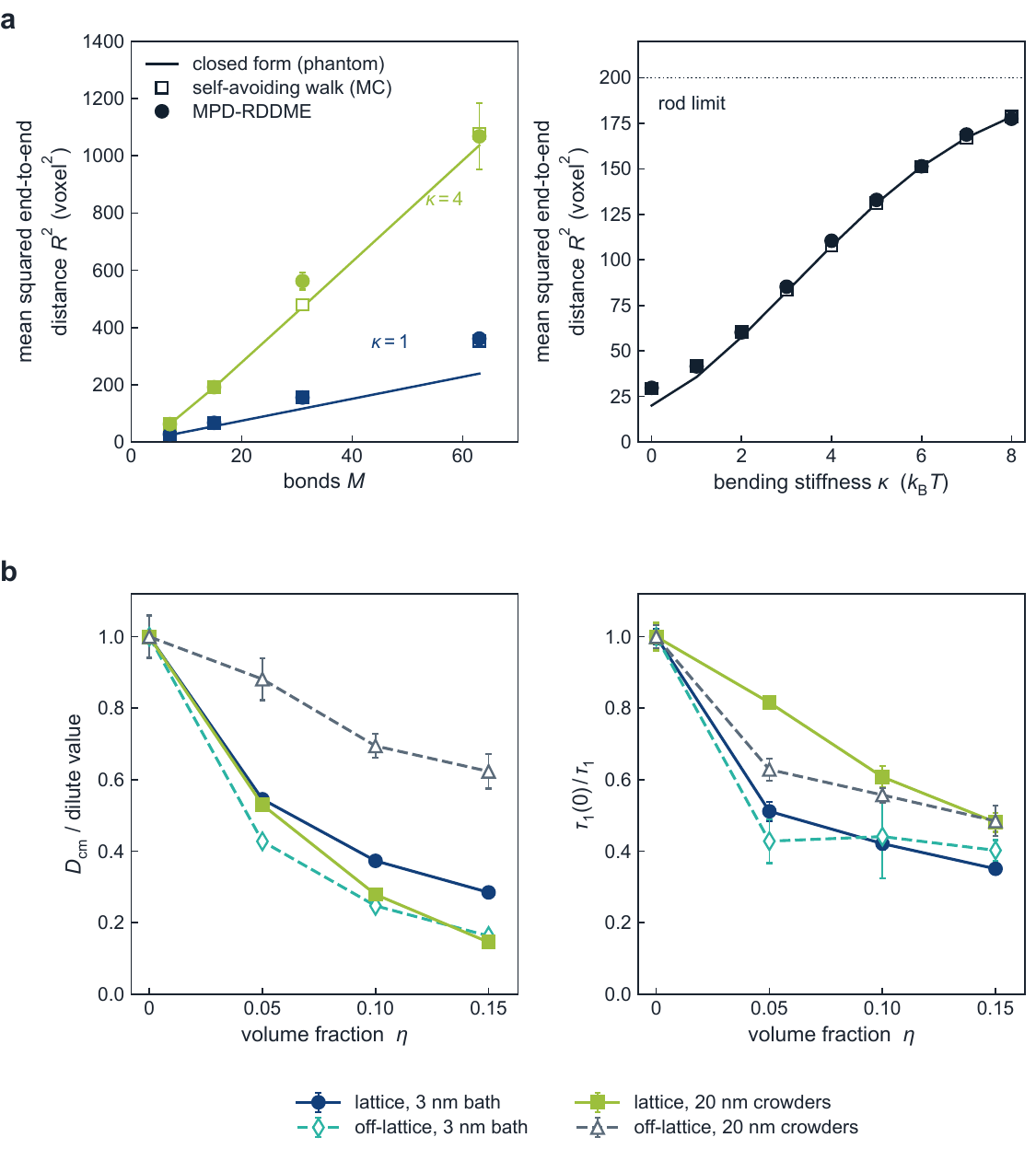}
    \caption{Lattice polymers. (a) Chain size $\langle R^2\rangle$ without crowders. Points are the lattice simulation,
    lines the closed form for a phantom fcc chain, and open squares self-avoiding walks; the gap between the two
    reference curves is the effect of excluded volume. (b) Center-of-mass diffusion $D_{\mathrm{cm}}$ and Rouse first
    relaxation mode $\tau_1$ (see \eqref{eq:rouseTimes}), each divided by its value without crowders. Lower means
    slower. Solid lines are the lattice and dashed lines the off-lattice reference, error bars standard errors over
    runs.}
    \label{fig:polymer}
\end{figure}

For polymer dynamics, the lattice and the reference ReaDDy cannot be directly matched on a microscopic stiffness. Their
bending energies have different functional forms; that is, one in the cosine of the turn and one quadratic in
the turn itself, and the lattice never samples bends below $35^\circ$. Instead we match four
emergent quantities. 

The first is the bond length. On the edge orbit it is fixed at $\sqrt{2}\,h$. 
The second is the persistence length, which is a function of $\kappa$ by way
of Eq.~\ref{eq:fccLp} and the reference through its own bending constant. 
The third is the diffusion coefficient $D_1$ of a single subunit, which
on the lattice is the bare hop rate times $h^2$ and in the reference is an input. The fourth is the
bead diameter, which sets the excluded volume in both. Once these four agree, the entire relaxation spectrum 
can be dictated by Rouse theory~\citep{rouse1953theory}.

\begin{figure}[htbp]
    \centering
    \includegraphics[max width=\linewidth,max height=0.72\textheight]{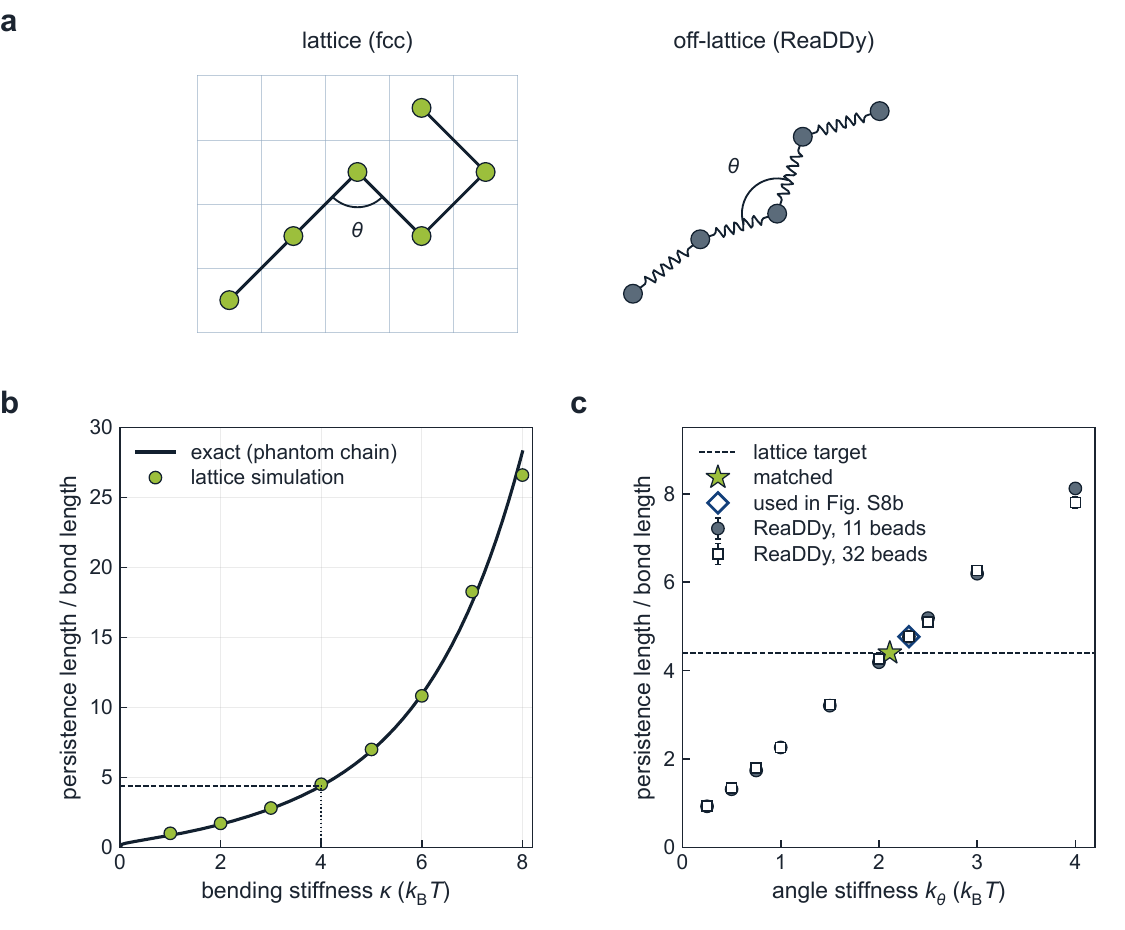}
    \caption{Matching the persistence length of the off-lattice reference chain to the lattice chain. (a) The two
    chains, with $\theta$ the angle at a joint ($\theta=\pi$ for a straight chain): on the lattice, bonds of length
    $\sqrt2\,h$ on one fcc sublattice meet at discrete $\theta$ with bending energy $\kappa(1+\cos\theta)$, the term
    of Eq.~\ref{eq:structPotential} with the turn $\theta_{ijk}=\pi-\theta$; off the lattice, harmonic bonds with angle
    energy $\kappa(\theta-\pi)^2$.
    (b) Lattice: $L_p/\langle b\rangle$ against $\kappa$, the closed form for the phantom chain (line) and the simulated
    lattice chain (points); the dashed line is the $\kappa=4$ value. (c) Reference: $L_p/\langle b\rangle$ against
    $\kappa$, for chains of $11$ (circles) and $32$ beads (squares), each point eight runs started as coils and as straight chains, with standard errors smaller than the
    symbols. The star is the matched $\kappa=2.11$; the open diamond is $\kappa=2.31$, the value used for \Cref{fig:polymer}b. The dotted line is the lattice target at $\kappa=4$.}
    \label{fig:lp-match}
\end{figure}

The persistence length match is an inversion of a measured relation (\Cref{fig:lp-match}). Both chains are measured
with the same estimator, $L_p/\langle b\rangle=-1/\ln\langle\cos\theta_{ijk}\rangle$ from the turn $\theta_{ijk}$
between neighboring bonds. On the lattice, $L_p$ follows from $\kappa$ by Eq.~\ref{eq:fccLp}, and the simulated chain reproduces that
relation (\Cref{fig:lp-match}b); at $\kappa=4$, the stiffness of \Cref{fig:polymer}b, the target is
$L_p/\langle b\rangle=4.40$. The reference relation has no closed form, so we scanned the angle stiffness $\kappa$ of
the $11$-bead chain of \Cref{fig:polymer}b on a grid, with eight runs per point, and inverted it by local interpolation
(\Cref{fig:lp-match}c). A root finder stepping on the difference of two stochastic evaluations converges on whichever
pair of draws straddled the target and reports a precision it does not have, and a global polynomial fit through the
grid is biased by eight percent at the stiff end, because the relation has curvature a quadratic cannot follow over two
decades. The reference also has an equilibration problem of its own: at the stiff target, a run of $4\times10^5$ steps
from a coil start stops short and looks converged. Each run therefore lasted $3.2\times10^6$ steps and was started both
as a coil and as a straight chain, which approach equilibrium from opposite sides; the two starts agree within one
standard error at every $\kappa$. The match is $\kappa=2.11$.

The reference runs of \Cref{fig:polymer}b used $\kappa=2.31$, which an earlier match had obtained from shorter runs
started as coils only. At that stiffness, the converged scan gives $L_p/\langle b\rangle=4.77\pm0.05$, which is $8\%$
above the lattice target (\Cref{fig:lp-match}c). Rerunning the dilute chain (four runs) and the densest
crowder point ($\eta=0.15$, three runs of $2$~s) at the matched $\kappa=2.11$ leaves the ratios of \Cref{fig:polymer}b
unchanged within their standard errors: at $\eta=0.15$, $D_{\mathrm{cm}}$ relative to its dilute value moves from
$0.62\pm0.05$ to $0.64\pm0.01$, and $\tau_1(0)/\tau_1$ from $0.49\pm0.04$ to $0.48\pm0.01$. The relation does not depend
on chain length: a $32$-bead chain gives $L_p/\langle b\rangle=4.77\pm0.06$ at $\kappa=2.31$ and matches the target at
$\kappa=2.08$ (\Cref{fig:lp-match}c).

The dynamical comparison is made on the Rouse modes. For a chain of $N$ subunits at positions
$\mathbf{R}_n$ with free ends, they are the discrete cosine transform
\begin{align}
\mathbf{X}_p = \frac{1}{N}\sum_{n=1}^{N} c^{(p)}_n\,\mathbf{R}_n ,
\qquad
c^{(p)}_n = \cos\!\left[\frac{p\pi\left(n-\tfrac{1}{2}\right)}{N}\right],
\qquad p = 0,1,\dots,N-1 ,
\label{eq:rouseModes}
\end{align}
with $\mathbf{X}_0$ the center of mass.

The relaxation times are where the match earns its keep. In the Rouse model~\citep{rouse1953theory} the modes are
independent and each decays exponentially,
$\langle \mathbf{X}_p(t)\cdot\mathbf{X}_p(0)\rangle = \langle \mathbf{X}_p^2\rangle e^{-t/\tau_p}$,
with
\begin{align}
\tau_p = \frac{\langle b\rangle^{2}}{12\,D_1 \sin^{2}\!\left(\dfrac{p\pi}{2N}\right)}
\;\xrightarrow[p\ll N]{}\;
\frac{N^{2}\langle b\rangle^{2}}{3\pi^{2}D_1\,p^{2}} ,
\label{eq:rouseTimes}
\end{align}
obtained by eliminating the bead friction in favor of $D_1=\kT/\zeta$ and the spring constant in
favor of the bond length. Every quantity on the right is one of the four matched above, so
Eq.~\ref{eq:rouseTimes} contains nothing that was fitted to the reference. Comparing the measured mode
autocorrelations between the lattice and the reference therefore tests the dynamics directly: the
relaxation times test whether the lattice moveset relaxes the same conformations on the same clock.

\section{GPU implementation of MPD-RDDME}\label{sec:si-gpu}

This section describes how the simulation framework runs on the GPU. The physics of the model is
given in the main text.

\subsection{Input-Output and Time step Loop}\label{sec:si-gpu-io}

As in Lattice Microbes~\cite{roberts2013lattice}, the user specifies a box of
$n_x\times n_y\times n_z$ cubic subvolumes of side $h$, regions in which reaction and diffusion rules
can differ, the species with their initial populations and diffusion coefficients $D_s$, the
reactions with their stoichiometries and rate constants, and the run settings (timestep $\Delta t$,
run length, random seed, and output intervals). MPD-RDDME adds the
species-dependent external potentials $\Phi_s^{\mathrm{ext}}(\mathbf r,t)$ and the hard-sphere
diameters $\sigma_s$, from which the solver evaluates the excess free energy $\beta F_v$ of every
subvolume [Eq.~\eqref{eq:betaFV}]. Large spheres and lattice polymers are optional inputs.

\begin{figure}[htbp]
    \centering
    \includegraphics[max width=\linewidth,max height=0.74\textheight]{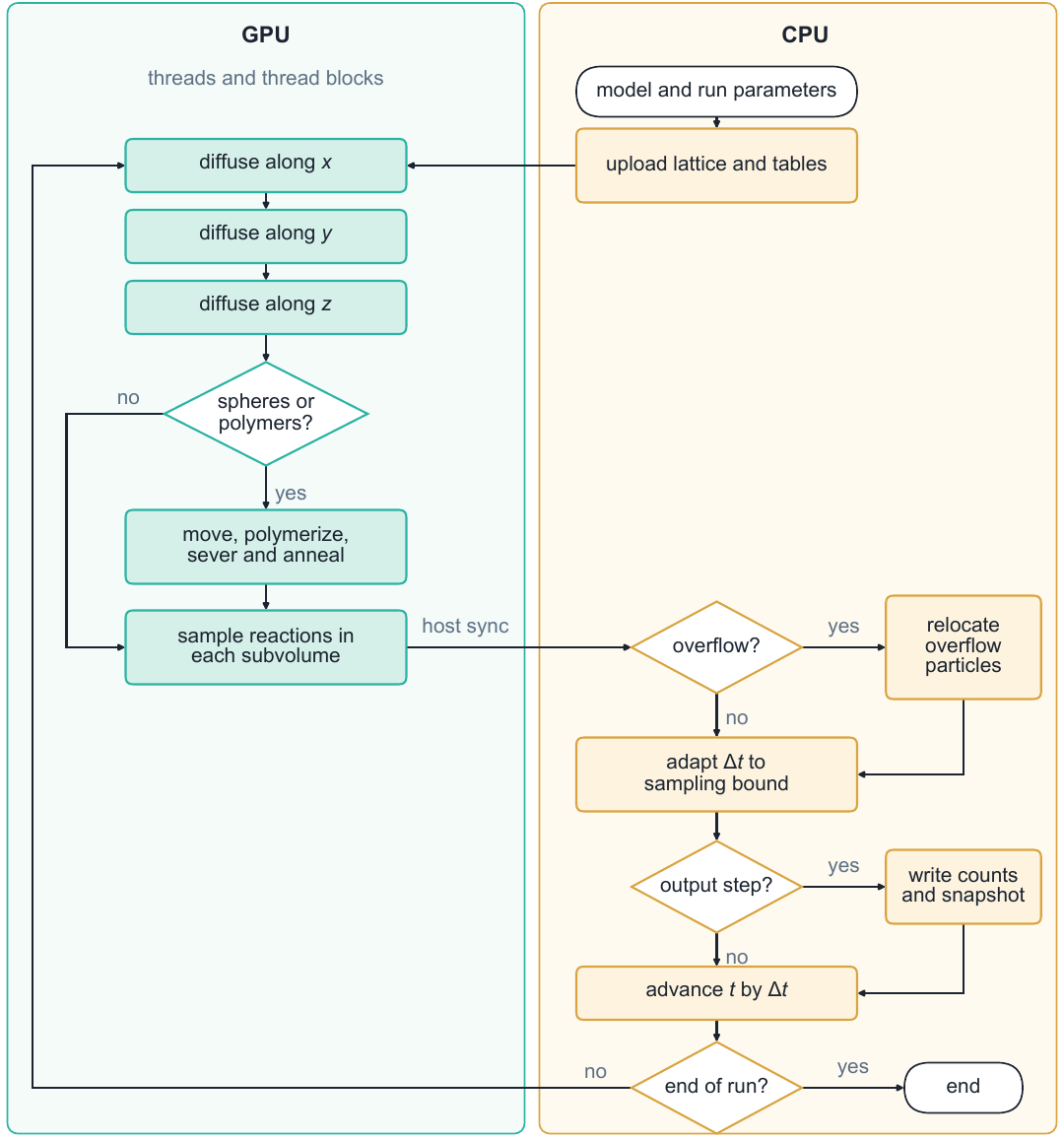}
    \caption{One timestep of MPD-RDDME. GPU steps (left) run with one thread per subvolume for the
    diffusion passes and the reactions, and with one thread block per sphere or chain for the structure
    update, which runs only when spheres or polymers are present. The CPU (right) then reads the
    overflow list and the step summary. Particles that did not fit into a full subvolume (64 slots) are
    relocated after the lattice is copied to the host, and the lattice is copied back. With adaptive
    time stepping, $\Delta t$ is reduced or increased by $10\%$ according to the largest $P_++P_-$ of
    the step [Eq.~\eqref{eq:samplingBound}]. Species counts and lattice snapshots are written at the
    output intervals.}
    \label{fig:si-gpu-io}
\end{figure}

Each timestep applies the $x$, $y$, and $z$ diffusion passes, the structure update when spheres or
polymers are present, and the reaction operator, and then advances $t\gets t+\Delta t$
(Fig.~\ref{fig:si-gpu-io}). All of these operators run on the GPU, and the lattice stays in GPU
memory for the whole run. After each step the CPU handles the particles that overflow a full
subvolume, adjusts the timestep, and writes the species counts and lattice snapshots at the chosen
intervals (Sec.~\ref{sec:si-gpu-host}).

\subsection{Data in GPU Memory}\label{sec:si-gpu-data}

A particle is stored only as the one-byte index of its species, so a model can contain up to 255
species and individual particles are not tracked. Each subvolume has 64 particle slots, which is a
compile-time setting. Four slots are packed into one 32-bit word, so a subvolume occupies 16 words.
The lattice is stored word-major: word $j$ of subvolume $v$ sits at word address $jV+v$, where
$v=x+n_x(y+n_y z)$ and $V$ is the number of subvolumes. Threads that handle consecutive subvolumes
therefore read consecutive addresses. Two copies of this lattice in GPU
global memory serve as the read and write buffers of the diffusion passes.

Global memory also holds the site type of each subvolume (one byte), the transition table of species-
and site-dependent hop factors $D_s\Delta t/h^2$, the external-potential fields, the reaction tables
(reactants, rates, and the site types where each reaction may occur), and a $256\times4$ packing table
of the factors $(\pi/6h^3)\sigma_s^n$, $n=0,\dots,3$. Summing a table row over the particles of a
subvolume gives $N$, $\eta$, and the moments $M_n$ of Eq.~\eqref{eq:fmtMoments}. For structures, a
per-subvolume overlay records the sphere or chain that owns the subvolume, the bond direction to the
next subunit, and the role of each subunit, and a registry holds the center of each sphere and the
ends of each chain. A bond direction is an index into a fixed table of the 26 neighbor offsets, of
which only the 12 fcc edge directions are used for bonds in this work. No neighbor lists are stored:
the face neighbors of $v$ are $v\pm1$, $v\pm n_x$, and $v\pm n_xn_y$. Particle hops are periodic at
the box faces, whereas structure moves treat them as walls. Two small buffers, the overflow list and
a per-step summary (the largest hop probability and the number of topology events), are in host
memory mapped into the GPU address space, so the CPU reads them without a copy.
Table~\ref{tab:si-gpu-memory} lists where each of these is stored.

\begin{table}
 \caption{Placement of the solver data in GPU memory.}
 \label{tab:si-gpu-memory}
 \newcommand{\mcell}[2]{\parbox[t]{#1}{\raggedright #2\strut}}
 \begin{tabular}{@{}lll@{}}
  \toprule
  Memory & Scope and lifetime & Contents \\
  \midrule
  Global & \mcell{0.26\linewidth}{All threads; persists across kernels} &
   \mcell{0.50\linewidth}{Read and write copies of the particle lattice; site types;
   external-potential fields; structure overlay and registry; large-sphere footprint;
   packing table} \\ \addlinespace
  Constant & \mcell{0.26\linewidth}{All threads, read-only; cached, 64~KiB} &
   \mcell{0.50\linewidth}{Lattice dimensions; species and reaction counts; reaction and
   transition tables; scalar parameters; pointers to global arrays} \\ \addlinespace
  Shared & \mcell{0.26\linewidth}{One thread block; lifetime of the block} &
   \mcell{0.50\linewidth}{Lattice tile with single-subvolume apron; site types;
   hop choices; per-subvolume $\eta$, FMT moments and free energy} \\ \addlinespace
  Registers & \mcell{0.26\linewidth}{One thread} &
   \mcell{0.50\linewidth}{Unpacked word; running FMT moment sums; hop energies and
   probabilities; random numbers} \\ \addlinespace
  Mapped host & \mcell{0.26\linewidth}{Host and GPU} &
   \mcell{0.50\linewidth}{Overflow list; per-step summary (largest hop probability,
   number of topology events)} \\
  \bottomrule
 \end{tabular}
\end{table}

\begin{figure}[htbp]
    \centering
    \includegraphics[max width=\linewidth,max height=0.70\textheight]{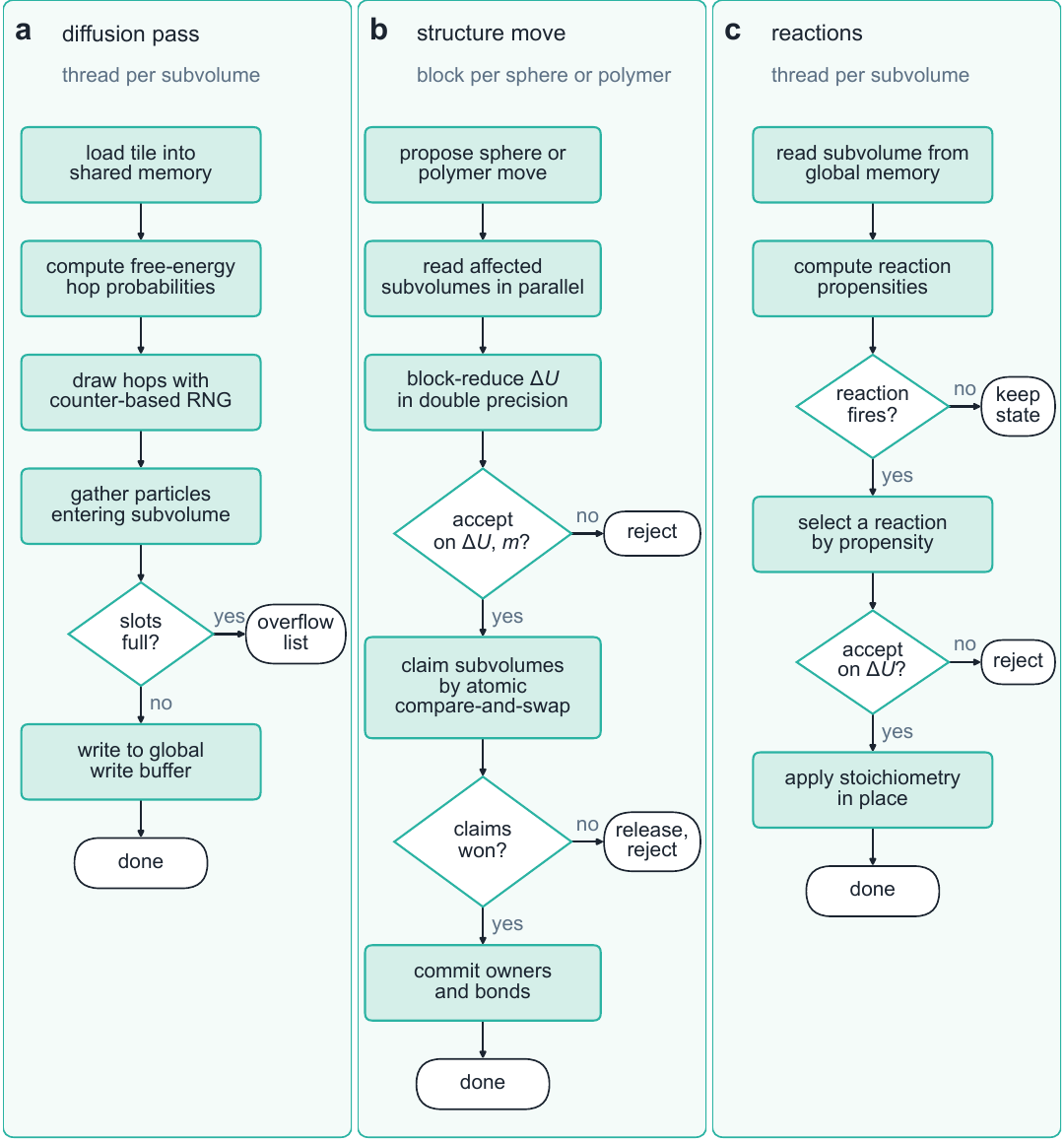}
    \caption{The three operators operating on the GPU. (a) One diffusion pass along one axis, one thread per subvolume.
    A tile of subvolumes and its apron are staged in shared memory. Hops are drawn
    with hashed random numbers. Particles exceeding the cap are appended to the overflow list, and each thread writes to a second copy
    of the particle lattice (Sec.~\ref{sec:si-gpu-diffusion}). (b) One structure move, one thread block for each multi-lattice structure  (Sec.~\ref{sec:si-gpu-structures}). (c) Reactions,
    one thread per subvolume (Sec.~\ref{sec:si-gpu-reactions}).}
    \label{fig:si-gpu}
\end{figure}

\subsection{Diffusion passes}\label{sec:si-gpu-diffusion}

Each of the $x$, $y$, and $z$ passes uses one thread per subvolume (Fig.~\ref{fig:si-gpu}a). A thread
block copies a tile of subvolumes along the pass axis, with a one-subvolume apron on each side, from
the read buffer into shared memory. Each thread then evaluates the hop probabilities of its subvolume,
draws an outcome for every particle there and in its neighbors, and collects the particles that end in
its subvolume. The read buffer does not change during a pass, so every thread sees the same state.

A particle in an apron is evaluated twice: by the block that owns its subvolume and by the
neighboring block that it may enter. The random number of each hop decision is a hash of the
subvolume index, the particle's slot, the timestep, the pass axis, and the run seed, so both blocks
obtain the same number and make the same decision, and no particle is lost or duplicated at a tile
border. Each thread writes only its own subvolume to the write buffer. A particle that would exceed
the capacity of its destination is written to the overflow list instead.

\subsection{Structures}\label{sec:si-gpu-structures}

Large spheres and polymers are updated by one thread block per sphere or chain
(Fig.~\ref{fig:si-gpu}b). The threads of a block read the subvolumes that the move involves in
parallel, block reduction involves summing the energy change in double precision. The move is accepted
with $\min(1,e^{-\Delta U})$ times the mobility (friction) factor $m$. An atomic compare-and-swap operation is performed when moving into a new subvolume; if another structure occupies it, the move is rejected.

\begin{figure}[htbp]
    \centering
    \includegraphics[max width=\linewidth,max height=0.74\textheight]{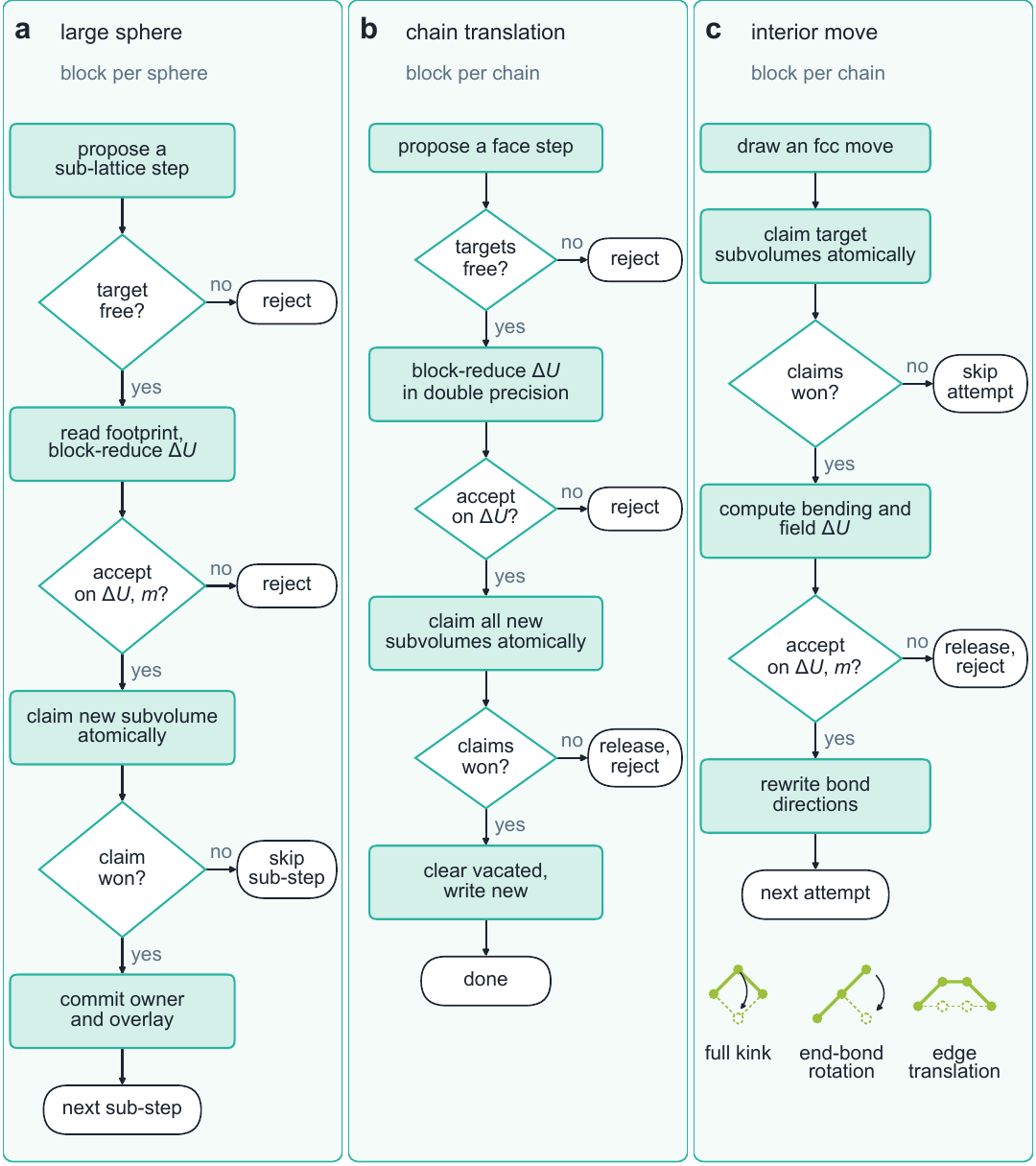}
    \caption{Large-sphere and polymer moves (Sec.~\ref{sec:si-gpu-structures}). (a) One
    sub-step of a large sphere, one thread block per sphere. (b) Whole-chain translation by
    one face step ($\pm \hat{x}, \pm \hat{y}, \pm \hat{z}$), one thread block per chain. (c) One interior move of the fcc move set
    (Sec.~\ref{sec:fcc-moveset}), one thread block per chain: a full kink, an end-bond rotation, or an
    edge translation of a segment between two fixed anchors (illustrated on the chart).}
    \label{fig:si-gpu-moves}
\end{figure}

\paragraph{Large spheres} (Fig.~\ref{fig:si-gpu-moves}a). The center is a real position that moves on a sub-lattice of spacing
$\delta=h/Q$. The timestep is divided into the smallest number of sub-steps
$n_{\mathrm{rep}}\in\{1,2,4,\dots,64\}$ for which the move probability per sub-step is at most $1/2$,
and each sub-step proposes a step of $\delta$ along $\pm x$, $\pm y$, or $\pm z$
[Eq.~\eqref{eq:structure-hop-propensity}]. The step is rejected if the center would leave the box or
enter a subvolume owned by another structure.

\begin{figure}[htbp]
    \centering
    \includegraphics[max width=\linewidth,max height=0.78\textheight]{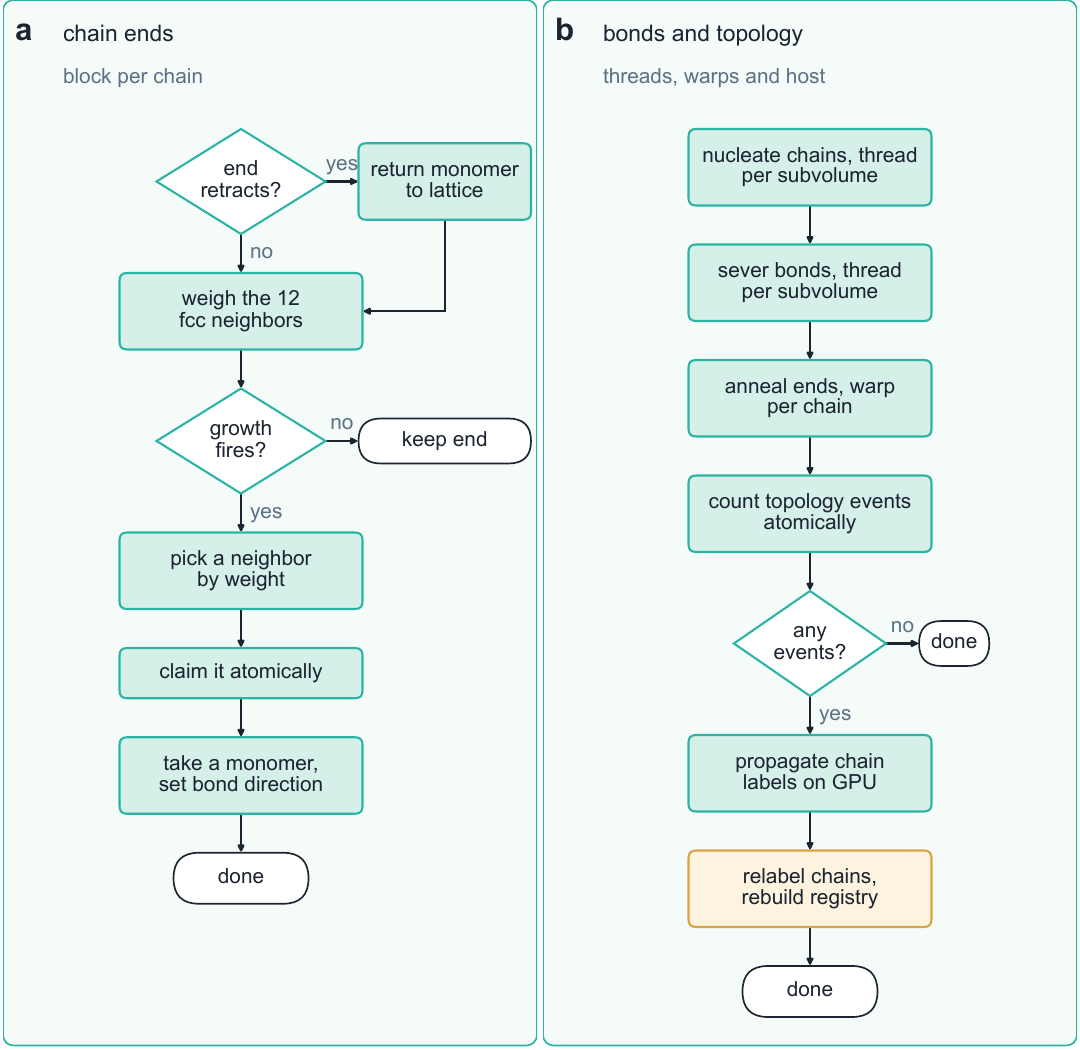}
    \caption{Polymer chemistry. (a) Chain ends, one thread block per chain. An end subunit that is the
    last one in its subvolume retracts with probability
    $1-\exp(-k_{\mathrm{off}}e^{-\varepsilon_{\mathrm{bond}}}\Delta t)$. After retraction, the corresponding monomer is returned to 
    the particle lattice. Tip growth weighs the 12 fcc neighbors by $w$, fires with probability
    $1-\exp(-k_{\mathrm{on}}\Delta t\sum w)$, claims the chosen neighbor by atomic compare-and-swap, consumes
    one monomer from its slots, and sets bond direction. (b) Nucleation and severing use one thread
    per subvolume and annealing one warp per chain. Every event increments a counter atomically; 
    when the counter is nonzero the chains are relabeled by label propagation on the GPU and the CPU
    rebuilds the registry.}
    \label{fig:si-gpu-chem}
\end{figure}

\paragraph{Polymers} Relaxation moves run, followed by chemistry
moves. A whole-chain translation (Fig.~\ref{fig:si-gpu-moves}b) shifts every subunit by one subvolume along $\pm x$, $\pm y$, or
$\pm z$, with a probability set by the chain's diffusion coefficient $D_T(N)=D_1/N$. The interior
moves (Fig.~\ref{fig:si-gpu-moves}c) are those of Sec.~\ref{sec:fcc-moveset}: half of the attempts propose an edge translation of a
segment between two fixed anchors, and the others an end-bond rotation or a full kink. A chain end (Fig.~\ref{fig:si-gpu-chem}a)
retracts with probability $1-\exp(-k_{\mathrm{off}}e^{-\varepsilon_{\mathrm{bond}}}\Delta t)$ and
releases its corresponding monomer to the particle lattice. For growth, each of the 12 fcc neighbors of the end is
weighted by $w=N_{\mathrm{mono}}\exp[-(E_{\mathrm{bend}}+E_{\mathrm{env}})]$, where $N_{\mathrm{mono}}$
is the number of free monomers there. Tip growth occurs with probability
$1-\exp(-k_{\mathrm{on}}\Delta t\sum w)$; the end claims
that subvolume and takes one monomer from its slots. Severing (Fig.~\ref{fig:si-gpu-chem}b) uses one thread per subvolume with an
interior bond, with probability $(1-e^{-k_{\mathrm{sev}}\Delta t})\min(1,e^{-E_{\mathrm{sever}}})$.
Annealing, its reverse, uses one warp per chain with one lane per edge direction and joins the $+$ end
of a chain to the $-$ end of another with probability
$(1-e^{-k_{\mathrm{ann}}\Delta t})\min(1,e^{-\Delta E_{\mathrm{form}}})$. Nucleation uses one thread per
subvolume. Every sever, join, or nucleation increments a counter. When it is nonzero at the end of the
step, connected polymers are identified by propagating labels on the GPU. The CPU then relabels chains and rebuilds the registry.

\subsection{Reactions}\label{sec:si-gpu-reactions}

The reaction operator uses one thread per subvolume (Fig.~\ref{fig:si-gpu}c). A thread reads the
particles of its subvolume, computes the reaction propensities, decides whether a reaction fires in
this timestep, selects one in proportion to its propensity, and accepts it with the factor of
Eq.~\eqref{eq:reactionEnergyChange}. An accepted reaction applies its stoichiometry to the subvolume in
place. Reactions are local to a subvolume, so no thread reads a neighbor, and no
shared memory or thread synchronizations are needed.

\subsection{Host coordination}\label{sec:si-gpu-host}

The CPU synchronizes with the GPU once per timestep, after the reaction operator
(Fig.~\ref{fig:si-gpu-io}). It also reads the
overflow list and the step summary. If the overflow list is not empty, the CPU copies the lattice to
host memory, relocates each listed particle, and copies the
lattice back. With adaptive time stepping, the CPU compares the largest $P_++P_-$ of the step with a
target below the bound of Eq.~\eqref{eq:samplingBound}. It reduces $\Delta t$ when the target is
exceeded by more than $10\%$, increases it by $10\%$ when the largest value is more than $10\%$ below
the target, and never exceeds the user's $\Delta t$. The lattice is copied to the host only for
output, user-function calls, and overflow relocation.

\section{Computational Performance}\label{sec:si-performance}

We time the lattice and the particle-based engine on the same systems at identical boxes and
particle counts. The simulation engines are ReaDDy on one pinned CPU core, with its SingleCPU kernel and the
detailed-balance reaction handler (DB) or its CPU kernel and the standard multithreaded handler
(best kernel, a different reaction model). For our lattice framework, we have 
the GPU solver on one NVIDIA B200; and a single-file
C++ implementation of the same lattice update rule on one pinned Xeon Platinum 8570 core. We use a 
separate C++ CPU implementation to have a direct fair comparison with the CPU ReaDDy simulations.

\begin{figure}[htbp]
    \centering
    \includegraphics[max width=\linewidth,max height=0.80\textheight]{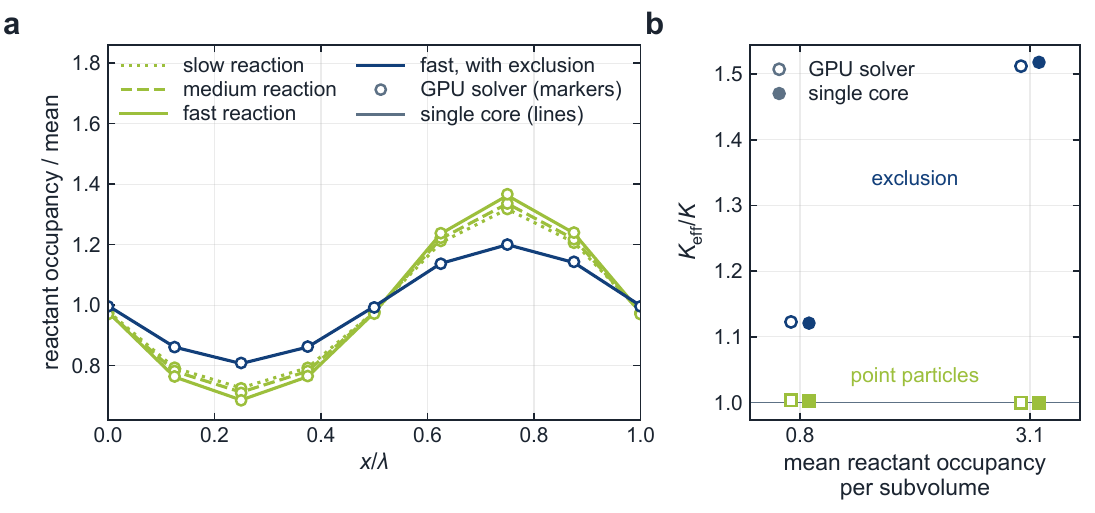}
    \caption{The single-core CPU implementation of the lattice update against the GPU solver. 
    (a) Steady-state reactant occupancy across one period of the sinusoidal field,
    normalized by its mean: markers are the GPU solver, lines the single-core implementation, green the point-particle simulation at three dissociation rates (slow, medium and fast: $k_{\mathrm{off}}=2\times10^2$, $2\times10^4$ and $5\times10^5$ s$^{-1}$) and navy the volume-excluded arm ($k_{\mathrm{off}}=5\times10^5$ s$^{-1}$). (b) The
    equilibrium-constant shift without a field at two mean occupancies, open symbols the GPU solver
    and filled symbols the single-core CPU implementation, circles with volume exclusion and squares without.
    Error bars are standard errors over replicates (seeds).}
    \label{fig:cpu-gpu-validation}
\end{figure}

This C++
single-core implementation was validated against the GPU solver (\Cref{fig:cpu-gpu-validation}).
In the reversible bimolecular reaction driven by a sinusoidal potential field, 
the steady-state reactant profiles of the two agree to within $1\%$ of the
profile amplitude at every reaction rate tested, $k_{\mathrm{off}}$ from $2\times10^{2}$ to
$5\times10^{5}$ s$^{-1}$, with and without exclusion. Without a field, over ten replicates (seeds) per
implementation, the equilibrium-constant shift $K_{\mathrm{eff}}/K$ agrees within two standard errors
in three of the four cases; at the higher density with exclusion the two differ by $0.4\%$
($1.518\pm0.0004$ against $1.512\pm0.0009$), which ten replicates (seeds) resolve. Both comparisons use a
$32^{3}$ lattice at $h=10$ nm, $K=200$ nm$^{3}$ and a $50$ $\mu$s step. Every point
is the median of three repetitions of a two-point measurement, wall$(n_2)-$wall$(n_1)$ over
$n_2-n_1$ steps, which cancels setup and one-time costs.

We note that the particle-based simulator ReaDDy has timesteps bounded by the stability of 
its integrator, set by its harmonic contact potential. The lattice simulator, on the other hand, 
can take larger timesteps set by the Wang-Peskin-Elston factor, at the expense of losing 
information on particle locations within each lattice site. We can then compare the performance 
of the two by a two-prong comparison: ratio of the wall-clock time per timestep for both methods (speed ratio), 
and the ratio of the largest timestep that each method can take at a specified accuracy tolerance (ceiling ratio). 
We subsume these two into a single ``advantage'' metric $A$, which we discuss in \Cref{sec:si-advantage}.

\begin{table}[htbp]
  \centering
  \caption{Speed ratio, ceiling ratio and the resulting advantage $A$ (Eq. \ref{eq:advantage}) for the two
  configurations of this paper, at a $32\times8\times8$ subvolume box. The speed ratio is the
  measured wall time per step of the particle engine over that of the lattice at the same box and
  the same particle content. The ceiling ratio is the largest step each method may take at a
  declared accuracy tolerance, with the particle engine quoted at its Euler stability limit rather
  than at the smaller step it is configured with. Matched physics means the detailed-balance
  reaction handler, which the particle engine implements for its single-core kernel only; best
  kernel means its multithreaded kernel with the standard handler, which is faster and is a
  different reaction model. The ternary mixture carries a sphere larger than a subvolume, which
  the single-file C++ CPU lattice reference does not implement, so it is reported on the device (GPU) side only.}
  \label{tab:advantage}
  \begin{tabular}{llrrrr}
  \toprule
  system & pairing & $c_{\mathrm{R}}/c_{\mathrm{L}}$ & $\tau_{\mathrm{L}}$ ($\mu$s) & $\tau_{\mathrm{L}}/\tau_{\mathrm{R}}$ & $A$ \\
  \midrule
  ternary mixture & GPU against one core & 899 & 53.2 & 133 & $1.2\times10^{5}$ \\
   & GPU against one socket & 127 &  &  & $1.7\times10^{4}$ \\
  \addlinespace
  sinusoidal, reversible binding & one core, matched physics & 7.37 & 53.1 & 133 & 978 \\
   & one core, best kernel & 7.66 &  &  & $1.0\times10^{3}$ \\
   & one socket each & 97.1 &  &  & $1.3\times10^{4}$ \\
   & GPU against one core & 180 &  &  & $2.4\times10^{4}$ \\
   & GPU against one socket & 146 &  &  & $1.9\times10^{4}$ \\
  \bottomrule
  \end{tabular}
\end{table}

The MPD-RDDME per-timestep cost is set
by the number of subvolumes and the particle method by the number of particles. FUrthermore, the former's 
step is set by the lattice spacing, a modeling choice, where ReaDDy's is set by a molecular contact
stiffness. The time to reach a simulated duration $T$ therefore costs $\Oh(V T/\tau)$ on the
lattice against $\Oh(N T/\tau_{\mathrm{p}})$ for the particle-based method, and the ratio grows with the
concentration $N/V$ and the timestep gap (ceiling ratio) $\tau/\tau_{\mathrm{p}}$. GPU ports of particle-based
engines, reported at one to two hundred times serial Smoldyn~\citep{dematte2012smoldyn} and up to
two orders of magnitude over one core for the original ReaDDy~\citep{biedermann2015readdymm},
would reduce the measured ratios without changing that scaling. Despite the GPU port of Smoldyn 
reported in the literature, we were not able to find the source code for the implementation, and so did not add it to our comparison tables.

\subsection{Ceiling Ratios and the Advantage Metric}\label{sec:si-advantage}

We define the advantage metric as follows:
\begin{align}
    A = \underbrace{\frac{c_{\mathrm{R}}}{c_{\mathrm{L}}}}_{\text{speed}}
    \times \underbrace{\frac{\tau_{\mathrm{L}}}{\tau_{\mathrm{R}}}}_{\text{ceiling}},
    \qquad
    \tau_{\mathrm{L}} = \min\!\left(\tau_{\mathrm{samp}},\ \tau_{\mathrm{rxn}}\right),
    \label{eq:advantage}
\end{align}
where $c_{\mathrm{R}}$ ($c_{\mathrm{L}}$) is the wall-clock time per timestep taken by the particle-based (lattice) simulator, 
and $\tau_{\mathrm{L}}$ ($\tau_{\mathrm{R}}$) is the timestep ceiling of the lattice (particle-based) simulator. 

We can estimate $\tau_L$ as follows. 
The Wang-Peskin-Elston factor of Eq.~\ref{eq:wpe-factor} is unbounded on
energetically downhill moves, so the pair of one-step probabilities along an axis is controlled by the largest
value the factor takes anywhere on the lattice,
\begin{align}
    \mathcal{B}^{\max} \equiv \max_{v,\mathbf{e}} \mathcal{B}\!\left(\Delta U_{v\to v+\mathbf{e}}\right),
    \qquad
    \tau_{\mathrm{samp}}(\varepsilon) = \frac{\varepsilon h^{2}}{D\,\mathcal{B}^{\max}} ,
    \label{eq:tausamp}
\end{align}
at a specified accuracy tolerance $\varepsilon$ on the per-pass hop propensity. The reaction operator applies
at most one event per subvolume per step, so for a propensity held constant across the step the
probability that the step should have fired more than once is
$1-(1+a_v\tau)e^{-a_v\tau} = \tfrac{1}{2}(a_v\tau)^2 + \Oh((a_v\tau)^3)$. Holding this below a
specified $\varepsilon_r$ gives $\tau_{\mathrm{rxn}} = \varepsilon_r/\max_v a_v$. Hence, 
\begin{align}
    \tau_L = \min\left(\tau_{\text{samp}}, \tau_{\text{rxn}}\right) = \min\left(
    \frac{\varepsilon h^{2}}{D\mathcal{B}^{\max}}, \; 
    \frac{\varepsilon_r}{\max_v a_v} 
    \right).
    \label{eq:taul}
\end{align}

The particle engine's timestep ceiling is the linear stability limit of its integrator. ReaDDy integrates
overdamped Langevin dynamics by explicit Euler-Maruyama~\citep{ermak1978brownian,kloeden1992numerical}.
Linearized about contact, its harmonic repulsion $U_{\mathrm{c}}(r)=\tfrac{1}{2}k(r-d)^{2}$ $(r<d)$ makes
the overlap of a contacting pair an Ornstein-Uhlenbeck process with relaxation rate $\beta D_{\mathrm{rel}}k$,
$D_{\mathrm{rel}}=D_i+D_j$, and explicit Euler-Maruyama on that process is stable in the mean and in mean square
only for $\beta D_{\mathrm{rel}}k\tau<2$~\citep{higham2000mean,higham2001algorithmic,leimkuhler2015molecular}.
A particle against a partner that does not move gives the largest such step,
\begin{align}
    \tau_{\mathrm{R}} = \frac{2}{\beta D k},
    \label{eq:tauR}
\end{align}
We quote Eq.~\ref{eq:tauR}, so the ceiling ratio, and with it the
advantage, is a lower bound. With $D=10^{5}$ nm$^{2}$s$^{-1}$ and $\beta k = 50$ nm$^{-2}$,
$\tau_{\mathrm{R}} = 4.0\times10^{-7}$ s; the engine is configured at $1.6\times10^{-7}$ s,
$\beta Dk\tau = 0.8$, inside the two-particle bound as well. The diffusion coefficient cancels against
that of Eq.~\ref{eq:tausamp} when the two ceilings are divided.

\subsection{Advantage as a Function of Reaction Counts}

\begin{figure}[htbp]
    \centering
    \includegraphics[max width=\linewidth,max height=0.80\textheight]{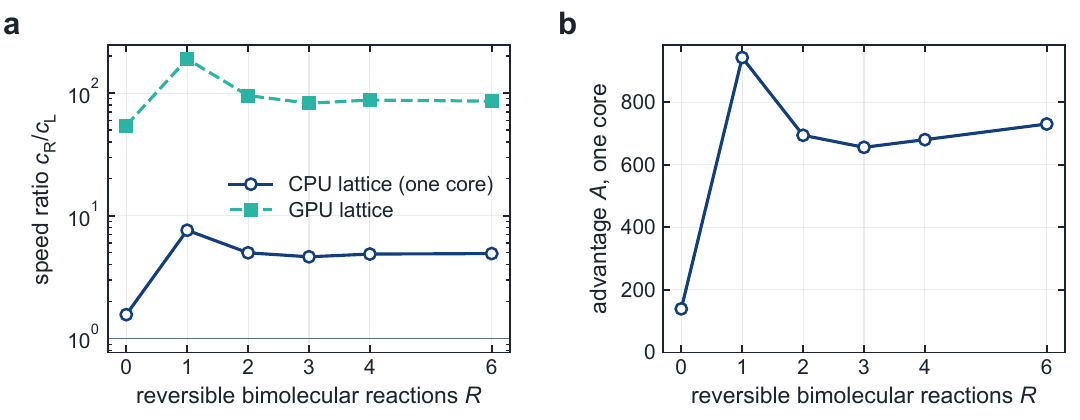}
    \caption{Speed ratio and advantage against the number of reversible bimolecular reactions $R$
    of Eq.~\ref{eq:network}, at fixed packing fraction. (a) The speed ratio, with the best
    single-core particle kernel in the numerator and each lattice simulation in turn in the denominator;
    the horizontal line marks parity. (b) The advantage of Eq.~\ref{eq:advantage} at one core with
    matched physics. $R=0$ is diffusion alone. Medians of three repetitions.}
    \label{fig:advantage}
\end{figure}

\Cref{fig:advantage} holds the packing fraction at $\eta=0.09$, that of the reversible-binding
system of \Cref{sec:reversible}, and varies the number $R$ of reversible bimolecular reactions,
\begin{align}
    \mathsf{A}_i+\mathsf{B}_i \rightleftharpoons \mathsf{C}_i, \qquad i = 1,\dots,R,
    \label{eq:network}
\end{align}
with $R=0$ being diffusion alone. The same $13\,038$ particles are split evenly over the $R$ pairs
in a $32\times8\times8$ subvolume box. Every species pair has diameters the same as those described in \Cref{sec:reversible}, 
and each pair's binding constant is $R$ times the single-pair value at
$k_{\mathrm{off}}=200$ s$^{-1}$. This keeps the bound fraction the same for each pair and the
total event rate equal at every $R$, so we're isolating the effects of reaction counts. 
The axis therefore measures the cost of carrying more
reaction channels, $2R$, and more species, $3R$, and not the cost of firing more events. The rate
is chosen where the lattice's reaction operator is within its single-firing ceiling.

The lattice cost does not grow with $R$. One core takes $3.24$ to $3.26$ ms per step at every
$R\geq1$, and the GPU $0.13$ ms at $R=1$ and $0.17$ to $0.19$ ms for $R$ from $2$ to $6$. The
particle engine's cost is set by its reactions: $4.0$ ms per step without them, $23.6$ ms for one
pair, and $16$ to $19$ ms once the reactants are split over two or more pairs, where each particle
has fewer reactive partners within the reaction radius. The single-core speed ratio is therefore
largest for a single pair, $7.3$, and holds at $5.1$ to $5.7$ with ReaDDy DB handler from $R=2$ to
$6$. The speed ratio for the GPU against ReaDDy's CPU kernel (standard handler) on one core holds at $83$ to $95$. 
The advantage at one core with
matched physics is $140$ for diffusion alone, $940$ for one pair and $660$ to $730$ from $R=2$ to
$6$.

\section{Additional Results}\label{sec:si-additional}

\subsection{Binary Hard-Sphere Mixture}\label{sec:si-twospecies}
We compare the subvolume occupancy distributions of purely diffusive, volume-excluded
particles in a binary mixture (both particles smaller than the lattice subvolumes) against ReaDDy~\citep{schoneberg2013readdy,hoffmann2019readdy2}. For the ReaDDy simulations, we run with harmonic
repulsions whose contact distance is set so that the Barker-Henderson effective
diameter~\citep{barker1967square,barker1967successful} equals
each $\sigma_{ij}$, and against the point-particle limit of the same lattice model.

\begin{figure}[htbp]
    \centering
    \includegraphics[max width=\linewidth,max height=0.80\textheight]{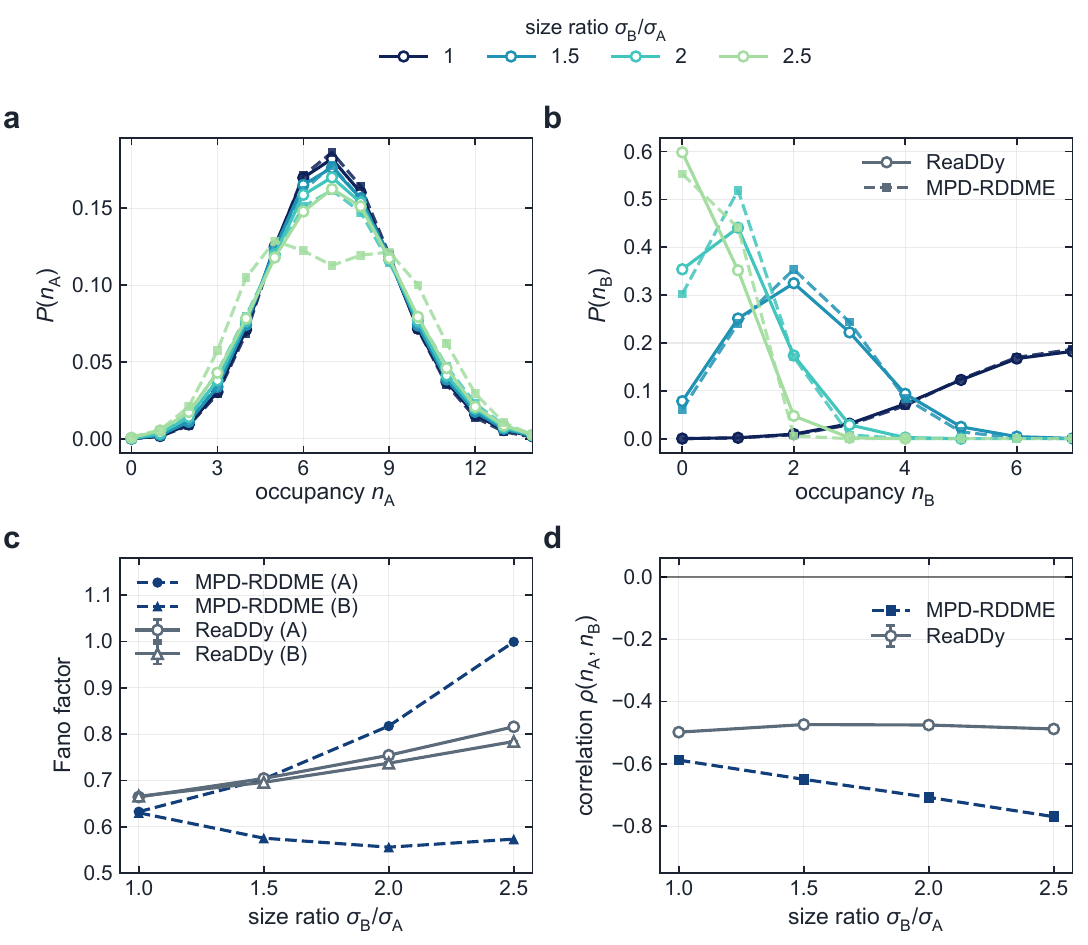}
    \caption{Equilibrium subvolume occupancy of freely diffusing binary hard-sphere
    mixtures at total crowder density $\eta=0.20$ split equally by volume, $\sigma_\mathsf{A}=3$ nm,
    $h=10$ nm, at four size ratios. Open symbols and solid lines are ReaDDy at the
    Barker-Henderson effective diameters~\citep{barker1967square,barker1967successful} ($16$ replicates per ratio, equilibrated frames), filled
    symbols and dashed lines MPD-RDDME. (a), (b) Marginal occupancy of each species. (c) Fano
    factor of each species against the size ratio. (d) Correlation
    $\rho(n_\mathsf{A},n_\mathsf{B})$ of the two occupancies of the same subvolume, which
    vanishes for point particles and is negative for hard spheres.}
    \label{fig:two-eq-readdy}
\end{figure}

The reference statistics are strongly sub-Poissonian at every composition. The point-particle limit is also structurally incapable
of the anti-correlation that crowding produces. That is, its Fano factor is one and its cross-species
correlation zero at every ratio. The time-resolved relaxation, with a third species larger than the subvolume added, 
is in the main text.

\subsection{Occupancy Distributions of the Ternary Mixture}\label{sec:si-three-eq}

\begin{figure}[htbp]
    \centering
    \includegraphics[max width=\linewidth,max height=0.80\textheight]{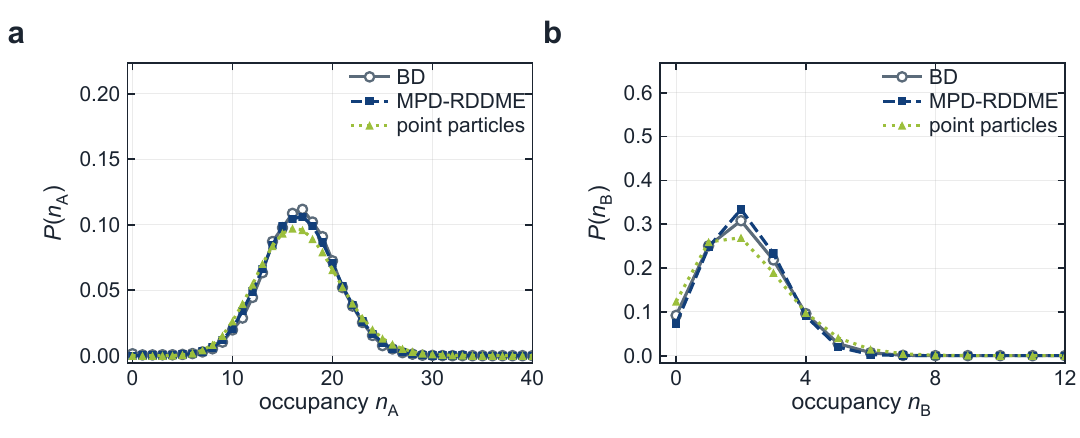}
    \caption{Subvolume occupancy distributions at $\tau=20$ for the ternary mixture of \Cref{fig:three-relax-readdy}.
    (a) $\mathsf{A}$ ($2$ nm); (b) $\mathsf{B}$ ($4$ nm). Open circles are the BD reference, filled symbols the
    volume-excluded (MPD-RDDME) and point-particle lattices; $\tau=D_\mathsf{A}t/h^2$.}
    \label{fig:three-eq}
\end{figure}

\Cref{fig:three-eq} shows the subvolume occupancy distributions of $\mathsf{A}$ and of $\mathsf{B}$ at the end of the
relaxation of \Cref{fig:three-relax-readdy} ($\tau=20$). For an ideal gas with a fixed total particle number, voxel
occupancy follows a binomial distribution that approaches a Poisson distribution when the voxel occupies a small
fraction of the box. The volume-excluded simulations departed from this near-Poisson baseline, as observed in earlier
lattice simulations~\citep{cianci2016molecular}, and followed the reference more closely than the point-particle
lattice. The divergence of these distributions from the reference over the whole relaxation is in the main text
(\Cref{fig:three-relax-readdy}b).

\subsection{Cross-Species Correlation}\label{sec:si-rho}
The divergence of the occupancy distributions from the reference and the body's mobility during the relaxation of the
ternary mixture are in the main text (\Cref{fig:three-relax-readdy}); the distributions at $\tau=20$ are in \Cref{fig:three-eq}. Here we show the cross-species correlation of the same runs.

\begin{figure}[htbp]
    \centering
    \includegraphics[max width=\linewidth,max height=0.80\textheight]{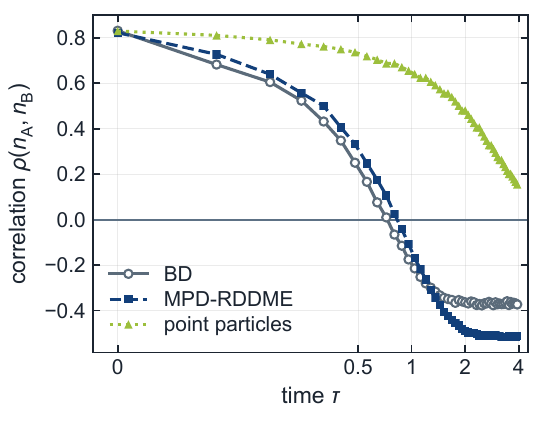}
    \caption{Cross-species occupancy correlation of the ternary relaxation of
    \Cref{fig:three-relax-readdy}, over all subvolumes. Both engines start at $+0.83$ from the slab
    initial condition. The volume-excluded lattice crosses zero within $0.08$ of the reference and
    settles $0.14$ below it; the point-particle limit never becomes negative. Band, the standard error over forty reference replicates; $\tau=D_\mathsf{A}t/h^2$.}
    \label{fig:three-rho}
\end{figure}

\Cref{fig:three-rho} follows $\rho(n_\mathsf{A}, n_\mathsf{B})$, taken over all subvolumes so
that the sign change the slab start produces is visible. The correlation begins at $+0.83$ in every
arm, because a start that puts both species in the same half of the box correlates them by
construction rather than by any interaction. What happens next separates the descriptions. In the
reference the correlation crosses zero at $\tau=0.80$ and settles at $-0.37$: once the box is filled,
a subvolume crowded with $\mathsf{A}$ has less room for $\mathsf{B}$. The volume-excluded lattice
crosses at $\tau=0.88$ and settles at $-0.51$, reproducing the sign, the timing and the shape of the
approach while overshooting the depth of the anticorrelation. 
The point-particle limit decays monotonically toward zero and never reaches the negative
branch at all, because without exclusion the two species compete for nothing 
and are completely independent.

\subsection{Equilibrium Constant of a Reversible Reaction at Fixed Rates}\label{sec:si-rxK}

\begin{figure}[htbp]
    \centering
    \includegraphics[max width=\linewidth,max height=0.80\textheight]{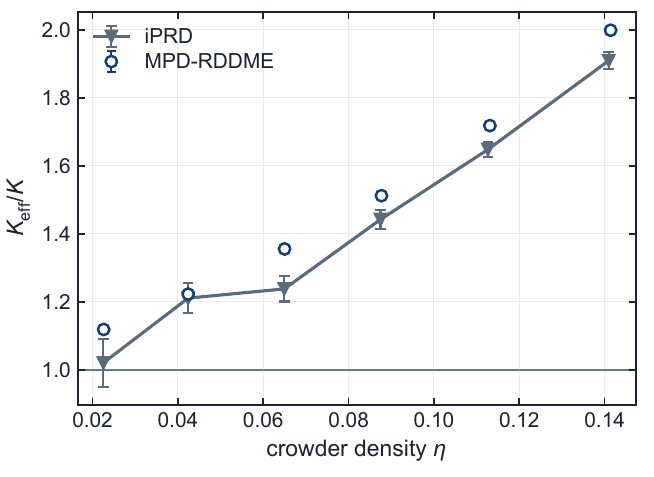}
    \caption{Shift of the apparent equilibrium constant of
    $\mathsf{A}+\mathsf{B}\rightleftharpoons\mathsf{C}$ with crowder density. Open circles are MPD-RDDME, the
    ratio of the exclusion run to the point-particle run at identical rates, so the horizontal line at
    one is the point-particle lattice; the triangles are the reference, interacting-particle
    reaction dynamics (ReaDDy), labeled iPRD.
    Every point of every simulation is the mean of
    ten replicates, with error bars showing the standard error of the mean (where not visible, they
    are smaller than the symbols).
    $\sigma_\mathsf{A}=\sigma_\mathsf{B}=3$ nm, $\sigma_\mathsf{C}=3.78$ nm, no external field;
    total crowder density is conserved by the reaction since
    $\sigma_\mathsf{C}^{3}=\sigma_\mathsf{A}^{3}+\sigma_\mathsf{B}^{3}$.}
    \label{fig:rx-K-main}
\end{figure}

We replicate the analysis of ~\citep{hoffmann2019readdy2} in showing how the 
equilibrium constant changes due to finite-size effects. 
\Cref{fig:rx-K-main} shows the shift in the equilibrium constant as a function of the initial
crowder density at fixed rate constants with no external field. The apparent equilibrium constant
$K_{\mathrm{eff}} = \langle N_\mathsf{C}\rangle V/(\langle N_\mathsf{A}\rangle\langle
N_\mathsf{B}\rangle)$ is plotted against values computed from ReaDDy at the same densities.
Volume exclusion roughly doubles the
equilibrium constant by $\eta = 0.14$, and the simulation scheme closely tracks the values predicted
by ReaDDy for the whole sweep, within a few percent from $\eta = 0.09$ up. This also
is a demonstration of the Le Chatelier's principle, where the forward reaction is favored because
of reduced particle particle count in the product side. The same reaction under a resolved
sinusoidal field is in the main text.

\subsection{Composition of the Reversible Reaction in the Sinusoidal Field}\label{sec:si-rxfield-counts}

\begin{figure}[htbp]
    \centering
    \includegraphics[max width=\linewidth,max height=0.80\textheight]{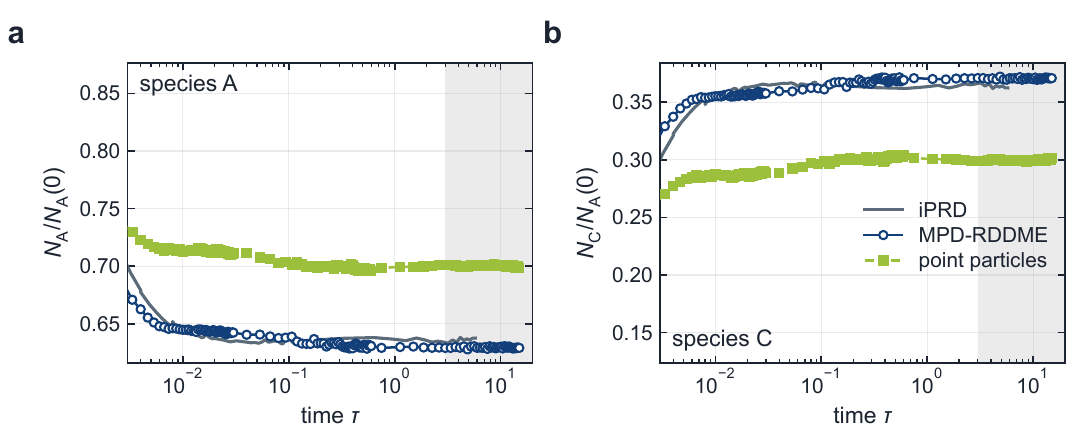}
    \caption{System composition of $\mathsf{A}+\mathsf{B}\rightleftharpoons\mathsf{C}$ in the
    sinusoidal field of \Cref{fig:rx-field-kl}. (a) The reactant $\mathsf{A}$; (b) the product $\mathsf{C}$. Each count is divided by the initial reactant count. The shaded bands
    indicate the late-time windows whose means are quoted in the text; $\tau=Dt/h^2$.}
    \label{fig:rx-field-counts}
\end{figure}

Shown in \Cref{fig:rx-field-counts} are the ratios over time of the reactant and product $\mathsf{A}$ and $\mathsf{C}$, respectively, from 
the reversible reaction
$\mathsf{A}+\mathsf{B}\rightleftharpoons\mathsf{C}$ in the sinusoidal field of \Cref{sec:reversible}. 
These counts are normalized by the initial reactant count $N_\mathsf{A}(0)$, against $\tau = Dt/h^{2}$. 
The reference averages $48$ runs in each of its three windows.
Both lattice runs reached their steady composition by $\tau\approx1$. Averaged over $\tau\geq3$, the
volume-excluded lattice holds $N_\mathsf{A}/N_\mathsf{A}(0)=0.630$ and
$N_\mathsf{C}/N_\mathsf{A}(0)=0.370$ against the reference's $0.636$ and $0.364$, while the
point-particle lattice holds $0.700$ and $0.300$.

\subsection{Divergence from the Reference During the Relaxation}\label{sec:si-rxfield-kl}

\begin{figure}[htbp]
    \centering
    \includegraphics[max width=\linewidth,max height=0.80\textheight]{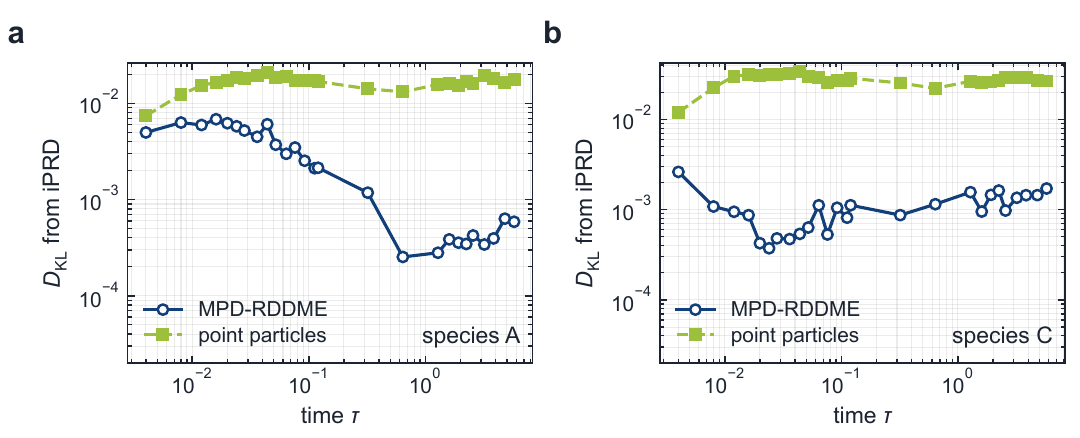}
    \caption{Kullback-Leibler divergence of the subvolume occupancy distribution from the reference against time,
    for the reversible reaction in the sinusoidal field of \Cref{fig:rx-field-kl}. (a) The reactant $\mathsf{A}$; (b) the product $\mathsf{C}$. Lower is closer to the reference; $\tau=Dt/h^2$.}
    \label{fig:rx-field-kl-time}
\end{figure}

\Cref{fig:rx-field-kl-time} shows the Kullback-Leibler divergence of the subvolume occupancy distribution of each
species from the reference, at every time of the relaxation of \Cref{fig:rx-field-kl}. The point-particle lattice
stays near $2\times10^{-2}$ for $\mathsf{A}$ and $3\times10^{-2}$ for $\mathsf{C}$ through the whole relaxation. The
volume-excluded lattice is lower at every time.

\subsection{Timestep Error of the Operator Splitting}\label{sec:si-splitting}
Each timestep applies the operators of Algorithm~\ref{alg:rddme} of the main text in sequence. First, the
drift-diffusion operator moves the particles along $x$, then along $y$, then along $z$. Then the multi-lattice handler
moves the bodies and the polymers. Then the reaction operator changes the species in each subvolume. Each operator
starts from the state that the previous operator made. In one operator, all decisions use the state at the start of
that operator.

This sequence has two types of error that are first order in $\tau$. The first type comes from the order of the
operators, and is the splitting error. When the order is fixed, this error has the same sign at each step, and it adds
up over many steps. When the order is random at each step, this error has the two signs equally often, and it cancels
on average~\citep{strang1968construction}. 

To measure the discrepancies between these two scehems, we run a simulations with the five operators in a random order at each step. The five
operators are the three drift-diffusion passes, the multi-lattice handler and the reaction operator. 

We test with the reversible binding of \Cref{sec:reversible} in its sinusoidal field, with volume exclusion. We used
$\Delta t = 0.08$, $0.04$, $0.02$ and $0.01\ \mu$s. These timesteps are much smaller than those used in the ternary relaxation 
experiment, because the
dissociation rate $k_{\mathrm{off}} = 5\times10^{5}$ s$^{-1}$ makes the chemical relaxation fast, approximately
$2\ \mu$s. The limit of one reaction per subvolume per step is a good approximation only when
$k_{\mathrm{off}}\Delta t \ll 1$; here $k_{\mathrm{off}}\Delta t$ is $0.005$ to $0.04$. We ran each initial seed (replicate) to
$\tau = 0.008$ to record the relaxation, and to $\tau = 0.6$ to record the response to the field.

We also test with the ternary mixture relaxation of \Cref{sec:finitesize}. We used $\Delta t = 50$, $25$,
$12.5$ and $6.25\ \mu$s, and ran each replicate to $\tau = 4$. In this system, the bath, the bodies and the coupling between
them are in different operators. Thus the mean squared displacement of the bodies at $\tau = 4$ shows the effect of the
order on this coupling.

\begin{figure}[htbp]
    \centering
    \includegraphics[max width=\linewidth,max height=0.80\textheight]{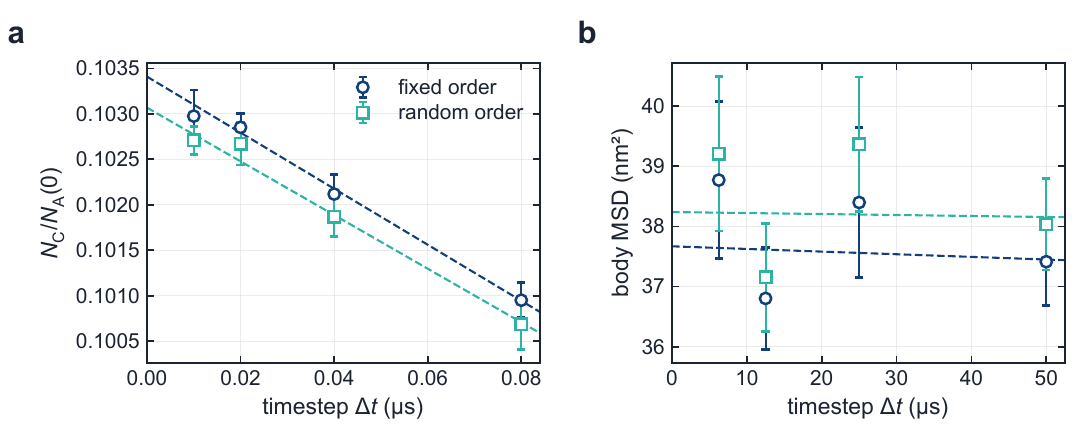}
    \caption{Timestep error of the operator splitting, for the fixed and the random order of the operators.
    (a) System 1: the product fraction $N_\mathsf{C}/N_\mathsf{A}(0)$ at $0.4\ \mu$s, early in the chemical relaxation.
    (b) System 2: the mean squared displacement of the bodies at $\tau = 4$. Each point
    is the mean of $16$ replicates, with its standard error. Each dashed line is a fit linear in $\Delta t$; its intercept is
    the limit $\Delta t \to 0$.}
    \label{fig:tau-ladder}
\end{figure}

\Cref{fig:tau-ladder} shows the results. In the reversible bimolecular reaction setup, the error is first order:
$N_\mathsf{C}/N_\mathsf{A}(0)$ at $0.4\ \mu$s is $0.1010$ at $\Delta t = 0.08\ \mu$s and $0.1030$ at $0.01\ \mu$s, and
the linear extrapolation to $\Delta t \to 0$ gives $0.1034$. Thus the error is $2.3\%$ at $0.08\ \mu$s and $0.4\%$ at
$0.01\ \mu$s. In the ternary mixture relaxation setup, the mean squared displacement of the bodies is $37$ to $39$ nm$^{2}$. It does not change with
the timestep, within its statistical error of $3\%$.

The order of the operators has no effect that $16$ replicates can measure. In the first, the random order and the fixed order
agree within approximately one standard error at each timestep. Their slopes are almost equal, $-3.1$ and
$-3.0\times10^{-2}$ per $\mu$s, and their extrapolated limits differ by $0.3\%$. In system 2, the two orders also agree
within approximately one standard error at each timestep. The steady state also does not change with the order:
$K_{\mathrm{eff}}/K$ of system 1 goes from $1.490$ at $0.08\ \mu$s to $1.486$ to $1.488$ at $0.01\ \mu$s for the two
orders. 

\subsection{Polymer-Crowder Demixing from Other Initial Conditions}\label{sec:si-demix-starts}

\begin{figure}[htbp]
    \centering
    \includegraphics[max width=\linewidth,max height=0.77\textheight]{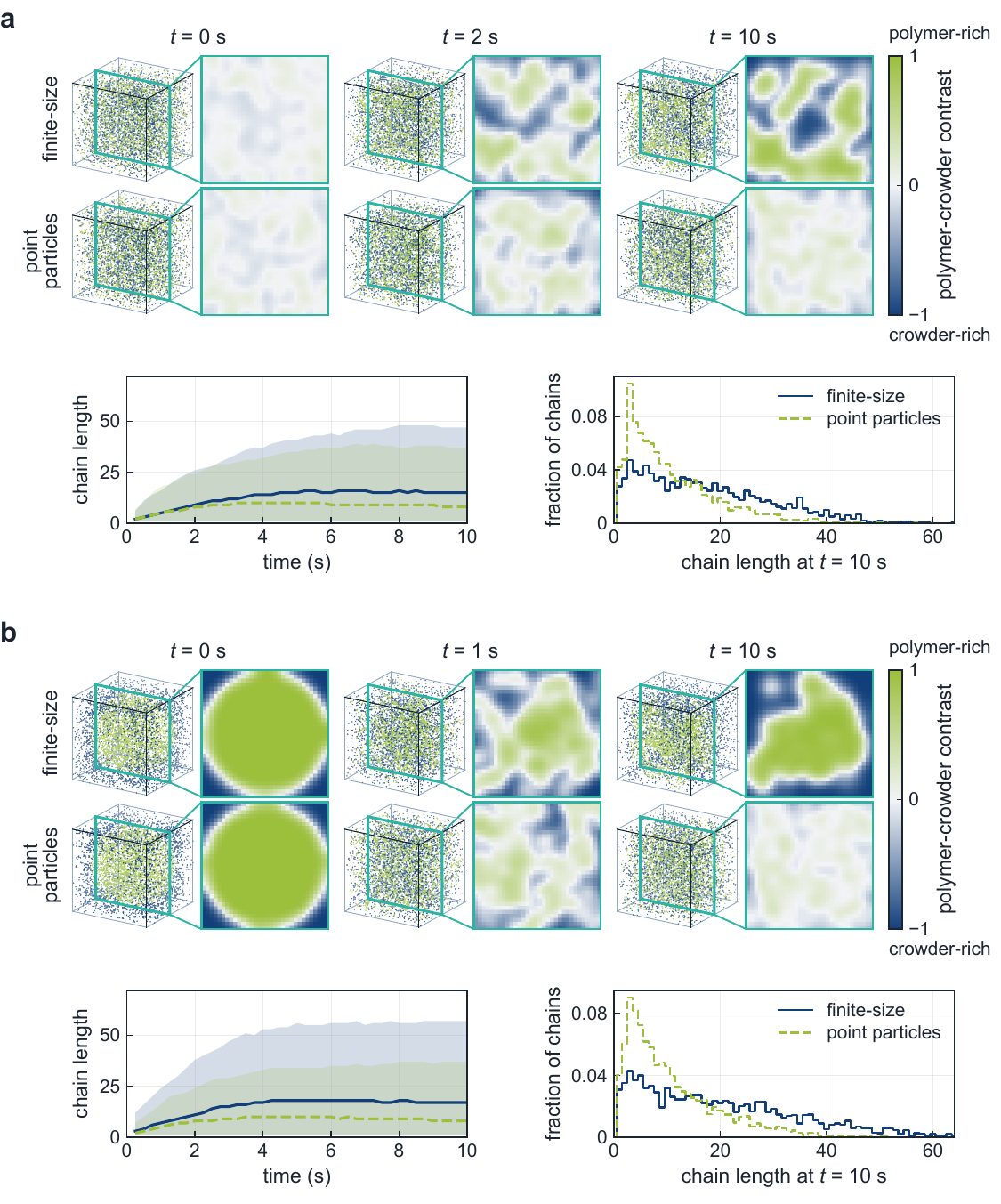}
    \caption{Polymer-crowder demixing from two other initial conditions than \Cref{fig:phase}\textup{c}, drawn as there: the box in
    3D beside a slice through its middle (outlined), colored as in \Cref{fig:phase}\textup{c}, and beneath, the chain lengths over time (median; band, 1st to 99th percentile) and their distribution at $t=10$~s. The contrast is $(p-c)/(p+c)$. (a) Uniform start: monomers and crowders at random positions. (b) Droplet
    start: monomers in a central sphere, crowders outside it. At $t=0$ the polymer is free monomer; afterwards it is the
    chain subunits.}
    \label{fig:demix-starts}
\end{figure}

\Cref{fig:phase}\textup{c} starts the polymer and the crowders in two halves of the box. \Cref{fig:demix-starts} shows the
same system from two other starts. In \Cref{fig:demix-starts}\textup{a}, the monomers and the crowders start at random positions
in the whole box. With volume exclusion, polymer-rich and crowder-rich domains form within $2$~s and grow to $10$~s;
without it, the mixture stays uniform. At $10$~s the volume-excluded run holds $2650$ chains of median length $15$
subunits, against $3166$ of median length $8$ in the point-particle control. In \Cref{fig:demix-starts}\textup{b}, the monomers
start in a central sphere that holds half of the box volume and the crowders outside it. With volume exclusion, the
polymer stays in one domain near the center: $71\%$ of the chain subunits are inside the starting sphere at $1$~s and
$78\%$ at $10$~s, against $58\%$ and $59\%$ for point particles, which mix within $1$~s.

In both demixed starts, the crowders fill their half of the box at twice their average density, a volume fraction of
$0.40$. Random placement cannot pack them that densely on the lattice, so they start on a random subset of the
subvolumes whose three coordinates are all even; the crowders move off this packing as the run starts. The chains in this implementation do not
cross the periodic boundary of the box, so the box faces act as walls for the chains. In the slab start of
\Cref{fig:phase}\textup{c} the polymer half touches one face, and the polymer collects against it; the droplet start touches
no face.

\section{Application: Receptor Transport in an Immunological Synapse}\label{sec:synapse}

Here we record a second biological application, showing the effects of both drift and volume exclusion in 
simulating the formation of the immunological synapse.
When a T cell contacts
an antigen-presenting cell, T-cell receptors (TCRs) bind peptide-major-histocompatibility complexes (pMHCs).
The resulting complexes are carried inward by actin retrograde flow and dynein-driven transport,
accumulating in a central region of the contact zone~\citep{dharanfarago2017,siokis2018}. The
unstructured-mesh reaction-drift-diffusion master equation of Ref. \cite{isaacsonzhang2024crddme}
reproduces that accumulation, and states that it must omit the size-based segregation forces
acting between the complexes because it carries no two-body interactions. In our setup, 
we added the effects of volume exclusion, but disregard the formation of microclusters and other 
two-body interactions with other components of the synapse.

\begin{figure}[htbp]
    \centering
    \includegraphics[max width=\linewidth,max height=0.77\textheight]{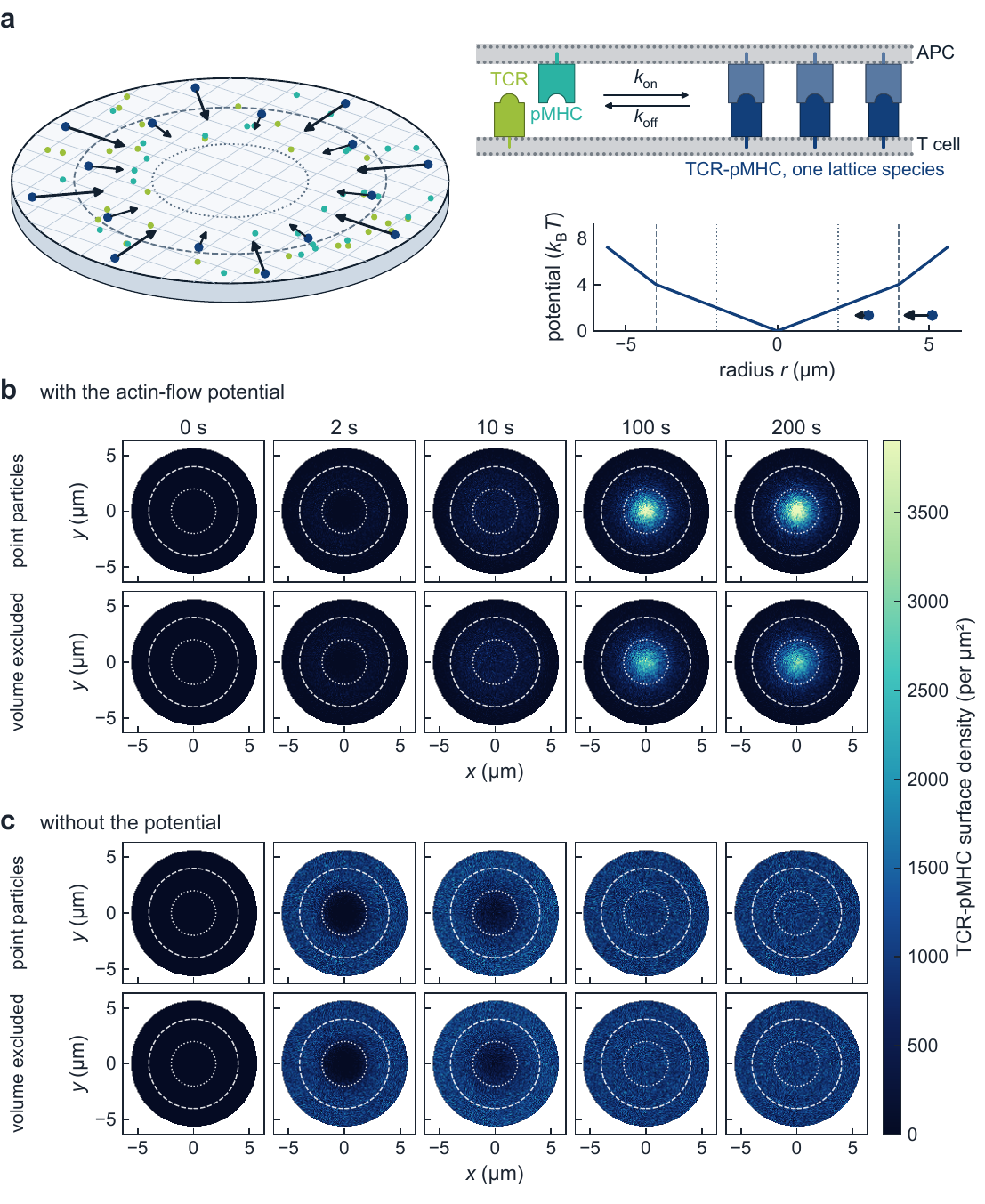}
    \caption{The immunological-synapse application. (a) The model: TCR and pMHC bind
    reversibly and the lattice carries the bound pair as one species; the contact zone, with the
    drive acting on the product alone; and the potential of Eq.~\ref{eq:synapse-ramp}.
    (b) Complex surface density at five times with the actin-flow potential ($f_0=1\ k_\mathrm{B}T\,\mu\mathrm{m}^{-1}$; the arrows in (a) mark the forces $f_0$ and $2f_0$), without volume exclusion above and with it below. (c) The same with the drive off ($f_0=0$). b and c share one color axis.}
    \label{fig:synapse-maps}
\end{figure}

We used Ref. \cite{isaacsonzhang2024crddme}'s configuration and parameters: 
a disk of radius $5.6\ \mu$m with no-flux boundaries, an
actin-depletion region inside $2\ \mu$m and the edge of the peripheral supramolecular activation
cluster at $4\ \mu$m, resolved here on a Cartesian lattice of $40$ nm subvolumes one layer thick.
Free receptors and ligands, TCRs and pMHCs, diffused at $0.10\ \mu$m$^{2}$s$^{-1}$ and feel no active force; the
bound complex, TCR-pMHC, diffused at $0.06\ \mu$m$^{2}$s$^{-1}$ and felt an effective 
centripetal potential ramp of slope $f_0$ inside the cluster edge and $2f_0$ beyond it, 
as a model for retrograde actin flow~\citep{dharanfarago2017}.

\begin{align}
    \Phi(\mathbf{r}) = \begin{cases}
        f_0 \|\mathbf{r}\| & \|\mathbf{r}\| \leq R_p \\
        f_0 R_p + 2f_0 (\|\mathbf{r}\| - R_p) & \|\mathbf{r}\| > R_p,
    \end{cases}
    \label{eq:synapse-ramp}
\end{align}

where $R_p = 4\ \mu$m is the edge of the peripheral supramolecular activation
cluster and $f_0=1\ k_{\mathrm{B}}T\ \mu$m$^{-1}$. Binding was modeled as the reversible reaction
$\mathsf{TCR}+\mathsf{pMHC}\rightleftharpoons\mathsf{TCR}\textrm{-}\mathsf{pMHC}$ at
$k_{\mathrm{on}}=2\times10^{4}$ M$^{-1}$s$^{-1}$ and $k_{\mathrm{off}}=0.1$ s$^{-1}$, with a
reaction radius of $15$ nm, the mean length of the complex. Receptors and ligands started at
$1410\ \mu$m$^{-2}$ outside the actin-depletion region and no complexes were present at $t=0$.
Densities were reported per unit area, and the maps were block-summed to $80$ nm display cells.
In \Cref{fig:synapse-maps} the dotted and dashed circles mark the actin-depletion region and the
cluster edge, the two arrows on the potential are in the ratio of its two slopes, and the times
shown are $t=0$, $2$, $10$, $100$ and $200$ s.

\Cref{fig:synapse-maps} follows the complex profile from the initial condition, in which no
complexes exist and the receptors and ligands are initiated outside the actin-depletion region.
Both simulations accumulated complexes in the central region over tens of seconds. They differed in
how far that accumulation persisted: at $200$ s the point-particle limit reached a peak
surface density of $6.4\times10^{3}\ \mu$m$^{-2}$ against $4.4\times10^{3}\ \mu$m$^{-2}$ with
volume exclusion.

We also show how the same setup evolves when the driving potential is absent i.e., no actin flow. With $f_0=0$
(\Cref{fig:synapse-maps}c) the complexes formed wherever the reactants were, the core filled by
diffusion alone, and by $100$ s the disk was uniform; the two descriptions were then
indistinguishable, their peak densities agreeing to one percent at $200$ s against a factor of
$1.4$ under the ramp.

\section{Calculating Local Polymer and Crowder Amounts in the Polymer Phase Separation Visualization}\label{sec:si-local-amount}

The slices of \Cref{fig:phase}\textup{c} and \Cref{fig:demix-starts} compare how much polymer and how much crowder sits at each
place, each relative to its own average. Here, we detail the calculation of these local amounts.

Let $n_X(\mathbf{v})$
be the amount of species class $X\in\{P, C\}$ in subvolume $\mathbf{v}$: for the polymer the number
of chain subunits it holds, and for the crowders the number of subvolumes a finite-size crowder
covers there or, in the point-particle simulation, the number of crowder particles it holds. The amounts are
summed over cells $b$ of $2\times2\times2$ subvolumes ($20$ nm), smoothed with a Gaussian kernel
$G_s$ of width $s=1.5$ cells ($30$ nm), and averaged over the two cells ($40$ nm) of depth $j_0$
and $j_0+1$ at the middle of the box along the slicing axis,
\begin{align}
    N_X(b) = \sum_{\mathbf{v}\in b} n_X(\mathbf{v}), \qquad
    \bar N_X(i,k) = \frac{1}{2}\sum_{j=j_0}^{j_0+1} \big(G_s * N_X\big)(i,j,k).
\end{align}
Each field is then divided by its own mean over the slice, and the two are compared through
\begin{align}
    p(i,k) = \frac{\bar N_P(i,k)}{\langle \bar N_P\rangle}, \qquad
    c(i,k) = \frac{\bar N_C(i,k)}{\langle \bar N_C\rangle}, \qquad
    \chi(i,k) = \frac{p(i,k)-c(i,k)}{p(i,k)+c(i,k)},
    \label{eq:localAmount}
\end{align}
which is $0$ where the mixture holds its slice-average composition and $\pm1$ where only one class or phase
is present. The local ratio is $p/c = (1+\chi)/(1-\chi)$, so $\chi=0.75$ is seven times more polymer
than crowder relative to the averages.

\renewcommand\refname{Supplemental References}
\makeatletter
\immediate\closeout\SI@auxfile
\def\NAT@bibsetnum#1{\setlength{\topsep}{\z@}\NATx@bibsetnum{\ref{LastBibItemSI}}}%
\def\endthebibliography{\auto@bib@innerbib\edef\@currentlabel{\arabic{NAT@ctr}}%
  \label{LastBibItemSI}\endNAT@thebibliography\aftergroup\auto@bib@empty}%
\IfFileExists{SIrefs.bbl}{
}{%
  \PackageWarningNoLine{main}{SIrefs.bbl not found: run `bibtex SIrefs'}}
\immediate\write\@auxout{\gdef\string\@extra@binfo{}}
\makeatother

\end{document}